\documentclass[12pt]{iopart}
\usepackage{iopams}

\expandafter\let\csname equation*\endcsname\relax
\expandafter\let\csname endequation*\endcsname\relax

\usepackage{amsmath}
\usepackage{amssymb,mathrsfs,framed,esint,slashed}
\usepackage{graphicx}
\usepackage[colorlinks]{hyperref}
\usepackage{color}
\usepackage[all]{xy}

\newcommand{\dee}{{d}}
\newcommand{\del}{d} 

\newcommand\setlist[1]{\left\{#1\right\}}
\newcommand\sign[1]{\mathrm{ sign}\,#1} 
\renewcommand\Im[1]{ { \mathrm{Im} \, #1} }

\newcommand\abs[1]{\left| #1 \right|}

\newcommand\avg[1]{\left< #1 \right>}

\newcommand\conjugate[1]{{{#1}}^*}

\newcommand\Operator[1]{{\widehat{#1}}}
\newcommand\Commutator[2]{\left[#1, #2\right]}
\newcommand\AntiCommutator[2]{\left\{#1, #2\right\}}

\newcommand\InnerProduct[2]{\left<#1 | #2 \right>  }

\newcommand\ket[1]{\left| #1 \right> }
\newcommand\braket[2]{\left< #1 \right.\left| #2 \right> }
\newcommand\BraKet[2]{\left<\left< #1 \right.\left| #2 \right>\right> }
\newcommand\ketbra[2]{\left| #1 \right>  \left< #2 \right| }
\newcommand\braOpket[3]{\left< #1 \left| #2 \right| #3 \right>}

\newcommand\PoissonBracket[2]{\left\{ #1 , #2 \right\}_{PB}}

\newcommand{\ceil}[1]{\lceil #1 \rceil}

\def\aindex{a}
\def\bindex{b}
\def\iindex{i}
\def\jindex{j}
\def\kindex{k}
\def\mindex{m}
\def\nindex{n}

\def\nuindex{\nu}

\def\kcoord{k}
\def\pcoord{p}
\def\tcoord{t}

\def\xcoord{x}
\def\ycoord{y}
\def\zcoord{z}

\def\Jcoord{J}

\def\phat{\hat p}

\def\CO{\mathcal{O}}

\def\LieDerivative{{\mathfrak L}}

\def\Naturals{\mathbb{N}}
\def\Integers{\mathbb{Z}}
\def\Reals{\mathbb R} 
\def\Complex{\mathbb C}

\def\UnitaryGroup{U}

\def\dim{d}

\def\Manifold{{\mathcal M}}
\def\ConfigurationSpace{{\mathcal M}}
\def\Manifold{\ConfigurationSpace}

\def\SC{\mathrm{SC}}
\def\mod{{\SC}}
\def\sc{{\rm sc}}

\def\CS{\mathrm{CS}}
\def\PS{\mathrm{PS}}

\def\mass{m}

\def\func{f}
\def\prob{f}
\def\pdf{f}

\def\Energy{E}

\def\action{J}
\def\angle{\theta}
\def\maslov{\mu}

\def\Gaussian{G}

\def\velocity{v}

\def\velocityPS{{\dot \zcoord}}

\def\WaveFunc{\Psi}

\def\WavePhase{S}
\def\WavePhaseEig{\tilde \WavePhase} 

\def\LagrangeForm{\alpha}
\def\LagrangeTensor{{\del \LagrangeForm}}

\def\Lag{ L }
\def\Ham{H}

\def\PoincareForm{  \alpha  }
\def\PoissonMatrix{\Pi}
\def\PoissonTensor{\boldsymbol{\PoissonMatrix}}
\def\RouthianFunc{W}
\def\RouthianForbidden{\overline \RouthianFunc}

\def\WaveAmp{R}

\def\waveFunc{\psi}
\def\waveConj{\conjugate{\waveFunc}}
\def\basisFunc{\phi}

\def\BasisFunc{\Phi}

\def\densityMatrix{\rho}

\def\phat{\hat \pcoord}
\def\qhat{\hat \xcoord}

\def\LVE{ {LVE}}

\def\Hhat{\hat \Ham}

\def\AOp{\Operator{A}}
\def\BOp{\Operator{B}}
\def\Propagator{U}

\def\Observable{O}
\def\ObservableOp{\Operator{\Observable}}

\def\KvHOp{\hat K}

\def\pOp{\Operator{\pcoord}}
\def\xOp{\Operator{\xcoord}}

\def\WaveActionPrinciple{{\mathcal A}}

\def\HamOp{\Operator{\Ham}}

\def\HamOp{\Operator{\Ham}}

\def\LieDerivative{{\mathfrak L}}

\def\TotalWavePhase{ \WavePhase}
\def\TotalWaveAmp{ \WaveAmp}
\def\SobolevSpace{\mathcal W}
\def\HilbertSpace{{\mathcal H}}
\def\SobolevSpace{W}
\def\HilbertSpace{{ H}}
\def\pdeOrder{m}

\begin{document}

\title[Semiclassical Theory and the Koopman-van Hove Equation ]{Semiclassical Theory and 
the Koopman-van Hove Equation}

\author{Ilon Joseph}

\address{Lawrence Livermore National Laboratory, Livermore CA, USA}
\ead{joseph5@llnl.gov}
\vspace{10pt}
\begin{indented}
\item[]\today
\end{indented}

\begin{abstract}
The phase space Koopman-van Hove (KvH) equation can be derived from the asymptotic semiclassical analysis of partial differential equations.
Semiclassical theory yields the Hamilton-Jacobi equation for the complex phase factor and the transport equation for the amplitude.
These two equations can be combined to form a nonlinear semiclassical version of the KvH equation in configuration space.
There is a natural injection of configuration space solutions into phase space and a natural projection of phase space solutions onto configuration space. 
Hence, every solution of the configuration space KvH equation satisfies both the semiclassical phase space KvH equation and the Hamilton-Jacobi constraint.
For configuration space solutions, this constraint resolves the paradox that there are two different conserved densities in phase space.
For integrable systems, the KvH spectrum is the Cartesian product of a classical and a semiclassical spectrum.
If the classical spectrum is eliminated, then, with the correct choice of Jeffreys-Wentzel-Kramers-Brillouin (JWKB)
matching conditions, the semiclassical  spectrum satisfies the Einstein-Brillouin-Keller quantization conditions which include the correction due to the Maslov index.
However, semiclassical analysis uses different choices for boundary conditions,  continuity requirements, and the domain of definition. 
For example, use of the complex JWKB method allows for the treatment of tunneling through the complexification of phase space.
Finally, although KvH wavefunctions include the possibility of interference effects, interference is not observable when all observables are 
approximated as local operators on phase space.
Observing interference effects requires consideration of nonlocal operations, e.g. through higher orders in the asymptotic theory. 
\end{abstract}

%
%
%
%
%

\section{Introduction \label{sec:intro} }

A number of authors have recently considered the evolution of Koopman wavefunctions through the Koopman-van Hove (KvH) equation on phase space  \cite{Bondar19prsa, Joseph20prr, Tronci21jpp}. 
The KvH equation is an improvement over the Koopman-von Neumann (KvN) equation in the sense that it describes the
nontrivial evolution of  
 the complex phase of the wavefunction.
In fact, the KvH phase evolution agrees with Feynman's prescription for the path integral.
One interesting application of the phase factor is to allow classical and quantum systems to be coupled together self-consistently \cite{Bondar19prsa}.
Another interesting application is the ability to use quantum computers to efficiently simulate the evolution of nonlinear dynamics by simulating the linear evolution of the KvH wavefunction \cite{Joseph20prr, Joseph23pop}.  
In fact, this work was originally motivated by the question of how to develop quantum algorithms  for simulating semiclassical systems.

In this work, the precise relation between the semiclassical theory of quantum mechanics \cite{LandauQMBook, Maslov81book, Guillemin90Book, Littlejohn92jsp, BatesBook, CvitanovicChaosBook, HallBook} and the classical ``pre-quantum'' Koopman-van Hove (KvH) equation \cite{Bondar19prsa, vanHovePhD, Kostant72preprint} will be derived.
We first show that a nonlinear $\dim$-dimensional configuration space version of the KvH equation can be derived from Jeffreys-Wentzel-Kramers-Brillouin (JWKB) semiclassical
theory.  
We then derive a  linear $2\dim$-dimensional semiclassical version of the phase space KvH equation that generates solutions for the configuration space version.
Finally, we prove that if the classical spectrum is eliminated and the correct JWKB matching conditions are applied, then the phase space KvH eigenfunctions reproduce the semiclassical spectrum.
Thus, in a certain sense, the semiclassical KvH equation simultaneously contains both classical and semiclassical physics.

For JWKB analysis \cite{LandauQMBook, SakuraiBook}, one considers the limit in which Planck's constant  vanishes, $\hbar\rightarrow 0$, but still employs a small but finite value of Planck's constant to obtain the final result.
In this case, there is still a clear notion of superposition of states, interference effects, quantization of eigenvalues, and tunneling through classically forbidden regions.
However, it is well known that JWKB solutions typically have an essential singularity in $\hbar$ and that taking this limit is perhaps better stated as considering the limit in which the quantum numbers of the semiclassical system become large, so that there are rapid phase oscillations.
As $\hbar\rightarrow 0$, the complex phases vary more and more rapidly, which makes the method of stationary phase quite accurate, but the limit does not exist in the usual sense.
Because the JWKB approximation reduces the order of the PDE, the result is an asymptotic perturbation expansion for the solution.
In this work, we explore the hypothesis that the first two orders of this asymptotic expansion define the ``semiclassical limit.''

According to semiclassical theory, at lowest order in Planck's constant $\hbar$, the complex phase of the wavefunction is given by $\WavePhase/\hbar$, where the action, $\WavePhase$, must satisfy the Hamilton-Jacobi equation (HJE, Eq.~\ref{eq:HJE}), which is analogous to the eikonal equation of geometric optics.
At the next order, the probability density in configuration space must satisfy a configuration space version of the Liouville equation (LVE, Eq.~\ref{eq:LVE_config-space}), sometimes referred to as the Vlasov or collisionless Boltzman equation.
The LVE implies the configuration space amplitude transport equation (ATE, Eq.~\ref{eq:ATE_config-space}) for the wave amplitude.
A configuration space version of the KvH equation (Eq.~\ref{eq:KvH_config-space}) can be derived by combining the HJE and ATE into a single equation for the complex  wavefunction. 
This semiclassical KvH equation is naturally posed in configuration space, and, hence, is closely related to semiclassical JWKB theory.
The semiclassical KvH equation generates a unitary evolution of the wavefunction, but it has an intrinsically nonlinear dependence on the complex phase factor. 

While the natural domain of the wavefunction is the $\dim$-dimensional configuration space, labeled by coordinates, $\xcoord$, 
consideration of the initial data required to solve the HJE, which requires one to specify the gradient of the phase function, motivates the introduction of the canonically conjugate momentum coordinates, $\pcoord =\partial_\xcoord\WavePhase$, and an extension of the domain to the full $2\dim$-dimensional phase space.
On phase space, the Lagrangian acts as a source of action for the phase (HJE, Eq.~\ref{eq:HJE_phase-space}).
If the density is advected along the flow via the usual LVE (Eq.~\ref{eq:LVE_phase-space}), this implies that the amplitude is advected via the phase-space ATE (Eq.~\ref{eq:ATE_phase-space}).
Combining these equations leads to the standard phase space KvH equation (Eq.~\ref{eq:KvH_phase-space}) for the complex wavefunction.
The phase space KvH equation is both linear and unitary.
A modified phase space KvH equation  (Eq.~\ref{eq:KvH_phase-space_modified})  generates configuration space solutions after integration over momentum space.  
While this equation is not unitary on phase space, the dynamics it generates  are unitary on configuration space.
Another point of view is that integration of the solutions of the standard phase space KvH equation over momentum space with a specially chosen weight (Eq.~\ref{eq:KvH_phase-space_projection}) generates solutions of the configuration space version.
Moreover, if one is willing to introduce the ``square-root of the delta function,'' as in 
\ref{sec:square_root_delta}, then, as shown in 
\ref{sec:alternate_mapping_norm}, to lowest order, the standard phase space KvH eigenfunctions reproduce the configuration space results when using the \emph{phase space inner product}.

The propagators for the configuration space and phase space KvH equations have a natural correspondence with each other.
However, like traditional JWKB theory, the approximations that lead to the configuration space version of the KvH equation become invalid near caustics and classical turning points, where the configuration space Jacobian becomes singular.
Although the standard phase space version of KvH suffers no problems in these regions, the configuration space and phase space KvH solutions have different dependence on the configuration space Jacobian that ultimately leads to different quantization conditions.

The solutions of the configuration space KvH equation
satisfy the  semiclassical phase space KvH equation and vice-versa as long as the phase factor satisfies the Hamilton-Jacobi constraint.
Hence, for any initial condition that satisfies the Hamilton-Jacobi constraint, the phase space and configuration space solutions are in one-to-one correspondence.
For example, semiclassical eigenfunctions satisfy this constraint because they are confined to Lagrangian submanifolds that are invariant under temporal evolution \cite{BatesBook}.
In general, there are multiple branches or ramifications of Lagrangian submanifolds that must be glued together in order to satisfy the boundary conditions, e.g. glued along interfaces defined by the caustics of the PDE.
In phase space, the Hamilton-Jacobi equation is 
satisfied along each branch of the Lagrangian submanifold.
In configuration space, the eigenfunction is a superposition of these different branches of the Lagrangian submanifolds, and the Hamilton-Jacobi equation approximately holds, within the stationary phase approximation.
Satisfaction of the boundary conditions leads to well-known semiclassical ``quantum'' effects such as the 
quantization conditions and tunneling through classically forbidden regions.

The semiclassical analysis provides an answer to the interesting paradox first observed by Refs.~\cite{Bondar19prsa, Joseph20prr, Tronci21jpp} as to how to interpret the fact that, in phase space, the charge  density differs from wave action  density, while, in configuration space, all such densities are in agreement. 
Because semiclassical wavefunctions are composed of a superposition of solutions that always satisfy the Hamilton-Jacobi constraint, the difference between these current densities vanishes identically.

For integrable dynamics, the KvH spectrum is the Cartesian product of the classical spectrum and a semiclassical approximation to the quantum spectrum. 
The classical spectrum corresponds to  amplitude transport while the 
semiclassical spectrum corresponds to phase evolution. 
Note, however, that in order to satisfy the constraint that the amplitude varies more slowly than the phase, semiclassical theory typically enforces the time-independent ATE. 
In fact, it is more accurate to say that the asymptotic form of the wavefunction is completely determined by the HJE and the Van Vleck determinant \cite{Froman65book}.
In this case, the parts of the spectra associated with the ATE are eliminated, which results in the usual semiclassical JWKB spectrum.
With this restriction, the JWKB eigenstates are no longer complete over phase space, but are complete over configuration space.
For the standard phase space KvH equation, the semiclassical spectrum  satisfies the Bohr-Sommerfeld quantization conditions.
For the semiclassical KvH equation, the semiclassical spectrum  satisfies the Einstein-Brillouin-Keller (EBK) quantization conditions, if the JWKB connection formulae are used.

For continuous initial data, the solutions of the original PDE evolve continuously in time. 
If the approximate JWKB solutions of the phase space KvH equation are to maintain continuity in space, then the classical trajectories must be allowed to enter classically forbidden regions.
This can be accomplished using the techniques of \emph{complex JWKB theory} \cite{Froman65book, Knoll76ap, Maslov94book}  
and \emph{ exact WKB theory} \cite{Balian74, Voros83apt, Sueishi20jhep} which allow both the phase and amplitude of the wavefunction to vary rapidly at lowest order, to analytically continue the phase space trajectories into the complex domain.
This is simply related to the traditional method of using analytic continuation  to derive JWKB connection formulae that allow the wavefunction to tunnel through classically forbidden regions.
Thus, the semiclassical KvH problem differs from the classical KvH problem in using different boundary conditions, demanding continuity in configuration space, and in the fact that the natural domain is actually the complexification of phase space.
This gives the semiclassical version the ability to describe tunneling effects.

The next section derives the semiclassical KvH equation in both configuration space and phase space as well as  the  mapping from phase space to configuration space.
Then, in Section~\ref{sec:KvH_analysis}, we analyze the propagator
of the KvH equation and the spectrum of eigenvalues and eigenfunctions for integrable systems.
Section~\ref{sec:semiclassical_analysis} reviews semiclassical analysis   in order to explain the similarities and differences with the KvH equation.
Section~\ref{sec:examples} illustrates the basic principles with a review of the JWKB solutions of the harmonic oscillator in both phase space and configuration space, as well as the generalization to a potential well.

Finally, Sec.~\ref{sec:interference} discusses the possibility of observing interference between KvH eigenfunctions. 
Solutions of the KvH equation can potentially display interference effects in the sense that one solves for the wavefunction rather than for the probability distribution function (PDF).
\emph{However, for the lowest order asymptotic theory, where all observables are treated as local in phase space, interference effects are not observable.}
This is because, when all observables are local in phase space, there is a unique basis in which the laws of classical probability apply to all observables.
In order to observe interference effects, one must observe nonlocal operators, e.g. by going to higher order in perturbation theory, so that the nonlocal properties of observables such as the momentum operator can be observed.
Conclusions are summarized in the final section.

\section{Semiclassical KvH Theory \label{sec:semiclassical_KvH}}

In this section, the derivation of JWKB theory from asymptotic perturbation theory is reviewed for any linear self-adjoint  partial differential equation (PDE) that is first order in time.
Let the PDE be specified on a $\dim$-dimensional configuration space manifold, $\Manifold$, with local coordinates $\xcoord =\setlist{ \xcoord^i} $ in each chart.
Partial derivatives with respect to $\xcoord$ will be denoted
$\partial_\xcoord:=\setlist{\partial/\partial\xcoord^i}$.
In addition, define the position operator $\xOp$ as multiplication by $\xcoord$ and the  conjugate momentum operator $\pOp =-i\hbar \partial_\xcoord$ where $\hbar$ is Planck's constant. 
Extension of these arguments to higher order in time is easily performed by appending the time coordinate to the spatial coordinates using the extended phase space formalism, but is an unnecessary complication for the discussion here.

Consider the following action for a complex scalar field $\waveFunc:\Manifold \rightarrow \Complex$
\begin{align} \label{eq:quantum-action-integral}
\WaveActionPrinciple[\waveFunc,\waveConj;\tcoord,\xcoord]:= -\int\left[\waveConj \HamOp(\tcoord, \xOp,\pOp)\waveFunc + \Im { \hbar \waveConj  \partial_\tcoord \waveFunc} \right]  d^\dim\xcoord d\tcoord
\end{align}
where the Hamiltonian operator $\HamOp(\tcoord, \xOp,\pOp)$ is assumed to be Hermitian.
Due to the canonical commutation relations, $\Commutator{\xOp}{\pOp}=i\hbar$, it is important to explicitly specify the operator ordering in the definition of the Hamiltonian operator $\HamOp(\tcoord,\xOp,\pOp)$. 
The variational principle requires the solution to be an extremum of the action, so that the Fr\'echet derivative, $\delta \WaveActionPrinciple/\delta \waveFunc^*=0$, vanishes.
This leads to the generalized Schr\"odinger equation 
\begin{align}
i\hbar\partial_\tcoord  \waveFunc(\tcoord,\xcoord)  = \HamOp(\tcoord, \xOp,\pOp) \waveFunc(\tcoord,\xcoord)  .
\end{align}
If the Hamitonian is an $\pdeOrder$-th order differential operator in space, then solutions are typically sought in a Sobolev space, $\HilbertSpace^k(\Manifold):=\SobolevSpace^{k,2}(\Manifold)$, where  $k=\ceil{\pdeOrder/2}$ weak derivatives of the solution exist and have finite $L^2$ norm.
Hence, the space of solutions is contained within Hilbert space, $\HilbertSpace^k\subset\HilbertSpace^0=L^2 $.

\subsection{Configuration Space KvH \label{sec:KvH_config-space}}

The JWKB or eikonal approximation consists of considering the wavefunction to have an asymptotic perturbation expansion of the form 
\begin{align} \label{eq:asymptotic-expansion}
\waveFunc(\tcoord,\xcoord)=\exp{\left(i\epsilon^{-1} \sum_{\jindex=0}^\infty  \TotalWavePhase_\jindex(\tcoord,\xcoord) \epsilon^\jindex/\hbar \right)}  
\simeq e^{i \left(\TotalWavePhase_0 /\epsilon  + i \TotalWavePhase_1   +  i \epsilon \TotalWavePhase_2 + \dots  \right)/\hbar} \simeq  e^{ i  \TotalWavePhase_0 /\epsilon\hbar   } \TotalWaveAmp_1  \TotalWaveAmp_2 \cdots.
\end{align}
Here, each term $\TotalWaveAmp_j:=\exp{\left( i \epsilon^{j}\TotalWavePhase_{j }/\epsilon \hbar\right)}$, is considered to be complex and independent of the others.
Derivatives are assumed to be weak, of order $\epsilon$, so that the lowest order expression for the $\pdeOrder$-th order derivative of the wavefunction is 
\begin{align}\label{eq:asymptotic-derivatives}
(\epsilon \phat)^m\waveFunc \simeq  (\partial_x \TotalWavePhase_0)^m \waveFunc
+\dots = \pcoord^m \waveFunc+\dots
\end{align}
where the gradient of the lowest order term for the phase, $\pcoord:=\partial_x\TotalWavePhase_0$, defines the ``classical momentum.'' 
The action of the Hamiltonian on the asymptotic expansion generates a perturbation series for the generalized Schr\"odinger equation that can then be solved order by order in $\epsilon$.
The first two orders, explicitly derived in 
\ref{sec:semiclassical_derivation}, are the Hamilton-Jacobi equation (HJE) and the amplitude transport equation (ATE) \cite{VanVleck1928pnas}.
In the following, the phase and amplitude refer to the two lowest order terms in the asymptotic expansion $\WavePhase:=\TotalWavePhase_0$ and $\WaveAmp:=\TotalWaveAmp_1=\exp{(i\TotalWavePhase_1/\hbar)}$, 
but recall that these are complex functions.
Since we shall only work to
lowest order in $\epsilon$, to ease the notation, we now set $\epsilon=1$.

The most expedient route to deriving the first two leading order equations   is to substitute the asymptotic approximation into the variational principle in Eq.~\ref{eq:quantum-action-integral}.
For the purpose of using the variational principle to derive the HJE and ATE, $\WavePhase$ and $\WaveAmp$ are assumed to be real functions.
 After the derivation is performed, they are analytically continued to the complex domain in order to treat tunneling regions \cite{Balian74, Knoll76ap, Voros83apt, Maslov94book, Sueishi20jhep}.
As shown in 
\ref{sec:semiclassical_derivation}, substituting 
the zeroth order expression, $\psi=\WaveAmp \exp(i \WavePhase/\hbar)$, into the variational principle and only retaining the lowest order term for the derivative $\phat^m\waveFunc \simeq    \pcoord^m \waveFunc +\dots $, yields the semiclassical action
\begin{align} \label{eq:config-space_action}
\WaveActionPrinciple[\waveFunc,\waveConj;\tcoord,\xcoord]:=- \int\left[  \abs{\waveFunc}^2 \Ham(\tcoord, \xcoord,\partial_\xcoord \WavePhase)
+\Im {\hbar \waveConj \partial_\tcoord \waveFunc}
\right]  d^\dim\xcoord d\tcoord =
- \int  \left( \Ham +\partial_\tcoord \WavePhase\right)
 \WaveAmp^2  d^\dim\xcoord d\tcoord
.
\end{align}
Here, the classical Hamiltonian is defined as the leading order term that results from simply replacing the $\xOp$ and $\pOp$ operators with their classical commuting expressions:  $\Ham(\tcoord,\xcoord,\pcoord) :=\HamOp(\tcoord,\xcoord,\pcoord)$.
As shown in 
\ref{sec:semiclassical_derivation}, because we are only considering the two lowest orders of the asymptotic expansion, higher order corrections due to operator ordering considerations are not required.
Note that this Hamiltonian is highly nonlinear in the phase of the wavefunction due to the dependence on $ \pcoord(\tcoord,\xcoord)=\partial_\xcoord \WavePhase(\tcoord,\xcoord)$. 
Variations are to be taken with respect to $\waveFunc$, considered to be a complex function, which is equivalent to taking the variation with respect to  and $\waveFunc$ and $\waveConj$ independently.
Because $\WavePhase$ and $\WaveAmp$ represent 
independent degrees of freedom, variations must be 
taken with respect to these quantities, independently.

Due to the asymptotic expansion, the action integral only depends on the first derivatives of $\waveFunc$ in space and time. 
(Higher order terms, such as the Bohm potential for Schr\"odinger's equation, would need to be retained in order to reproduce the original action integral in Eq.~\ref{eq:quantum-action-integral}.)
In order for the asymptotic expansion of the action integral (Eq.~\ref{eq:config-space_action}) to be finite, solutions for $\WaveAmp$ should be sought within the Sobolev space, $\SobolevSpace^{1,2}=\HilbertSpace^1 \subset \HilbertSpace^0= L^2$, and, 
if the original Hamiltonian operator is $\pdeOrder$-th order in spatial derivatives, then solutions for the phase should be sought so that each component of $\partial_\xcoord \WavePhase$ is an element of $\SobolevSpace^{0,\pdeOrder}$. 
Yet, in Sec.~\ref{sec:KvH_analysis} it will be shown that an explicit propagator can be constructed for the resulting PDE that significantly widens the range of solutions to essentially arbitrary initial data.

\subsubsection{Configuration Space KvH Derivation \label{sec:derivation_config-space}}
Taking the variation of the action in Eq.~\ref{eq:config-space_action} with respect to $\abs{\waveFunc}^2 $ yields the nonlinear Hamilton-Jacobi equation (HJE)
\begin{align} \label{eq:HJE}
\Energy = -\partial_\tcoord \WavePhase = \Ham(\tcoord,\xcoord,\pcoord)
\end{align}
where the Hamilton-Jacobi constraint is
\begin{align} \label{eq:HJE-momentum}
\pcoord=\partial_\xcoord \WavePhase.
\end{align} 
Taking the variation of the action with respect to $\WavePhase$ yields the Liouville-Vlasov equation (LVE)
\begin{align}   \label{eq:LVE_config-space}
\partial_\tcoord \abs{\waveFunc}^2  + \partial_\xcoord\cdot \left(  \abs{\waveFunc}^2 \partial_{\pcoord } \Ham\right)= 0 
\end{align}
for the probability distribution function $\pdf=\abs{\waveFunc}^2$.
 Partial derivatives with respect to $ \pcoord =\setlist{ \pcoord_\iindex}$ are denoted $\partial_\pcoord := \setlist{ \partial/\partial \pcoord_\iindex }$.
The inner product symbol denotes a contraction such as $ \pcoord\cdot \xcoord:=\pcoord_\iindex \xcoord^\iindex $, with the Einstein summation convention implied.
Defining the group velocity, $\dot\xcoord:=\partial_\pcoord \Ham$, the LVE can be rewritten as 
\begin{align}
-(d/d\tcoord)^\dagger\pdf = \partial_\tcoord \pdf  + \partial_\xcoord\cdot (\dot\xcoord \pdf) =0
\end{align}
  where, here,  $(d/d\tcoord)^\dagger$ is defined as the adjoint of $d/d\tcoord=\partial_\tcoord+\dot \xcoord\cdot \partial_\xcoord$ over the space-time measure $\int d\tcoord d^\dim\xcoord$.
The amplitude transport equation (ATE) follows directly from the LVE \cite{Joseph20prr}
\begin{align}   \label{eq:ATE_config-space}
\partial_\tcoord \WaveAmp  + \tfrac{1}{2}\AntiCommutator{ \partial_\xcoord }{\,\cdot \, \,  \partial_{\pcoord}\Ham } \WaveAmp= 0 
\end{align}
where 
\begin{align}
\AntiCommutator{\AOp}{\BOp}=\AOp\BOp +
\BOp\AOp
\end{align} 
 is the anticommutator.
While some authors might also refer to the configuration space version of the LVE as the amplitude transport equation, here we are emphasizing the distinction between the transport of different geometric objects that represent probability densities vs. probability amplitudes, as well as the close relation between the configuration space and phase-space formulations of KvH. 

The  characteristic curves of the LVE are defined by the group velocity  
\begin{align} \label{eq:dxdt}
\dot \xcoord :=  d\xcoord /d\tcoord=  \left.  \partial_\pcoord  \Ham   \right|_{\tcoord,\xcoord}
.
\end{align} 
If the velocity, $\partial_\pcoord  \Ham(\tcoord,\xcoord,\pcoord)$, was independent of $\pcoord=\partial_x\WavePhase$, then this would be an ODE that can simply be integrated along each trajectory. 
However,  the velocity generically depends on $\pcoord$, and, hence, on the phase function. Thus, in order to solve for the evolution of the configuration space trajectory in time, one must also specify how $\pcoord=\partial_\xcoord \WavePhase(\tcoord,\xcoord)$ evolves in time.

The evolution of the momentum field $\pcoord$ is completely determined by the Hamilton-Jacobi constraint $\pcoord=\partial_\xcoord \WavePhase$.
The result is straightforward to derive   \cite{CvitanovicChaosBook}
\begin{align}
d\pcoord/d\tcoord &= \partial_\tcoord \pcoord +  \dot \xcoord \cdot \partial_\xcoord\pcoord 
			= -\left(\partial_\xcoord\Ham + \partial_\pcoord\Ham\cdot \partial_\xcoord\pcoord \right) +  \dot \xcoord \cdot \partial_\xcoord\pcoord
\end{align}
using the HJE and the fact that $\left. \partial_\xcoord\Ham\right|_\tcoord=\left. \partial_\xcoord\Ham\right|_{\pcoord,\tcoord}+\left. \partial_\pcoord\Ham\right|_{\xcoord,\tcoord} \cdot \partial_\xcoord\pcoord $.
Hence, along the trajectories defined by the group velocity, the rate of change of momentum   is determined by   the generalized forces
\begin{align} \label{eq:dpdt}
\dot \pcoord := d\pcoord/d\tcoord = -\left. \partial_\xcoord\Ham  \right|_{\tcoord,\pcoord} .
\end{align}
Now we have arrived at Hamilton's equations of motion (Eqs.~\ref{eq:dxdt} and \ref{eq:dpdt}) which are necessary to completely specify the characteristics of the LVE and ATE. 
Hamilton's equations can be solved by integrating the equations along each trajectory independently for any initial condition, $\xcoord_0:=\xcoord(\tcoord=0)$ and $\pcoord_0:=\pcoord(\tcoord=0)$.
Thus, in the context of semiclassical theory, Hamilton's equations are not just ODEs on phase space that determine the classical trajectories, they are  generated by PDEs on configuration space that determine the amplitude and phase of the wavefunction.
Note that this also implies that the trajectory remains on the dispersion surface \cite{TracyBook} defined by $\Energy(\tcoord) = \Ham(\tcoord,\xcoord,\pcoord)$, so that
\begin{align}
d\Ham/d\tcoord =\partial_\tcoord \Ham + \dot \xcoord \cdot \partial_\xcoord \Ham + \dot \pcoord \cdot \partial_\pcoord \Ham = \partial_\tcoord\Ham.
\end{align}

Along the trajectory, the phase evolves as a scalar field with a source
\begin{align}
\left. d\WavePhase/d\tcoord\right|_{\xcoord_0}&:=\left. \partial_\tcoord \WavePhase \right|_{\xcoord} +  \dot \xcoord  \cdot \partial_\xcoord\WavePhase.
\end{align}
Substitution of the Hamilton-Jacobi relations yields
\begin{align}\label{eq:HJE_config-space} 
\left(d/d\tcoord\right) \WavePhase= -\Ham +\dot \xcoord  \cdot \pcoord
= \pcoord \cdot \partial_\pcoord \Ham  -\Ham=: \Lag_\Ham
\end{align}
where $\Lag_\Ham(\tcoord,\xcoord,\pcoord)$ is the Lagrangian associated with the Hamiltonian, $\Ham(\tcoord,\xcoord,\pcoord)$.
This becomes a function of configuration space after substitution of the Hamilton-Jacobi constraint $\pcoord=\partial_\xcoord \WavePhase$.
Given the conservation of wave action in Eq.~\ref{eq:LVE_config-space}, the conservation of wave amplitude  satisfies the standard Koopman-von Neumann (KvN) equation
\begin{align} \label{eq:KvN}
\tfrac{1}{2} \left(d/d\tcoord-d/d\tcoord^\dagger\right) \WaveAmp=\partial_\tcoord \WaveAmp+ \tfrac{1}{2} \AntiCommutator{\partial_{\xcoord }} {\, \cdot \,\,\dot \xcoord }   \WaveAmp  =0.
\end{align}
This equation states that the amplitude, and, therefore, the wavefunction $\waveFunc$ itself, transforms as the square root of a probability density, called a \emph{half-form} in the language of geometric quantization \cite{HallBook}.
If the temporal evolution of the coordinates along the trajectories is specified as 
$\xcoord=\xi(\tcoord;\tcoord_0,\xcoord_0,\pcoord_0)$, then a half-volume form transforms as the square root of the  determinant of the coordinate transformation
$\WaveAmp(\tcoord,\xcoord)=\WaveAmp(\tcoord_0,\xcoord_0)\left(\det{ \partial \xcoord_0/\partial \xcoord} \right)^{1/2}$.

 The Lie derivative of a half form is defined as 
\begin{align} \label{eq:Lie-derivative_half-form}
 \LieDerivative_\velocity \waveFunc := 
 \tfrac{1}{2} \AntiCommutator{\partial_{\xcoord^\jindex} }{ \velocity^\jindex }   \waveFunc
 = \tfrac{1}{2} \AntiCommutator{\partial_{\xcoord } }{ \, \cdot \,\, \velocity  }   \waveFunc.
\end{align}
 Combining the amplitude and phase equations yields the semiclassical {\bf configuration space  KvH  equation}
\begin{align} \label{eq:KvH_config-space} 
 \tfrac{1}{2}i\hbar \left(d/d\tcoord-d/d\tcoord^\dagger\right) \waveFunc=
i\hbar\left(\partial_\tcoord + \LieDerivative_{\dot \xcoord} \right) \waveFunc=i\hbar \left( \partial_\tcoord + \tfrac{1}{2} \AntiCommutator{\partial_{\xcoord}  }{\, \cdot \,\, \dot \xcoord  }\right)  \waveFunc  =-\Lag_\Ham  \waveFunc.
\end{align}
Introducing the configuration space \emph{Kostant-van Hove Hamiltonian} operator
\begin{align} \label{eq:KvH-Ham_config-space} 
\KvHOp^\CS_\Ham   :=-i\hbar \LieDerivative_{\dot \xcoord}- \Lag_\Ham   
\end{align}
the configuration space KvH equation simply becomes  
\begin{align} \label{eq:KvH-Schrodinger_config-space} 
i\hbar \partial_\tcoord \waveFunc = \KvHOp^\CS_\Ham  \waveFunc.
\end{align}
This  form is similar to the Schr\"odinger equation in the sense that it results from a Hermitian Hamiltonian operator.  
However, because this equation is first order in both space and time, it is much more closely related to the HJE and LVE in its mathematical properties than the usual Schr\"odinger equation, which is first order in time but second order in space. 
The semiclassical analysis of the KvH equation is similar to standard JWKB theory and suffers from the same issues.
The approximations that lead to the derivation of the KvH equation break down near caustics generated by classical turning points.

\subsubsection{General Configuration Space Ansatz \label{sec:general_config-space}}

The semiclassical ansatz breaks down whenever there is a classical turning point in the region of interest.
Hamilton's equations of motion specify the evolution of the phase space coordinates, $\zcoord=\setlist{\xcoord,\pcoord}$, as a function of time, starting from the initial data for the position, $\xcoord_0=\xcoord(\tcoord=0)$, and for the gradient of the phase, $\pcoord_0=\partial_\xcoord \WavePhase (\tcoord=0,\xcoord)$.
Given the solution, $\zcoord=\zeta(\tcoord;\tcoord_0,\zcoord_0)$, one can directly integrate the change in action along a trajectory to find 
\begin{align}
\Delta \WavePhase(\tcoord; \tcoord_0,\xcoord_0,\pcoord_0)=\int  \Lag_\Ham(\tcoord,\zeta(\tcoord;\tcoord_0,\zcoord_0)) d\tcoord.
\end{align}
However, when there are multiple initial values for $\pcoord_0$ that lead to the same final position $\xcoord=\xi(\tcoord; \tcoord_0,\zcoord_0)$, then the original semiclassical ansatz, which assumes that there is a single phase function, $\WavePhase$, must break down.

Although the original PDE was linear, the asymptotic representation in configuration space is nonlinear and leads to the nonlinear Hamilton-Jacobi equation. In order to retain the property that any superposition of semiclassical solutions is again a solution, one must postulate an alternate form of the wavefunction as a sum over solution branches, indexed by $\Jcoord$.
While $\Jcoord$ is often treated as if it is discrete, in general, the freedom in choosing $\Jcoord$ represents the ability to choose an arbitrary initial condition for the gradient of the phase, $\pcoord_0=\partial_\xcoord \WavePhase(\tcoord=0,\xcoord,\Jcoord)$.
Hence, if the configuration space manifold has dimension $\dim$, then the $\Jcoord$ index can take values within the conjugate momentum space, a real manifold of dimension $\dim$. 
Moreover, if the original PDE is $\pdeOrder$-th order in space, then the Hamiltonian is a $\pdeOrder$-th order polynomial in $\pcoord=\partial_\xcoord\WavePhase$, so there are generically $\pdeOrder$  distinct roots for the Hamilton-Jacobi constraint, labeled by $\sigma$.
Thus, there are $\pdeOrder$ different solution branches that potentially contribute to the configuration space solution.
In phase space, each of these branches represents a different branch of the Lagrangian submanifold that makes an independent contribution in configuration space.
This is because, for each branch or ramification of the submanifold, the momentum is locally expressed as a function of configuration space coordinates.
Now the ansatz for the wavefunction has the form
\begin{align} \label{eq:wavefunction_branches}
\waveFunc(\tcoord,\xcoord) = \sum_\sigma \int \WaveAmp_\sigma(\tcoord,\xcoord, \Jcoord) e^{i\WavePhase_\sigma(\tcoord,\xcoord, \Jcoord)} d^\dim\mu[\Jcoord]
\end{align}
where $d^\dim\mu[\Jcoord]$ represents an appropriate spectral measure \cite{ReedSimonBookI} in $\Jcoord$ associated with the relevant eigenproblem and $\sigma$ indexes the different branches of the Lagrangian submanifold at each point in $\xcoord$.
To ease the notation, we will often suppress the $\sigma$ indices, but indicate the sum where necessary.

As described more fully in Sec.~\ref{sec:KvH_propagator}, the fact that Hamilton's equations uniquely specify the evolution of the trajectories, given the initial phase, allows one to uniquely describe the propagation of the wavefunction forward in time.
 Hamilton's equations generate a symplectic coordinate transformation and the change in action is actually a mixed variable generating function, $\WavePhase(\tcoord,\xcoord,\pcoord_0)$, that determines how canonical coordinates on phase space transform as time progresses \cite{LanczosBook,GoldsteinBook,LichtenbergBook}.
The evolution of the conjugate variables are determined via $\pcoord=\partial_\xcoord\WavePhase$ and $\xcoord_0=\partial_{\pcoord_0} \WavePhase$.
Because the wavefunction is a half-form in configuration space, the amplitude transforms as a scalar times  the square root of the Van Vleck determinant \cite{VanVleck1928pnas}
\begin{align}
\WaveAmp(\tcoord,\xcoord)/\WaveAmp(\tcoord_0,\xcoord_0) =\left(\det \left. \partial \xcoord_0/\partial \xcoord\right|_{\pcoord_0} \right)^{1/2}
= \left(\det \left. \partial \pcoord/\partial \pcoord_0 \right|_{\xcoord}\right)^{1/2}  
= \left(\det \partial^2\WavePhase/\partial \pcoord_0 \partial \xcoord \right)^{1/2} .
\end{align}

The general configuration space ansatz in Eq.~\ref{eq:wavefunction_branches} is motivated by the fact that a \emph{complete solution} of the Hamilton-Jacobi equation is required in order to completely solve Hamilton's equations of motion \cite{LanczosBook, GoldsteinBook, LichtenbergBook}.
The complete solution depends on a $\dim$-dimensional set of constants of the motion, $\Jcoord$.
In this case, the action, $\WavePhase(\tcoord, \xcoord,\Jcoord)$, can be viewed as a generating function for a symplectic transformation of phase space to \emph{action-angle coordinates}, $\setlist{ \Jcoord,\angle }$, where the Hamiltonian, $\Ham(\Jcoord)$, depends on $\Jcoord$ alone. 
Hence, the angle evolves linearly in time $\angle=\angle_0 +\omega\tcoord$ where $\omega=\partial_\Jcoord\Ham$.
In this case, the amplitude factor transforms as the square root of the Van Vleck determinant 
\begin{align} \label{eq:VanVleck_determinant}
\WaveAmp(\tcoord,\xcoord,\Jcoord) /a(\xcoord_0,\Jcoord) 
= \left(\det \left.\partial\angle/\partial \xcoord \right|_{\Jcoord} \right)^{1/2} 
 =  \left(\det \left. \partial\pcoord/\partial \Jcoord \right|_\xcoord \right)^{1/2}
  =  \left(\det   \partial^2 \WavePhase/\partial \Jcoord \partial \xcoord   \right)^{1/2}
\end{align}
where $a(\xcoord_0,\Jcoord)$ is a scalar function.

When there are multiple branches, one can still use Eq.~\ref{eq:config-space_action} to generate the same equations of motion, if one uses the stationary phase approximation to eliminate cross-terms between different branches.  
After using the stationary phase approximation, the action becomes
\begin{align} \label{eq:config-space_action_generalized}
\WaveActionPrinciple[\waveFunc,\waveConj;\tcoord,\xcoord]:=- \sum_\sigma \int \WaveAmp_\sigma(\tcoord,\xcoord,\Jcoord)^2\left[  \Ham(\tcoord, \xcoord,\partial_\xcoord \WavePhase_\sigma(\tcoord,\xcoord,\Jcoord) )
+ \partial_\tcoord \WavePhase_\sigma (\tcoord,\xcoord,\Jcoord)
\right]  d^\dim\xcoord d^\dim\mu[\Jcoord] d\tcoord
.
\end{align}
This is equivalent to assuming that the wavefunction is constrained to have the form in Eq.~\ref{eq:wavefunction_branches} and then using the variational principle in Eq.~\ref{eq:config-space_action} for each term independently.

\subsection{Standard Phase Space KvH \label{sec:KvH_phase-space} }

Phase space is defined as the cotangent bundle, 
$\mathcal{ T^* M}$, of dimension $2\dim$, associated to the configuration space manifold.
In any local region of phase space, one can define canonical coordinates, $\zcoord^\aindex =\setlist{\xcoord^\iindex, \pcoord_\iindex}$, where $\xcoord\in \Manifold$ and $\pcoord\in \mathcal{ T^* M}_\xcoord$.
As explained in the previous section, for a complete solution of the Hamilton-Jacobi equation, both the phase, $\WavePhase$, and the amplitude, $\WaveAmp$, must actually be considered to be functions of phase space.  
In addition to configuration space coordinates, $\xcoord$, they must depend on a  $\dim$-dimensional functionally independent set of constants of the motion, $\Jcoord(\tcoord,\zcoord )$, such as the initial conditions $\xcoord_0$ or $\pcoord_0$, or a set of adiabatic invariants. 
For a complete solution, the phase, amplitude, and wavefunction become phase space functions by substitution, $\WavePhase(\tcoord,\zcoord):=\WavePhase(\tcoord,\xcoord, \Jcoord(\tcoord,\zcoord))$ and $\WaveAmp(\tcoord,\zcoord):=\WaveAmp(\tcoord,\xcoord, \Jcoord(\tcoord,\zcoord))$.
For a solution in the form of the general configuration space ansatz (Eq.~\ref{eq:wavefunction_branches}), for each branch of the phase, there is an inclusion map defined by the Hamilton-Jacobi constraint, $\pcoord=\partial_\xcoord\WavePhase$, that embeds the configuration space solution within phase space.

The fact that the KvH PDE on configuration space can also be interpreted as a PDE on phase space is straightforward to understand.
Because the KvH equation is defined through a first-order differential operator in space and time, the characteristics of this operator should uniquely specify the evolution. 
However, the velocity in configuration space, $\dot\xcoord=\partial_\pcoord\Ham$, depends on the gradient of the phase, $\pcoord=\partial_\xcoord\WavePhase$, and, hence, the equations of motion for the characteristics cannot be closed without also specifying $\dot\pcoord=-\partial_\xcoord\Ham$.
Now since the trajectories for the characteristic curves follow Hamilton's equations of motion, starting from the initial data $\xcoord_0:=\xcoord(\tcoord=0)$ and $\pcoord_0:=\partial_\xcoord \WavePhase( \tcoord=0,\xcoord) $, the PDE on configuration space can actually be considered to be a PDE on phase space, if the momentum coordinates are properly taken into account.
As shown below, this argument is rigorously true for the generator of the Hamilton-Jacobi equation, $d\WavePhase/d\tcoord=\Lag_\Ham$.

However, the way in which one chooses the evolution for the amplitude is tied to the mapping between phase space and configuration space.
In this section, we will derive the standard phase space KvH equation without consideration of this mapping. 
The standard phase space KvH is important for connecting to previous literature, and, because the dynamics are unitary, it provides a foundation for many of the results that follow.
In Sec.~\ref{sec:KvH_phase-space_semiclassical}, we will derive a modified semiclassical version of the phase space KvH equation that is designed to work with a specific projection operation from phase space to configuration space.

\subsubsection{Phase Space KvH Derivation \label{sec:derivation_phase-space}}

Assume that the wavefunction is specified through a complete solution defined by the phase $\WavePhase(\tcoord,\xcoord, \Jcoord)$ and amplitude  $\WaveAmp(\tcoord,\xcoord, \Jcoord)$.
While this solution may not be available analytically, it can be constructed for finite time intervals by integration along the classical trajectories.
Due to the fact that $\dot \Jcoord=0$, the scalar advection operator on configuration space, where partial derivatives are taken at constant $\Jcoord$, is identical to the phase space advection operator with a simple use of the chain rule  
\begin{multline}
d/d\tcoord =\left. \partial_\tcoord \right|_{\xcoord,\Jcoord}+ \dot \xcoord  \cdot \left. \partial_\xcoord \right|_{\tcoord, \Jcoord} 
=\left. \partial_\tcoord \right|_{\xcoord,\Jcoord}+ \dot \xcoord  \cdot \left. \partial_\xcoord \right|_{\tcoord, \Jcoord} +\dot \Jcoord\cdot \left. \partial_\Jcoord\right|_{\tcoord, \xcoord}  \\
=\left.  \partial_\tcoord \right|_{\xcoord,\pcoord} + \dot \xcoord  \cdot \left. \partial_\xcoord \right|_{\tcoord,\pcoord} 
+  \dot \pcoord \cdot   \left. \partial_\pcoord  \right|_{\tcoord,\xcoord}= \left. \partial_\tcoord \right|_\zcoord +  \dot \zcoord \cdot \partial_\zcoord
.
\end{multline}
In phase space, the HJE takes the form
\begin{align}\label{eq:HJE_phase-space} 
\left(d/d\tcoord\right) \WavePhase
= \partial_\tcoord \WavePhase +  \dot \zcoord \cdot \partial_\zcoord\WavePhase =\partial_\tcoord \WavePhase +  \dot \xcoord \cdot \partial_\xcoord\WavePhase + \dot \Jcoord \cdot \partial_\Jcoord\WavePhase =-\Ham+\pcoord\cdot \partial_\pcoord \Ham = \Lag_\Ham
\end{align}
where now $\Lag_\Ham(\tcoord,\xcoord,\pcoord)$ is the usual phase space Lagrangian.

If we demand that $\pdf=\abs{\waveFunc(\tcoord,\zcoord)}^2$ is a probability density on phase space, then it must advected via the adjoint of the scalar advection operator $(d/d\tcoord)^\dagger  = -(\partial_\tcoord   +    \partial_{\zcoord}  \cdot \dot \zcoord)  $ over the phase space-time measure $\int d\tcoord d^\dim\zcoord $.
Hence, the LVE takes the form
\begin{align}\label{eq:LVE_phase-space} 
- \left(d/d\tcoord\right)^\dagger \pdf=   \left(\partial_\tcoord   +  \partial_{\zcoord} \cdot \dot \zcoord\right)   \pdf
  = 0
\end{align}
and the ATE takes the form
\begin{align}\label{eq:ATE_phase-space} 
\tfrac{1}{2}   \left( d/d\tcoord - d/d\tcoord^\dagger \right) \WaveAmp 
= \left(\partial_\tcoord+ \LieDerivative_{\dot\zcoord} \right) \WaveAmp=  \left(\partial_\tcoord   +\tfrac{1}{2}\AntiCommutator{  \dot \zcoord  }{\, \cdot \,\, \partial_{\zcoord }} \right)\WaveAmp 
  = 0
  .
\end{align}
In phase space, the amplitude evolves via $\WaveAmp(\tcoord,\zcoord)=\WaveAmp(\tcoord_0,\zcoord_0)\left(\det{\partial \zcoord_0/\partial \zcoord}\right)^{1/2}$.
Combining the HJE and ATE yields the {\bf phase space KvH equation}
\begin{align} \label{eq:KvH_phase-space} 
\tfrac{i}{2}\hbar  \left(d/d\tcoord - d/d\tcoord^\dagger\right)\waveFunc =
i\hbar \left(\partial_\tcoord+ \LieDerivative_{\dot\zcoord} \right) \waveFunc=
i\hbar\left( \partial_\tcoord+ 
 \tfrac{1}{2}\AntiCommutator{  \dot \zcoord }{\, \cdot \,\, \partial_{\zcoord }} \right)   \waveFunc  = -\Lag_\Ham \waveFunc .
\end{align}
Because the phase-space KvH equation is linear, like the original PDE, one does not need to include many different branches, as in Eq.~\ref{eq:wavefunction_branches}.  The additional momentum coordinates are already performing this role.
Note, however, that the standard phase space KvH equation differs from the configuration space version in the choice of the amplitude.
Hence, a modified semiclassical version is derived in Sec.~\ref{sec:KvH_phase-space_semiclassical} that yields the configuration space amplitude after projection to configuration space.

\subsubsection{Poisson Bracket Formulation}
A variational system derives from a Lagrangian 1-form, $\pcoord \cdot d\xcoord -\Ham d\tcoord$, that is actually a 1-form in space-time. 
The spatial part is the Poincar\'e 1-form $\PoincareForm$, and, in a canonical coordinate system $\PoincareForm= \pcoord\cdot d\xcoord$.
The exterior derivative of the Poincar\'e form yields the symplectic 2-form $d\PoincareForm$.
Hamilton's equations of motion can be written in terms of the Poisson bracket  \cite{SudarshanBook, Morrison98rmp}
\begin{align}
\PoissonBracket{A}{B} := \partial_{\xcoord }  A\cdot \partial_{\pcoord} B- \partial_{\pcoord}  A\cdot \partial_{\xcoord} B= dA \cdot \PoissonTensor \cdot d B.
\end{align}
For a non-degenerate Lagrange form, $\LagrangeTensor$, the Poisson tensor is the inverse of the Lagrange form $\PoissonTensor=\left(\LagrangeTensor\right)^{-1}$, and, in canonical coordinates it has the block form $\PoissonTensor=\left( \begin{array}{cc} 0 & 1\\ -1 &0 \end{array} \right)$.
In any other coordinate system, the Poisson tensor, $\PoissonTensor$, remains antisymmetric and must satisfy the Jacobi identity which implies that it is the inverse of a closed antisymmetric 2-form.
Using the Poisson tensor,
the
Hamiltonian vector field corresponding to Hamiltonian $\Ham$
can be written as
\begin{align}
\velocityPS^\aindex = \PoissonBracket{\zcoord^\aindex}{\Ham} =\PoissonMatrix^{\aindex\bindex} \partial  \Ham/\partial {\zcoord^\bindex} .
\end{align} 
The phase space velocity is explicitly divergence-free due to the antisymmetry of the Poisson tensor.
Hence, in a general coordinate system, the Lagrangian takes the form
\begin{align}
\Lag_\Ham = \PoincareForm \cdot \PoissonTensor \cdot d\Ham -\Ham.
\end{align}
 
This leads to the Poisson bracket form of the {\bf phase space KvH equation}
\begin{align} \label{eq:KvH_phase_space_PB} 
i\hbar\left(\partial_\tcoord + \LieDerivative_{\velocityPS}\right)  \waveFunc
= i\hbar \left(\partial_\tcoord \waveFunc -
 \PoissonBracket{\Ham}{\waveFunc} \right) = - \Lag_\Ham  \waveFunc .
\end{align}
Introducing the phase-space \emph{Kostant-van Hove Hamiltonian} operator
\begin{align} \label{eq:KvH_phase-space_PB} 
  \KvHOp^\PS_{ \Ham} \waveFunc:=-i\hbar \LieDerivative_{\velocityPS} \waveFunc - \Lag_\Ham\waveFunc
 =i\hbar \PoissonBracket{\Ham}{\waveFunc} - \Lag_\Ham\waveFunc
\end{align}
the phase space KvH equation can be written as 
\begin{align} \label{eq:KvH-Schrodinger_phase-space} 
i\hbar   \partial_\tcoord \waveFunc = \KvHOp^\PS_{ \Ham} \waveFunc .
\end{align} 
Note that the phase space KvH equation is a linear PDE. 

In his PhD thesis \cite{vanHovePhD}, van Hove discovered that these operators share a  Lie algebra property emphasized by Dirac
\begin{align}
\Commutator{  \KvHOp^\PS_{ A} }{ \KvHOp^{ \PS}_{ B} } =\KvHOp^\PS_{i\hbar \PoissonBracket{A}{B}}.
\end{align}
He proved that this algebra represents an essentially unique extension of the group of canonical transformations to the group of canonical transformations that also alters the $\UnitaryGroup(1)$ phase factor of a complex wavefunction.

\subsubsection{Phase Space Action and Resolution of the Phase Space Density Paradox \label{sec:density_paradox}}
The phase space KvH equation can be derived by taking variations of the semiclassical action
\begin{align} \label{eq:phase-space_action}
\WaveActionPrinciple[\waveFunc,\waveConj;\tcoord,\zcoord]:=\int\left[ \Lag_\Ham \abs{\waveFunc}^2- \Im \waveConj \hbar \left( \partial_\tcoord +\LieDerivative_{\velocityPS } \right)\waveFunc\right]  d^{2\dim} \zcoord d\tcoord.
\end{align}
Using integration by parts and neglecting boundary terms, this can be re-expressed as 
\begin{align} \label{eq:physical_phase-space_action}
\WaveActionPrinciple[\waveFunc,\waveConj;\tcoord,\zcoord]:=- \int \left( \Ham \pdf
+\Im {\hbar \waveConj  \partial_\tcoord \waveFunc }
\right) d^{2\dim} \zcoord d\tcoord
\end{align}
where the ``physical phase space density" is
\begin{align} \label{eq:physical_density}
\pdf(\tcoord,\zcoord) := \abs{\waveFunc}^2 +
 \partial_\zcoord\cdot  (\PoissonTensor \cdot \PoincareForm \abs{\waveFunc}^2) 
+ \Im{\hbar \PoissonBracket{\waveConj}{\waveFunc}} .
\end{align}
In canonical coordinates, this takes the simpler form
\begin{align}
\pdf = \abs{\waveFunc}^2 +
 \partial_\pcoord\cdot  \left(\pcoord \abs{\waveFunc}^2\right) 
+ \Im{\hbar \PoissonBracket{\waveConj}{\waveFunc}} .
\end{align}
In Refs.~\cite{Bondar19prsa,Joseph20prr} it proven that, if it is possible to set $\pdf=\abs{\psi}^2$ as an initial condition, then this relation holds for all time. 

Here, it is proven that the difference between these densities vanishes whenever the Hamilon-Jacobi constraint is satisfied.
For the mixed variable generating function, $\partial_\pcoord \left. \WavePhase\right|_{\xcoord,\Jcoord}=0$, and the difference in densities, $\delta \pdf$, becomes a divergence in momentum space
\begin{align}
\delta \pdf :=\pdf -\abs{\waveFunc}^2 =
 \partial_\pcoord\cdot  \left[ \abs{\waveFunc}^2( \pcoord-\partial_\xcoord \WavePhase)\right]  =  \sum_\jindex \partial_{\pcoord_\jindex}  \left[ \abs{\waveFunc}^2( \pcoord_\jindex-\partial_{\xcoord^\jindex} \WavePhase)\right] .
\end{align}
 Hence, $\int \delta \pdf d^\dim \pcoord=0$ because this is the divergence of a vector whose value vanishes on the boundary. 
Moreover,  the product of $  \delta  \pdf$ with any phase space function, $g(\tcoord,\zcoord)$, is also a  divergence in momentum space
\begin{align}
g   \delta   \pdf = \partial_\pcoord \cdot  \left[ g \abs{\waveFunc}^2( \pcoord-\partial_\xcoord \WavePhase)\right]- \abs{\waveFunc}^2( \pcoord-\partial_\xcoord \WavePhase) \cdot \partial_\pcoord g 
\end{align} 
because the second term vanishes identically. 
Once again, $\int g  \delta  \pdf d^\dim \pcoord =0$ because this is the divergence of a vector whose value vanishes on the boundary. 
Hence, if the Hamilton-Jacobi constraint is satisfied for any mixed variable generating function, then after integration over momentum space, the resulting configuration space scalar densities and current densities agree with each other.
Thus, whenever the configuration space version is the primary definition of the problem, the two different phase space densities and current densities will always agree on configuration space.

As discussed in Sec.~\ref{sec:KvH_phase-space_semiclassical}, any solution of the semiclassical phase space KvH equation will yield a solution of the configuration space KvH equation along any submanifold that is locally defined by the Hamilton-Jacobi constraint $\pcoord=\partial_\xcoord \WavePhase$. 
In general, there may be multiple branches associated to this constraint that patch together to form a \emph{Lagrangian submanifold}.
A Lagrangian submanifold is defined as a phase space submanifold of maximal dimension, $\dim$, whose tangent space annihilates the symplectic 2-form, $\LagrangeTensor$.
Hence, if the phase space solution can be foliated by an embedding of Lagrangian submanifolds, then a simple projection to configuration space can be performed.

Note that, even if multiple solution branches were to occupy the same Lagrangian submanifold, then the stationary phase approximation should be used and the phase space action in Eq.~\ref{eq:phase-space_action} would be modified to a form equivalent to Eq.~\ref{eq:config-space_action_generalized}.
As discussed previously, the KvH equation itself is only satisfied by each solution branch, $\waveFunc_\sigma$, labeled by the additional  index, $\sigma$, separately.
Upon using the stationary phase approximation, the action becomes
\begin{align} \label{eq:phase-space_action_generalized}
\WaveActionPrinciple[\waveFunc,\waveConj;\tcoord,\zcoord]:=\sum_\sigma  \int \left[ \Lag_\Ham \abs{\waveFunc_\sigma}^2- \Im \waveFunc_\sigma^* \hbar \left( \partial_\tcoord +\LieDerivative_{\velocityPS } \right)\waveFunc_\sigma \right]  d^{2\dim} \zcoord  d\tcoord.
\end{align}
Thus, conservation is satisfied by each solution branch separately, within the accuracy of the stationary phase approximation.

\subsection{Semiclassical Phase Space KvH
\label{sec:KvH_phase-space_semiclassical}
} 

Because the phase space and configuration space versions of KvH have identical phase space trajectories, the 
propagators for the two versions have a natural correspondence with each other.
In fact, this is  true even if a different weight of the divergence of the velocity is used for each equation
because this simply leads to a different power of the Jacobian in the amplitude factor -- as long as the Jacobian does not change sign.
However, if the Jacobian does change sign, then a fractional power of the Jacobian, such as the square root, will lead to multiple possible solution branches.
Given these considerations, it would be useful to derive a version of the phase space KvH equation that directly maps to the semiclassical configuration space version in a straightforward manner.

In this section, a modified ``semiclassical'' version of the phase space KvH equation is defined that generates solutions of the configuration space KvH equation. 
The key step is to define a projection operation that maps from phase space to configuration space in a universal manner.
In this section, we prove that, with the proper choice of amplitude dynamics, integration over momentum space yields a solution to the configuration space KvH equation
\begin{align} \label{eq:KvH_semiclassical_projection}
\waveFunc(\tcoord,\xcoord)=\int \waveFunc^\mod(\tcoord,\zcoord)d^\dim\pcoord.
\end{align}

Since the phase space KvH equation is essentially a space-time divergence with a source, integration of the divergence over momentum coordinates will yield a configuration space solution -- if the source is modified to the same form as the configuration space version.
In order to obtain the same configuration space Jacobian weight factor, one must actually modify the standard KvH  to the {\bf semiclassical phase space KvH equation}
\begin{align} \label{eq:KvH_phase-space_modified} 
i\hbar \left(\partial_\tcoord+   \partial_{\pcoord }\cdot  {\dot\pcoord} +\tfrac{1}{2} \AntiCommutator{\partial_{\xcoord }}{\,\, \cdot \, {\dot\xcoord} } \right) \waveFunc^\mod=
i\hbar \left(\partial_\tcoord+  \partial_{\zcoord }\cdot {\dot\zcoord} - \tfrac{1}{2} (\partial_{\xcoord }\cdot \dot\xcoord ) \right) \waveFunc^\mod= -\Lag_\Ham \waveFunc^\mod .
\end{align}
In this case, the general solution of $\psi^\mod$ is the same as that of the phase space KvH equation multiplied by a different Jacobian factor.
It is  a volume form in momentum space and the square root of a volume form (a half-form) in configuration space.
One can easily verify that this semiclassical evolution law leads to the configuration space evolution law (Eq.~\ref{eq:KvH_config-space}) after integration over momentum space via Eq.~\ref{eq:KvH_semiclassical_projection}. 
The term $\int d^\dim\pcoord \partial_\pcoord \cdot \left( {\dot\pcoord} \waveFunc^\mod \right)$ is a total divergence and, hence,  a surface integral that vanishes with suitable boundary conditions, e.g. periodic in $\pcoord$ or vanishing as $\pcoord\rightarrow\pm\infty$.

Although the modified KvH Eq.~\ref{eq:KvH_phase-space_modified} is not unitary on phase space, because the configuration space solutions are to be used with the configuration space Hilbert space inner product,  it does generate unitary dynamics in configuration space.
\emph{Morever, while configuration space wavefunctions must be in $L^2$, semiclassical phase space wavefunctions are only required to be in $L^1$, because this is all that is needed to generate a valid configuration space wavefunction after using the projection in Eq.~\ref{eq:KvH_semiclassical_projection}.}

The solution to the standard phase space KvH equation is a half-form in $\zcoord$, which implies that it propagates forward in time via
\begin{align}
\waveFunc   (\tcoord,\zcoord) =  \waveFunc (\tcoord_0,\zcoord_0)e^{i\Delta \WavePhase/\hbar}  \left(\det{ \partial \zcoord_0/\partial\zcoord}\right)^{1/2}.
\end{align}
The solution to the semiclassical phase space KvH equation is a full-form in $\pcoord$ and a half-form in $\xcoord$, which is equivalent to a full-form in $\zcoord$ and a minus half form in $\xcoord$.  
Hence, the amplitude must evolve via  
\begin{align}
\waveFunc^\mod  (\tcoord,\zcoord) 
&=\waveFunc^\mod(\tcoord_0,\zcoord_0) e^{i\Delta \WavePhase/\hbar}  
  \left(\det{ \partial \zcoord_0/\partial\zcoord  }\right) 
  \left(\det{\left.\partial \xcoord/\partial\xcoord_0 \right|_{\pcoord_0} }\right)^{1/2}
  .
\end{align}
The Jacobian factor can also be expressed as
\begin{align} \label{eq:phase_space_jacobian_factor}
\left(\det{ \partial \zcoord_0/\partial\zcoord}\right)  
\left( \det{ \left.\partial \xcoord/\partial\xcoord_0 \right|_{\pcoord_0}}\right)^{1/2}
=  \left(\det{ \partial \zcoord_0/\partial\zcoord}\right)
\left( \det{ \left.\partial \pcoord_0/\partial\pcoord \right|_{\xcoord}}\right)^{1/2} .
\end{align}
In the rest of this section, we simplify these expressions by assuming that $\zcoord$ is a set of canonical coordinates so that $\det{ \partial \zcoord_0/\partial\zcoord}=1$.
Note that, after using canonical coordinates, the phase space Jacobian factor in Eq.~\ref{eq:phase_space_jacobian_factor} is one over the Van Vleck determinant; i.e. it has the opposite power of that expected in configuration space from Eq.~\ref{eq:VanVleck_determinant}.

In Sec.~\ref{sec:KvH_eigenstates}, KvH eigenstates will be derived for integrable systems.
In this case, the phase, $\WavePhase$, represents a transformation to action-angle coordinates, $\setlist{\action,\angle}$, where each eigenstate is labeled by the action $\Jcoord$.
The configuration space solution is a superposition of eigenstates of the form
\begin{align}
\waveFunc  (\tcoord,\xcoord) &= \sum_\sigma \int a( \Jcoord) e^{i \WavePhase_\sigma/\hbar}   \left(\det{  \partial\angle/\partial\xcoord}\right)^{1/2}d^\dim \Jcoord 
= \sum_\sigma \int a( \Jcoord) e^{i \WavePhase_\sigma/\hbar}   \left(\det{  \partial\pcoord/\partial\Jcoord}\right)^{1/2}d^\dim \Jcoord
\end{align}
with an appropriate definition of $a( \Jcoord) $. 
Due to the difference in the topology of the $\pcoord$ and $\Jcoord$ surfaces, when integrating over $\Jcoord$ at constant $\xcoord$, one must explicitly sum over the different branches of the Lagrangian submanifold, labeled by $\sigma$.
After transforming the integral to momentum coordinates, one obtains
\begin{align}
\waveFunc  (\tcoord,\xcoord) 
&= \int a( \Jcoord) e^{i \WavePhase/\hbar}   \left(\det{  \partial\pcoord/\partial\Jcoord}\right)^{1/2} \abs{\det \partial \Jcoord/\partial \pcoord} d^\dim \pcoord
\\
&= \int a( \Jcoord) e^{i \WavePhase/\hbar}   \left(\det{  \partial\angle/\partial\xcoord}\right)^{1/2} \abs{\det \partial \xcoord/\partial \angle} d^\dim \pcoord
\end{align}
This form allows us to identify solutions of the semiclassical KvH equation as the integrand of these expressions.
The appearance of the opposite power of the Van Vleck determinant is simply due to the change of coordinates of the integrand.

The choice of the branch of phase for the square root of the determinant is not completely arbitrary. 
Because the semiclassical approximation breaks down near caustics, the solution should be correctly matched across the singular region by choosing the branch of the phase using the configuration space JWKB connection formulae.
In configuration space, the correct choice can be performed by using analytic continuation and choosing the principle branch of the phase function. 
As will be made clear by working through the examples in Sec.~\ref{sec:examples}, technically speaking, in order to obtain the correct Maslov index in classically allowed regions, the phase is set by taking the primary variable to be imaginary component of momentum, $-i\pcoord$.
Hence, if the semiclassical solution should be specified as 
\begin{align}
\waveFunc^\mod  (\tcoord,\zcoord) &= \waveFunc(\tcoord,\zcoord) \left|\det{ \partial \pcoord_0/\partial\pcoord}\right| \left(\det{ \partial \pcoord /\partial i\pcoord_0}\right)^{1/2}
 &= \waveFunc(\tcoord,\zcoord) \left|\det{ \partial \xcoord/\partial\xcoord_0}\right| \left(\det{ \partial \xcoord_0 /\partial i\xcoord}\right)^{1/2}
\end{align}
In action-angle coordinates, this takes the form
\begin{align} \label{eq:KvH_phase-space_projection}
\waveFunc(\tcoord,\xcoord)=\sum_\sigma \int \waveFunc(\tcoord,\zcoord)  \left(\det{\partial \pcoord/ \partial i\Jcoord}\right)^{1/2}  d^\dim\Jcoord 
= \sum_\sigma \int \waveFunc(\tcoord,\zcoord)  \left(\det{\partial^2 \WavePhase/ \partial i \Jcoord \partial \xcoord}\right)^{1/2} d^\dim\Jcoord .
\end{align}

This last observation shows that, whenever a Lagrangian submanifold for the configuration space solution is identified, then there is also a simple way to map the standard phase space KvH solution to configuration space -- simply multiply by the correct power of the determinant and then integrate over momentum space.
For example, if the mixed variable generating function is taken to be $\WavePhase(\tcoord,\xcoord,\Jcoord  )$, then the projection is 
\begin{align} \label{eq:KvH_phase-space_projection1}
\waveFunc(\tcoord,\xcoord)
= \int \waveFunc(\tcoord,\zcoord)  \left(\det{\partial \Jcoord/ \partial \pcoord}\right)^{1/2}  d^\dim\pcoord 
= \int \waveFunc(\tcoord,\zcoord)  \left(\det{\partial^2 \WavePhase/ \partial \Jcoord \partial \xcoord}\right)^{-1/2} d^\dim\pcoord 
,
\end{align}
again, assuming that the correct choice of phases is made.
Now the KvH dynamics is unitary on phase space, while the projection to configuration space carries the additional weight.
Interestingly enough,  the usual singularity at caustics in configuration space is converted to a weight that vanishes in phase space  (see Fig.~\ref{fig:wkb-cs}).

Yet another method of generating configuration space solutions directly from the standard phase space KvH solutions will be discussed in 
\ref{sec:alternate_mapping_norm}.
This method allows one to use the phase space inner product to lowest order.

\section{KvH Analysis \label{sec:KvH_analysis} }

The goal of this section is to construct the propagator for the various forms of the Liouville-Vlasov (LVE) and Koopman-van Hove (KvH) equations  as well as their eigenstates for integrable dynamical systems.
The next subsection shows that the propagator of the various forms of the KvH equation can be mapped to one another for times short enough that the sign of the determinant does not vanish.
Section~\ref{sec:KvH_eigenstates}  discusses the case of completely integrable systems, which have a complete set of  generalized eigenstates. In this case, the correspondence is global in the sense that it can be made for the entire spectrum.
Eigenstates, labeled by index $\Jcoord$, are defined to have separable dependence in space and time, and, hence, satisfy the relation
\begin{align}
i\hbar \left. \partial_\tcoord \basisFunc_\Jcoord \right|_\zcoord = \Energy_\Jcoord \basisFunc_\Jcoord
\end{align} 
where $\Energy_\Jcoord$ is constant; i.e. they have the form $\basisFunc_\Jcoord \propto\exp({-i\Energy_\Jcoord \tcoord/\hbar)}$.
For integrable dynamical systems, one can completely determine the KvH eigenstates and verify that they are orthogonal and complete over phase space.
In phase space, eigenstates have support on Lagrangian submanifolds of phase space, where multiple branches of submanifolds may need to be ``glued'' together in order to enforce the correct boundary conditions.
Because eigenstates select a particular value for the action $\Jcoord$, they satisfy the Hamilton-Jacobi constraint, and there is a one-to-one correspondence between semiclassical 
phase space eigenstates and configuration space eigenstates.

In action-angle coordinates, the KvH spectrum is the Cartesian product of classical and semiclassical eigenvalues.
The classical part of the spectrum is associated with amplitude transport due to angular motion. 
The semiclassical part of the spectrum is associated with the dependence of the Hamilton-Jacobi solution on action coordinates. 
Due to the ability to construct periodic functions with the classical spectrum, the
semiclassical spectrum is continuous and
extends over the entire range of classically allowed values of action. 
While inclusion of the classical spectrum allows the system to extend the dynamics to all of phase space, the co-existence of classical and semiclassical spectra is actually a departure from semiclassical results.
Once the classical spectrum is eliminated, then the semiclassical spectrum can develop a semi-discrete part, e.g. for integrable systems with bounded motion.
For standard KvH, the  discrete part satisfies the Bohr-Sommerfeld quantization conditions, but with JWKB matching conditions, the  discrete part of the spectrum satisfies the EBK quantization conditions.

There is clearly a difference in the natural Hilbert space inner product for configuration space and phase space.
The natural configuration space inner product is 
\begin{align} \label{eq:inner-product_config-space}
\braket{\eta}{\waveFunc} := \int \eta^*(\tcoord,\xcoord) \waveFunc(\tcoord,\xcoord) d^\dim\xcoord
\end{align}
while the natural phase space inner product is
\begin{align} \label{eq:inner-product_phase-space}
\BraKet{\Upsilon}{\WaveFunc} := \int \Upsilon^*(\tcoord,\zcoord) \WaveFunc(\tcoord,\zcoord) d^{2\dim}\zcoord
.
\end{align}
Orthogonality and completeness for KvH on each space should use the appropriate inner product.
However, it is important to note that, when the phase space version is being used to generate solutions for the configuration space version by integration over momentum space, then it is actually the configuration space inner product in Eq.~\ref{eq:inner-product_config-space} that should be used.
Moreover, semiclassical theory requires the configuration space inner product in order to systematically go to higher order  in the asymptotic expansion. 

The classical phase space KvH eigenproblem does not actually possess square integrable eigenfunctions.
Because the PDEs of interest are first order in space and time, they typically possess a continuous spectrum  \cite{HallBook,ReedSimonBookI}.
While proper eigenfunctions  in the discrete spectrum must be square-integrable,
 generalized eigenfunctions are defined as the limit of approximate eigenstates  in the continuous spectrum. 
 Hence, it does possess \emph{generalized eigenfunctions} that are defined in the sense of a rigged Hilbert space \cite{GelfandBook} (see Ref.~\cite{delaMadrid2005ejp} for a pedagogical introduction).
These generalized eigenfunctions use a Dirac delta function to confine their support to the relevant Lagrangian submanifold and have delta-function normalization, as in \cite{SakuraiBook}.

Interestingly enough, 
\ref{sec:alternate_mapping_norm} shows that one can 
directly use the phase space inner product to lowest order
if one is willing to modify the delta function to a \emph{generalized   distribution}  \cite{Egorov1990rms}, that we refer to as the ``square root of the delta function,'' defined in 
\ref{sec:square_root_delta}.

\subsection{KvH Propagator \label{sec:KvH_propagator}  }

\subsubsection{Phase Space Propagator \label{sec:propagator_phase_space}}

Let the general solution of the phase space equations of motion be specified as $\zcoord=\zeta(\tcoord; \tcoord_0,  \zcoord_0)$.
In phase space, this relationship can be uniquely inverted almost everywhere using  the inverse function $\zcoord_0=\zeta^{-1}(\tcoord_0;\tcoord, \zcoord)$.
The propagator for the scalar advection equation is simply
\begin{align}
\Propagator(\tcoord,\zcoord;\tcoord_0,\zcoord_0) 
	=\delta^{2\dim}(\zcoord_0-\zeta^{-1}(\tcoord_0;\tcoord,\zcoord)) 
	=\delta^{2\dim}(\zcoord-\zeta(\tcoord; \tcoord_0,\zcoord_0)) \abs{\det{(\partial\zcoord/\partial\zcoord_0)} }.
\end{align}
For any initial scalar function, $s(\tcoord_0,\zcoord )$, the phase space solution is given by
\begin{align}
s(\tcoord,\zcoord)
	=\int s(\tcoord_0,\zcoord_0)  \delta^{2\dim}(\zcoord_0-\zeta^{-1}(\tcoord_0;\tcoord,\zcoord)) d^{2\dim}\zcoord_0 
	=s(\tcoord_0, \zeta^{-1}(\tcoord_0;\tcoord,\zcoord)).
\end{align}

The propagator for the Liouville equation (LVE) carries a different weight for the Jacobian factor 
\begin{align}
\Propagator^{\mathrm \LVE} (\tcoord,\zcoord;\tcoord_0,\zcoord_0) 
	=\delta^{2\dim}(\zcoord-\zeta(\tcoord;\tcoord_0,\zcoord_0)) 
	= \delta^{2\dim}(\zcoord_0-\zeta^{-1}(\tcoord_0;\tcoord,\zcoord)) \abs{\det{(\partial\zcoord_0/\partial\zcoord)} }  .
\end{align}
For any initial probability distribution,  $\pdf(\tcoord_0,\zcoord)$, the phase space solution is given by
\begin{align}
\pdf(\tcoord,\zcoord)
	=\int \pdf (\tcoord_0,\zcoord_0)  \delta^{2d}(\zcoord_0-\zeta^{-1}(\tcoord_0;\tcoord,\zcoord)) d^{2\dim}\zcoord_0
	=\pdf (\tcoord_0, \zeta^{-1}(\tcoord_0;\tcoord,\zcoord)) \abs{\det{(\partial\zcoord_0/\partial\zcoord)} }  .
\end{align}

The propagator for the phase space KvH equation is constructed by using the correct weight of the Jacobian and multiplying by the correct phase factor.
If one defines the change in action by integrating $\Delta\WavePhase=\int \Lag_\Ham d\tcoord$ along a trajectory, then the propagator for the phase space KvH equation is simply
\begin{align}
\Propagator^\PS (\tcoord,\zcoord;\tcoord_0,\zcoord_0) 
	&= e^{i\Delta \WavePhase/\hbar}   \delta^{2\dim}(\zcoord_0-\zeta^{-1}(\tcoord_0;\tcoord,\zcoord))\left(\det{\partial\zcoord_0/\partial\zcoord}\right)^{1/2} 	
\\
		& = e^{i\Delta \WavePhase/\hbar}  \delta^{2\dim}(\zcoord-\zeta(\tcoord; \tcoord_0,\zcoord_0)) \left(\det{\partial\zcoord/\partial\zcoord_0}\right)^{1/2}  .
\end{align}
For any initial phase space wavefunction,  $\waveFunc(\tcoord_0,\zcoord)$, the phase space KvH solution is given by
\begin{align}
\waveFunc(\tcoord,\zcoord)
	&=\int \waveFunc (\tcoord_0,\zcoord_0)  e^{i\Delta \WavePhase/\hbar}  \delta^{2\dim}(\zcoord_0-\zeta^{-1}(\tcoord_0;\tcoord,\zcoord)) \left(\det{\partial\zcoord_0/\partial\zcoord}\right)^{1/2}  d^{2\dim}\zcoord_0 
	\\
	&= \waveFunc(\tcoord_0,\zeta^{-1}(\tcoord_0;\tcoord,\zcoord))  e^{i\Delta \WavePhase/\hbar}\left(\det{\partial\zcoord_0/\partial\zcoord}\right)^{1/2}   .
\end{align}
Since the phase space KvH evolution operator generates a canonical transformation of the coordinates, in canonical coordinates, the Jacobian is unity and the evolution is unitary.
Thus, a transformation to canonical coordinates can be used to determine the appropriate phase of the square root of the Jacobian.
However, this does not necessarily match the choice of phases required by semiclassical theory.

The propagator for the modified semiclassical KvH equation simply uses a different Jacobian factor. In canonical coordinates, the propagator is
\begin{align}
\Propagator^\SC(\tcoord,\zcoord;\tcoord_0,\zcoord_0)  
	&=e^{i\Delta \WavePhase/\hbar}   \delta^{2\dim}(\zcoord_0-\zeta^{-1}(\tcoord_0;\tcoord,\zcoord))\left(\det{\partial\pcoord_0/\partial\pcoord}\right)^{1/2}   
	\\	
	&= e^{i\Delta \WavePhase/\hbar}  \delta^{2\dim}(\zcoord-\zeta(\tcoord; \tcoord_0,\zcoord_0)) \left(\det{\partial\xcoord/\partial\xcoord_0}\right)^{1/2} .
\end{align}
For any initial phase space wavefunction,  $\waveFunc^\SC(\tcoord_0,\zcoord)$, the semiclassical phase space KvH solution is given by
\begin{align}
\waveFunc^\SC(\tcoord,\zcoord)&=\int \waveFunc^\SC(\tcoord_0,\zcoord_0) e^{i\Delta \WavePhase/\hbar}  
\delta^{2\dim}(\zcoord_0-\zeta^{-1}(\tcoord_0;\tcoord,\zcoord)) 
\left(\det{\partial\pcoord_0/\partial\pcoord}\right)^{1/2}   d^{2\dim}\zcoord_0 \\
&= \waveFunc^\SC (\tcoord_0, \zeta^{-1}(\tcoord_0;\tcoord,\zcoord))  e^{i\Delta \WavePhase/\hbar}
\left(\det{\partial\pcoord_0/\partial\pcoord}\right)^{1/2}   .
\end{align}
Because the square root of the configuration space Jacobian is needed in these expressions, if the configuration space Jacobian changes sign, one must use an appropriate set of JWKB matching conditions to determine the appropriate branch of the phase.

\subsubsection{Configuration Space Propagator   \label{sec:propagator_config_space} }

The configuration space version must be handled with care because solutions to the Hamilton-Jacobi equation develop caustics whenever multiple phase space trajectories can connect the same configuration space points.
Since  turning points, which generate caustics, can potentially be encountered at any time, if the phase function is to remain well-defined, the wavefunction must be represented as a sum over different solution branches.

The phase space embedding resolves the paradox of having multiple solution branches in configuration space.
Given the phase space propagator, it is clear that the initial conditions in phase space  completely specify the various solution branches.
This is because the phase-space KvH equation is linear and, aside from a set of measure zero, each trajectory, $\zcoord=\zeta(\tcoord;\tcoord_0,\zcoord_0)$, is uniquely determined by its initial condition $\zcoord_0:=\zcoord(\tcoord_0)$.
Although the configuration space KvH equation is not linear, the only nonlinearity appears in the phase factor and the solution for the phase, $\Delta \WavePhase=\int \Lag_\Ham d\tcoord$, is uniquely determined by integration along the trajectory for each initial condition.

Let the solution along a given trajectory be specified by $\xcoord=\xi(\tcoord;\tcoord_0,\zcoord_0)$ and $\pcoord=\partial_\xcoord\WavePhase=\eta(\tcoord;\tcoord_0, \zcoord_0)$.
Now, assume that there are multiple initial conditions, $\zcoord_{0,\Jcoord}=\setlist{\xcoord_{0,\Jcoord}, \pcoord_{0,\Jcoord}}$ indexed by $\Jcoord$,  that give rise to the same final configuration space position $\xcoord(\tcoord)$.
This implies that there are multiple branches for the inverse function $\xcoord_{0,\Jcoord}=\xi_{\Jcoord}^{-1}(\tcoord,\xcoord;\tcoord_0,\pcoord_{0,\Jcoord})$, also indexed by $\Jcoord$.
Then, the configuration space propagator  is defined by integration over all possible solution branches
\begin{align}
\Propagator^\CS(\tcoord,\zcoord;\tcoord_0,\zcoord_0) 
	&=e^{i\Delta \WavePhase/\hbar} \delta^\dim(\xcoord-\xi(\tcoord; \tcoord_0,\zcoord_0)) \left(\det{\partial\xcoord/\partial\xcoord_0}\right)^{1/2} \\
	&=\int e^{i\Delta \WavePhase/\hbar} \delta^\dim(\xcoord_0-\xi_\Jcoord^{-1}(\tcoord,\xcoord; \tcoord_0,\pcoord_{0,\Jcoord} )) 
	\left(\det{\partial\xcoord_0/\partial\xcoord}\right)^{1/2} d^\dim\mu[\Jcoord]
\end{align}
where, now, $d^\dim\mu[\Jcoord]$ is an appropriate measure over the different solution branches.
Hence, after evolving the initial wavefunction, $\waveFunc (\tcoord_0,\zcoord)$, in time, it becomes a superposition over solution branches:
\begin{align}
\waveFunc(\tcoord,\xcoord)&=\int \waveFunc (\tcoord_0,\xcoord_0) e^{i\Delta \WavePhase/\hbar} 
 \delta^{\dim}(\xcoord-\xi(\tcoord;\tcoord_0,\zcoord_0)) \left(\det{\partial\xcoord/\partial\xcoord_0}\right)^{1/2}    
 d^{\dim}  \xcoord_0    
\\
&=\int \waveFunc (\tcoord_0,\xi_\Jcoord^{-1}(\tcoord,\xcoord;\tcoord_0,\pcoord_{0,\Jcoord}))  
e^{i\Delta \WavePhase/\hbar} 
\left(\det{\partial\xcoord_{0,\Jcoord}/\partial\xcoord}\right)^{1/2}   d^\dim \mu[\Jcoord]
.
\end{align}
Depending on the initial conditions, at early times, there might be a finite number of branches, but the number of branches increases with time whenever there are multiple classical turning points.
In fact, the freedom in constructing the phase space wavefunctions makes it clear that the general solution for a configuration space wavefunction is an integral over the initial phase space wavefunction, and, hence, all possible $\pcoord_0$.

\subsubsection{Generalized Eigenstates on Extended Phase Space  }

If the Hamiltonian is time-dependent, then, in the generic case, the eigenfunctions will not be separable in space and time.
One method for understanding this case is to consider evolution on an extended phase space by introducing an additional parameter, $\tau$, where $d\tcoord/d\tau=1$.     
Then, solutions are sought for the extended KvH equation
\begin{align}
i\hbar (\partial_\tau + \partial_\tcoord)\waveFunc = \Hhat\psi.
\end{align}
Since $\Hhat$ does not depend on $\tau$, the eigenfunctions are separable in $\tau$. Solutions  that satisfy $\partial_\tau\waveFunc=0$ are  eigenfunctions of the extended equation as well as solutions of the original equation.

Because the evolution equations under consideration, e.g. LVE and KvH, are first order in space and time, the propagator itself defines a generalized eigenfunction of the extended phase space equation that satisfies $\partial_\tau\waveFunc=0$ given by
\begin{align}
\BasisFunc_{\tcoord_0,\zcoord_0} (\tcoord, \zcoord) := \Propagator(\tcoord,\zcoord;\tcoord_0,\zcoord_0) .
\end{align}
Due to the fact that the PDEs are first order in time and space, these generalized eigenfunctions are in $L^1$.
Due to the uniqueness of the solutions of the phase space ODEs, aside from a set of measure zero, any two states starting from different initial conditions, $\zcoord_0$, are orthogonal.

A  proper wavefunction in $L^2$ must be constructed by integration over an initial condition $ \psi(\tcoord_0,\zcoord)$  in $L^2$. 
The representation  is simply
\begin{align}
\psi(\tcoord,\zcoord)=\int \psi(\tcoord_0, \zcoord_0)\BasisFunc_{\tcoord_0,\zcoord_0} (\tcoord, \zcoord) d^{2\dim} \zcoord_0
\end{align}
in terms of the generalized basis functions.

\subsection{KvH Eigenstates \label{sec:KvH_eigenstates}  }

\subsubsection{Hamilton-Jacobi Solution for Integrable Dynamics}
In an integrable region of phase space, canonical action-angle coordinates, $\{\action,\angle\}$, can be constructed that ensure that the Hamiltonian, $\Ham(\action)$, is a function of $\action$ alone. 
This implies that the action is constant in time, $\action=\action_0$ and the angle evolves linearly in time, $\angle=\angle_0+\omega \tcoord$, where $\omega=\partial_\Jcoord\Ham$.
The canonical transformation to action-angle coordinates is given by Hamilton's principle function $\RouthianFunc $ which satisfies
\begin{align}
d\RouthianFunc  = \pcoord\cdot d\xcoord+\angle\cdot d\Jcoord.
\end{align}
Since $d\Jcoord/d\tcoord=0$, this implies that $\RouthianFunc$ can be determined by integration of the Routhian form 
along the trajectory
\begin{align}
 \RouthianFunc  =\int_{\xcoord_0}^\xcoord  \pcoord\cdot d\xcoord .
\end{align}
For a  closed trajectory, the action is defined via
\begin{align}
2\pi \Jcoord_i:=\oint \Jcoord_\iindex d\angle_\iindex =\oint d\Jcoord_i\wedge d\angle_i = \oint d\pcoord\wedge d\xcoord = \oint \pcoord \cdot d\xcoord 
\end{align}
where here, $d\xcoord$, is taken along the trajectory specified by $d\angle_i$ while all other $\angle_j$'s are held constant.
For an open trajectory, the range of $\xcoord$ and $\angle$ are infinite; hence, the action is defined by the average
\begin{align}
2\pi \Jcoord_i:=\lim_{T\rightarrow\infty} \int_{-T}^T \Jcoord_i  d\angle_i /\int_{-T}^T d \theta_i =\lim_{T\rightarrow\infty} \int_{-T}^T  \pcoord  \cdot d\xcoord /\int_{-T}^T  d \theta_i
.
\end{align}
When the energy is conserved, the solution to the Hamilton-Jacobi equation is 
\begin{align}
\WavePhase =\RouthianFunc  - \Energy \tcoord = \Jcoord\cdot\angle + \tilde \RouthianFunc - \Energy \tcoord.
\end{align} 
For the second form, $\tilde \RouthianFunc$ has vanishing average and, hence, is periodic in $\angle$.

\subsubsection{Phase Space Eigenstates }
Assume the dynamics are integrable and a set of action-angle coordinates, $\setlist{\action,\angle}$, can be found.
We will prove that, in this case, the KvH spectrum is the Cartesian product of a classical spectrum and a semiclassical spectrum.

First let us define the classical spectrum.
In action-angle coordinates, the Liouville-Vlasov equation (LVE) is equivalent to the Koopman-von Neumann (KvN) equation, so they both have the same classical spectrum.  The generalized eigenfunctions are
\begin{align}
\BasisFunc_{\Jcoord_0,\nuindex} (\tcoord,\zcoord)= e^{i\nuindex\cdot\theta_0}  \delta^\dim(\Jcoord-\Jcoord_0)/(2\pi)^{\dim/2} .
\end{align} 
Due to the Pontryagin duality theorem, if
the range of $\angle_\iindex$ is finite and normalized to $2\pi$, then $\nuindex_\iindex\in \Integers$; otherwise, $\nuindex_\iindex \in \Reals$.
The orthogonality relation is
\begin{align}
\BraKet{\BasisFunc_{\Jcoord_0,\nuindex_0}}{\BasisFunc_{\Jcoord_1,\nuindex_1}} 
&=
\int   \delta^\dim(\Jcoord-\Jcoord_0)  \delta^\dim(\Jcoord-\Jcoord_1) 
e^{i(\nuindex_1-\nuindex_0)\cdot\angle} d^\dim \Jcoord d^\dim \angle/(2\pi)^\dim
\\
&= \delta^\dim(\Jcoord_1-\Jcoord_0) \oint e^{i(\nuindex_1-\nuindex_0)\cdot\angle}  d^\dim \angle/(2\pi)^\dim\\
&= \delta^\dim(\Jcoord_1-\Jcoord_0)  \delta^\dim_{\nuindex_1-\nuindex_0}
\label{eq:phase-space_orthonormality}
\end{align} 
where $ \delta_{ \nuindex_1-\nuindex_0 }$ is defined as $\delta_{\nuindex_0,\nuindex_1}$ or  $\delta(\nuindex_1-\nuindex_0)$ if the range of $\angle_i$ is $2\pi$ or infinite.
If we define $\mu[\nuindex]$ as the spectral measure, then the completeness relation is 
\begin{align}
\int \BasisFunc_{\Jcoord,\nuindex}(\zcoord_1)  \BasisFunc^*_{\Jcoord,\nuindex}(\zcoord_0)  d^\dim \Jcoord d^\dim \mu[\nuindex]
&=
\int   \delta^\dim(\Jcoord-\Jcoord_0)  \delta^\dim(\Jcoord-\Jcoord_1) 
e^{i \nuindex \cdot(\angle_0-\angle_1)} d^\dim \Jcoord d^\dim \mu[\nuindex]
\\
&= \delta^\dim(\Jcoord_1-\Jcoord_0) \delta^\dim(\angle_1-\angle_0)
\end{align} 
for $\angle_1-\angle_0$ restricted to the domain of interest.
For the propagator, the solution is 
\begin{align}
\int \BasisFunc_{\Jcoord,\nuindex}(\tcoord_1, \zcoord_1)  \BasisFunc^*_{\Jcoord,\nuindex}(\tcoord_0, \zcoord_0)  d^\dim \Jcoord d^\dim \mu[\nuindex]
&=
\int   \delta^\dim(\Jcoord-\Jcoord_0)  \delta^\dim(\Jcoord-\Jcoord_1) 
e^{i \nuindex \cdot(\angle_	1-\angle_0-\omega  \Delta \tcoord)} d^\dim \Jcoord d^\dim \mu[\nuindex]
\\
&= \delta^\dim(\Jcoord_1-\Jcoord_0) \delta^\dim(\angle_1-\angle_0-\omega \Delta \tcoord)\\
&=\delta^{2\dim} (\zcoord-\zeta(\tcoord_1;\tcoord_0, \zcoord_0)) 
\end{align} 
where $\Delta \tcoord:=\tcoord_1- \tcoord_0$.
When converting this expression to an arbitrary coordinate system, one must multiply the final result by the correct weight of the Jacobian $\left( \det \partial \zcoord_0/\partial \zcoord\right)^a$, where $a=1$ for LVE and 1/2 for KvN.
For any of these equations, the eigenproblem is complete in phase space, and, thus, is equivalent to the initial value problem.
Note that the classical spectrum is identified with the dependence on $\exp(\pm i\nuindex\cdot \angle_0)$ which has both signs of frequency $\pm \nuindex\cdot\omega$.

For the KvH equation, this classical part of the spectrum still exists  due to the existence of homogenous solutions to $d\WavePhase/d\tcoord=\Lag_\Ham$. 
Now, there is also a semiclassical spectrum that is generated by the dependence of the energy on the action.
For integrable dynamics, one must use the complete solution to the Hamilton-Jacobi equation, the mixed variable generating function $\WavePhase(\tcoord,\xcoord,\Jcoord)$, to define the phase.
The semiclassical part of the KvH spectrum is continuous, because
 for a generic value of $\Jcoord_i$, the combination 
\begin{align}
\WavePhaseEig 
&=\WavePhase -\sum_{i}' \Jcoord_i \angle_{0i}+\Energy\tcoord_0
\end{align}
where $\sum_i'$ is a sum over angles with finite range, is a periodic function.
For example, if all angles have finite range, then
\begin{align}
\WavePhaseEig(\tcoord,\xcoord,\Jcoord)
 = \tilde \RouthianFunc+\Jcoord\cdot( \angle -\angle_0) -\Energy  (\tcoord -\tcoord_0)
&=  \tilde \RouthianFunc + (\Jcoord\cdot\omega - \Energy) ( \tcoord-\tcoord_0)
\end{align}
is periodic in $\angle$ because it only depends on $\tilde \RouthianFunc $.
Thus, the eigenfunctions for the phase space KvH equation are
 \begin{align}
\BasisFunc_{\Jcoord_0,\nuindex} (\tcoord,\zcoord)= e^{i\nuindex  \cdot\theta_0 +i\WavePhaseEig(\tcoord,\xcoord, \Jcoord )/\hbar} \delta^\dim(\Jcoord-\Jcoord_0)/(2\pi )^{\dim/2} .
\end{align} 
The classical spectrum is indexed by $\nuindex$ and the continuous semiclassical spectrum is indexed by $\Jcoord$.
The eigenfunctions for the modified semiclassical phase space KvH equation simply have an additional Jacobian factor
 \begin{align}
\BasisFunc^\SC_{\Jcoord_0,\nuindex} (\tcoord,\zcoord)= e^{i\nuindex \cdot\theta_0 +i\WavePhaseEig(\tcoord,\xcoord,\Jcoord )/\hbar} \delta^\dim(\Jcoord-\Jcoord_0) \left(\left. \det{\partial\Jcoord/\partial\pcoord} \right|_\xcoord \right)^{1/2}/(2\pi )^{\dim/2} .
\end{align} 
Thus, the spectrum is the same as for the standard KvH equation.

For the classical and 
semiclassical spectrum, this yields the same orthonormality and completeness relations as for KvN  because the phases exactly cancel. 
For orthonormality, the two delta functions ensure that the phase is evaluated at $\Jcoord=\Jcoord_1=\Jcoord_0$.
For completeness, there is the additional phase factor $\WavePhase(\tcoord_1,\xcoord_1,\Jcoord)-\WavePhase(\tcoord_0,\xcoord_0,\Jcoord)$.
However, for $\tcoord_1=\tcoord_0$, the 
integration over $d^\dim \mu[\nuindex]$
restricts the solution to $\angle_1=\angle_0$, and, hence $\xcoord_0=\xcoord_1$, so that, once again, this additional complex phase vanishes.  
For a finite time difference, the stationary phase approximation still yields $\delta^\dim(\angle_1-\angle_0-\omega \Delta \tcoord)$, so that one obtains the same result as before.
This proves both orthogonality and completeness of the classical eigenfunctions for the full phase space KvH equation.

Let us compare the KvH spectrum to the JWKB spectrum. 
As explained in Sec.~\ref{sec:semiclassical_analysis}, semiclassical analysis eliminates the classical spectrum, so that the only allowed value is $\nuindex=0$.
In this case, the KvH eigenspectrum becomes discrete, because  it is no longer possible to use non-integer $\nuindex$ to construct $ \WavePhaseEig$.
For the standard KvH equation, the eigenfunctions are
 \begin{align}
\BasisFunc_{\Jcoord_0} (\tcoord,\zcoord)= e^{ i\WavePhase(\tcoord,\xcoord,\Jcoord )/\hbar} \delta^\dim(\Jcoord-\Jcoord_0) /(2\pi )^{\dim/2} .
\end{align} 
Now, if angle $\angle_i$ has finite range, the eigenfunctions are only periodic for the specific  values of action, $\Jcoord_i/\hbar=\jindex_i\in \Naturals$, that correspond to the Bohr-Sommerfeld quantization conditions.
For the modified semiclassical KvH equation, the eigenfunctions are 
 \begin{align}
\BasisFunc^\SC_{\Jcoord_0} (\tcoord,\zcoord)= e^{ i\WavePhase(\tcoord,\xcoord , \Jcoord)/\hbar} \delta^\dim(\Jcoord-\Jcoord_0) \left(\left. \det{\partial\Jcoord/\partial\pcoord} \right|_\xcoord \right)^{1/2}/(2\pi )^{\dim/2} .
\end{align} 
As explained in Sec.~\ref{sec:semiclassical_analysis}, use of the correct JWKB connection formulae 
 yields the Maslov-Keller correction to the Bohr-Sommerfeld quantization conditions. 
This leads to the Einstein-Brillouin-Keller (EBK) quantization conditions,  $\jindex_i=\Jcoord_i/\hbar=\nindex_i+\mu_i/4$, where $n_i\in\Naturals$ and $\mu_i$ is the Maslov index.

\subsubsection{Configuration Space Eigenstates }
 
For the case of integrable dynamics, the phase space can be covered with a foliation of Lagrangian submanifolds that are invariant under the flow.
This yields a useful and essentially unique decomposition.
At lowest order in the asymptotic theory, different $\Jcoord$ will not interfere with each other and can be considered independently. 
This is clearly true for the phase space variational principle (Eq.~\ref{eq:phase-space_action}) and is asymptotically true for the configuration space variational principle (Eq.~\ref{eq:config-space_action}).

In order for the amplitude, $\WaveAmp_\Jcoord$, to satisfy the ATE, the configuration space eigenfunctions must be a superposition of terms in the Van Vleck form
\begin{align}
 \basisFunc_{\Jcoord_0,\nuindex} (\tcoord,\xcoord ) &=\int\BasisFunc_{\Jcoord_0,\nuindex}^\mod (\tcoord,\zcoord)d^\dim\pcoord\\
&=
\int e^{i\nuindex \cdot\theta_0+i\WavePhaseEig(\tcoord,\xcoord, \Jcoord )/\hbar}  \delta^\dim(\Jcoord-\Jcoord_0)  \left(\left. \det \partial \Jcoord/\partial \pcoord\right|_\xcoord\right)^{1/2} d^\dim \pcoord
 \\
 &=\sum_\sigma \int 
 e^{i\nuindex\cdot\theta_0+i\WavePhaseEig_\sigma (\tcoord,\xcoord, \Jcoord )/\hbar} \delta^\dim(\Jcoord-\Jcoord_0 ) \left(\left. \det \partial \pcoord/\partial \Jcoord\right|_\xcoord\right)^{1/2}  d^\dim \Jcoord
 \\
&=\sum_\sigma e^{i\nuindex\cdot\theta_0+i\WavePhaseEig_\sigma (\tcoord,\xcoord, \Jcoord_0)/\hbar} \left(\left. \det \partial \theta/\partial \xcoord\right|_{\Jcoord_0}\right)^{1/2}  /(2\pi)^{d/2}.
\end{align} 
 Hence, there is a unique mapping between the phase space and configuration space eigenfunctions that is determined by the phase factor for each term in the expansion.
As before, there is both a classical spectrum indexed by $\nuindex$ and a semiclassical spectrum that is continuous in $\Jcoord$.
Given this mapping, the configuration space eigenfunctions are actually complete over phase space and are overcomplete when considered over configuration space alone. 

If the classical spectrum is eliminated, as will be discussed in Sec.~\ref{sec:semiclassical_analysis}, then, once again, the eigenfunctions are restricted to the $\nu=0$ sector. In this case, the spectrum in $\Jcoord$ becomes discrete and the eigenfunctions become
\begin{align}
 \basisFunc_{\Jcoord} (\tcoord,\xcoord ) &=\sum_\sigma e^{ i\WavePhase_\sigma(\tcoord,\xcoord,\Jcoord)/\hbar} \left(\left. \det \partial \theta/\partial \xcoord\right|_\Jcoord\right)^{1/2}  /(2\pi)^{d/2}.
\end{align} 
If the phase of the Jacobian factor is neglected (e.g. the absolute value is used for all branches) and the angle is finite, then the allowed values of $\Jcoord$ satisfy the Bohr-Sommerfeld quantization conditions. 
If the phase is set through the JWKB matching conditions, then the allowed values satisfy the EBK quantization conditions.

Any foliation of phase space by Lagrangian submanifolds yields a well-defined mapping between phase space and configuration space.
This corresponds to treating the initial state as being defined by a particular integrable dynamical system.
However, if the Lagrangian submanifolds are not invariant under the dynamics, then their evolution in time must be tracked, including the generation of new branches whenever the submanifold develops folds and other types of singularites.
This highlights the fact that eigenstates, which do not change their form in time, must be supported on dynamically invariant Lagrangian submanifolds. 

\section{Semiclassical Analysis  
\label{sec:semiclassical_analysis}
}

Semiclassical analysis begins in configuration space, where the configuration space inner product should be used. 
Two of the most important conclusions of JWKB analysis are: (i) the allowed values of action are quantized for bound states and (ii) the wavefunction can tunnel through classically forbidden regions.
While JWKB analysis is covered by most textbooks on quantum mechanics \cite{LandauQMBook,SakuraiBook}, in this section, we remind the reader of the reasoning that leads to these conclusions precisely to illustrate the ways in which semiclassical  theory differs from classical KvH theory.
 
 First, the amplitude is not allowed to vary as strongly as the phase, and this is typically enforced using the time-independent ATE.
 This eliminates the classical part of the spectrum that is associated with the ATE and eliminates completeness in phase space over the angle coordinates.
Second, requiring the eigenfunction to satisfy periodic boundary conditions along any closed loop requires the allowed values of action through the loop to satisfy the semiclassical Einstein-Brillouin-Keller (EBK) quantization conditions.
For example, for bound states this typically restricts the action enclosed by the surface to be a half-integer multiple of $\hbar$.
This is a second reason for why the stationary eigenstates are not complete over phase space, now over the action coordinates.
An important conclusion is that a superposition of solutions to the semiclassical eigenproblem  provides a complete solution for the configuration space initial value problem, but not for the phase space initial value problem.

The classical pre-quantum limit, $\hbar\rightarrow 0^+$, eliminates the quantization conditions because all surfaces of constant $\Jcoord$ become generalized eigenstates. 
In this limit, tunneling along complex solution branches becomes negligible because the decay in such regions becomes infinitely rapid.

\subsection{Eliminating Classical Eigenstates }
The asymptotic semiclassical approximation results from assuming that the amplitude varies much more slowly than the phase \cite{VanVleck1928pnas,CvitanovicChaosBook}.
Although the general ATE solution allows an amplitude factor, $\WaveAmp$, that is determined by an arbitrary function of constants of the motion, one must assume that this variation is weak, otherwise, as in the complex JWKB method, the amplitude should be included as part of a self-consistent complex solution to the Hamilton-Jacobi equation \cite{Froman65book,Maslov94book}.
The problem is that the KvH equation mixes terms with two different orders in $\epsilon$ together. 
Yet, even the equation $d\WavePhase/d\tcoord=\Lag_\Ham$, has the full KvH spectrum, including the classical spectrum.  
From the phase space point of view, including the $\nuindex$ dependence is beneficial because it allows the  eigenfunctions  to be complete over phase space.
However, this adds many additional modes to the spectrum that are not present in the exactly quantized system.

Another deficiency is that the classical spectrum has both positive and negative 
frequency 
branches, $\pm\nu \cdot  \omega$, whereas the quantum spectrum is typically bounded from below.
The fact that both branches exists makes  sense if one interprets amplitude transport as the Liouville-von Neumann equation for a classical probability density.  In this case, it represents the possible energy differences between closely spaced  
energy levels. 

In order to restrict the solution to the  $\nuindex=0$ quantum sector, many authors propose using the time-independent amplitude transport equation
\begin{align}
\partial_\xcoord \cdot\left( \dot \xcoord \abs{\waveFunc}^2\right)=0.
\end{align}
Note, however, that this is not completely correct because it still allows zero frequency modes in the classical spectrum to play a role.
Zero frequency modes  occur whenever classical frequencies are commensurate (relatively rational); i.e. there exists a vector of integers, $m=\setlist{m_\jindex}$, such that $m\cdot\omega=0$.
For $\dim\geq 2$, in systems with  frequency variations ($  \partial_{\Jcoord_\jindex } \omega^\iindex \neq 0$) such modes  are accidental but ubiquitous.
 This allows  modes of the form $\exp{(im\cdot \angle)}$ to exist on so-called rational surfaces where $m\cdot\omega=0$.

Hence, it is more accurate to state that the asymptotic solution of the amplitude is given by the particular solution corresponding to the Van Vleck determinant.  Then, each eigenfunction is simply \cite{Froman65book}
\begin{align}
\waveFunc(\tcoord,\xcoord)= \int \waveFunc_\Jcoord  \basisFunc_{\Jcoord}(\tcoord,\xcoord)d^\dim\mu[\Jcoord]
=\sum_\sigma \int \waveFunc_\Jcoord e^{i\WavePhase_\sigma(\tcoord,\xcoord,\Jcoord)/\hbar}\left( \det\partial_\xcoord \partial_\Jcoord \WavePhase_\sigma \right)^{1/2} d^\dim\mu[\Jcoord]/(2\pi)^{\dim/2}
\end{align}
where $\WavePhase$ satisfies the Hamilton-Jacobi equation.
This can be performed in phase space as well via  the standard phase space version
\begin{align}
\waveFunc(\tcoord,\zcoord)=\int \waveFunc_\Jcoord \BasisFunc_{\Jcoord}(\tcoord,\zcoord)d^\dim\mu[\Jcoord]
=\int \waveFunc_\Jcoord e^{i\WavePhase(\tcoord,\xcoord,\Jcoord)/\hbar} \delta^\dim(\Jcoord-\Jcoord') d^\dim\mu[\Jcoord'] /(2\pi)^{\dim/2}
.
\end{align}
or the modified phase space version
\begin{align}
\waveFunc^\mod(\tcoord,\zcoord)=\int \waveFunc_\Jcoord^\mod \BasisFunc^\mod_{\Jcoord} (\tcoord,\zcoord)d^\dim\mu[\Jcoord]
=\int  \waveFunc_\Jcoord^\mod  e^{i\WavePhase (\tcoord,\xcoord,\Jcoord)/\hbar}
\left( \det\partial_\xcoord \partial_\Jcoord \WavePhase\right)^{1/2} \delta^\dim(\Jcoord-\Jcoord')  d^\dim\mu[\Jcoord'] /(2\pi)^{\dim/2} .
\end{align}
From the point of view of phase space KvH as a PDE, the cost is that this ansatz eliminates the completeness over  $\angle$ in phase space.
Of course, if quantization conditions are present, they restrict the allowed values of $\Jcoord$, which further reduces completeness in phase space.

One possible benefit of using the KvH equation is that the classical spectrum is useful for describing the evolution at large spatial scales, while the semiclassical spectrum is useful for obtaining approximate eigenfunctions at short spatial scales. In principle, this can be utilized by setting initial conditions that are a superposition of  smooth initial conditions and approximate semiclassical eigenstates.  From a numerical perspective, if one does not have enough spatial resolution to reach the quantum scale, then the {frequency of the classical modes will be restricted to be much less than the quantum energy levels. 
This has the benefit of obtaining the correct frequency response, but the drawback of requiring short time steps for accuracy and/or stability due to the need to resolve the rapid quantum eigenfrequencies.
Further numerical investigations into the utility of the KvH equation are warranted.

\subsection{Quantization  \& Maslov Index }
\def\nuindex{n}
The linear dependence of the phase on the angles, $\WavePhase/\hbar=\Jcoord\cdot\angle/\hbar + \dots $ indicates that, in order to be a periodic function of $\theta$, the values of $\Jcoord$ must be quantized.
If $\WavePhase$ is the only part that determines the complex phase factor, this would yield the Bohr-Sommerfeld quantization conditions $\Jcoord_i/\hbar = \nuindex_i\in \Naturals$.  

However, in configuration space, the square root of the Jacobian $\left(\det\partial \angle/\partial\xcoord\right)^{1/2}$ also plays a role in determining the complex phase.
For example, for a 1D system, the amplitude factor 
\begin{align}
 \partial^2\WavePhase/\partial \xcoord\partial\Jcoord= \left. \partial\theta/\partial \xcoord \right|_\Jcoord =\left.  \partial \pcoord/\partial \Jcoord \right|_\xcoord =  \left. \partial_\Jcoord \Ham\right|_\angle / \left. \partial_\pcoord \Ham \right|_\xcoord= \omega/  \dot \xcoord ,
\end{align}
 changes sign, and is in fact singular, whenever the group velocity, $\dot \xcoord $, vanishes.
More generally, if the system is separable, the momentum can be solved for in terms of constants of the motion. 
For the generic case of Hamiltonians that are nonlinear (quadratic and higher order) in the momentum, there will be multiple solution branches. 
From the  point of view of complex phase space, these branches form different sheets of complex Lagrangian submanifolds.
Because these branches  have  phase factors that vary on similar spatial scales they cause nontrivial interference patterns, and, hence, all branches must be included in order to satisfy a general set of boundary conditions. 
For example, in the common case of a real Hamiltonian that is quadratic in the momentum, there are two complex solutions.
If there is no linear momentum  term in the Hamiltonian, then the two branches are either purely real or purely imaginary, and can be labeled $\pcoord_\pm= \pm \sqrt{\pcoord^2}$.

The JWKB connection formulae \cite{LandauQMBook, Maslov81book, HallBook} dictate the manner in which the solution should be matched across turning points.
The JWKB matching condition is 
\cite{Kramers1926zp,LandauQMBook}
 \begin{align}
 \exp{\left(-\abs{\Delta \RouthianFunc}/\hbar\right)}/\abs{\pcoord}^{1/2} \xrightarrow[cw]{ccw} 
  \exp{\left(\pm i\left[\abs{\Delta \RouthianFunc}/\hbar-\pi/4\right]\right)}/ \abs{\pcoord}^{1/2}
 \end{align}
 where $\Delta \RouthianFunc$ is the change from the nearest turning point.
Extending the solution across the turning point can be performed by matching the JWKB solution to a more accurate approximation of the exact solution of the PDE near the turning points.
It can also be performed by analytic continuation to complex $\xcoord$ \cite{ LandauQMBook}. 
Yet another approach, developed by Maslov \cite{Maslov81book}, is to switch between configuration space and  momentum space before reaching the turning point.
 
Whenever the determinant passes through zero or infinity, as it must when encountering a classical turning point, the determinant changes sign.
This causes the square root to change phase by  $-\pi/4$ after passing halfway through the turning point.
The \emph{Maslov index} $\mu$ counts the number of half turns along the trajectory. 
In this case, the phase accumulates an extra phase $-\maslov \pi/4$ and this must be canceled by the $\Jcoord\cdot\theta$ term.
In general, this yields the Einstein-Brillouin-Keller (EBK) quantization conditions \cite{Keller58ap}
\begin{align}
 \jindex_i:= \Jcoord_i/\hbar  = \nuindex_i +  \mu_i/4
\end{align}
where $\nuindex_i\in \Naturals$.
For example, for a bound state with two turning points, the Maslov index is 2 and the quantization condition is  $\jindex_i:=\Jcoord_i/\hbar  = \nuindex_i+1/2$.
While this solution is generic, in special cases, the turning point might need to be resolved by a higher $\pdeOrder$-th order polynomial, in which case, the accumulated phase change would be $-\pi/2\pdeOrder$ and the Maslov correction would have the form $\mu/2\pdeOrder$ for non-negative integer $\mu$.

The projection to configuration space has important physical implications. 
If one were solving the phase space version without the projection, then one would derive the Bohr-Sommerfeld quantization conditions.
For the semiclassical KvH equation, the same weight factor is included that leads to the EBK  form of the quantization conditions.  
This causes  the semiclassical bound eigenstates to be  $\propto \cos{\left(  \abs{\Delta \RouthianFunc}/\hbar - \sum \mu_i \pi/4\right)}$.
Thus, the projection to configuration space via the modified KvH equation is an important step in obtaining the correct EBK quantization conditions.
Of course, in the limit $\nindex_i \gg 1 $, the Keller-Maslov correction becomes negligible and can be ignored.

\subsection{Orthonormality \& Completeness Relations}
Let us recall the orthonormality and completeness relations that apply to the discrete part of the JWKB spectrum. 
When integrated over $\xcoord$, the configuration space eigenfunctions are approximately orthogonal after usage of the stationary phase approximation. When $\Jcoord=\Jcoord'$, the determinants combine to approximately form  $ \det\partial\theta/\partial \xcoord$:
\begin{align}
\InnerProduct{\phi_\Jcoord}{\phi_{\Jcoord'}} 
&=  \oint e^{i(\WavePhase(\tcoord, \xcoord, \Jcoord)-i\WavePhase (\tcoord, \xcoord, \Jcoord'))/\hbar}\left( \left.\det \partial \theta /\partial \xcoord \right|_\Jcoord \right)^{1/2}\left(\left. \det \partial \theta/\partial \xcoord \right|_{\Jcoord'} \right)^{1/2}d^\dim \xcoord/(2\pi)^\dim
\\
&\simeq  \oint e^{i(\Jcoord-\Jcoord')\cdot \angle/\hbar} d^\dim \angle/(2\pi)^\dim  =\delta^\dim_{\Jcoord,\Jcoord'} .
\end{align}  
The final result is exact when $\Jcoord=\Jcoord'$ and 
 vanishes to $\CO(\epsilon)$ when $\Jcoord \neq \Jcoord'$.

In general, the 
Hamiltonian operator may possess a continuous part of the spectrum, in which case, the bounds of $\angle$ are infinite.
Moreover, as $\hbar \rightarrow 0$, one may want to consider the limit in which the discrete spectrum for $\Jcoord$ becomes ``continuous.''
This is similar to the limiting process that transforms the Fourier series to the Fourier transform, and, in a similar fashion,   the bounds of $\xcoord$ and $\angle$ should extend towards $\pm\infty$.
In this case, the configuration space inner product of two configuration space eigenfunctions yields
\begin{align}
\InnerProduct{\phi_\Jcoord}{\phi_{\Jcoord'}} =\int_{-\infty}^{\infty} \phi^*_\Jcoord (\xcoord) \phi_{\Jcoord'}(\xcoord)d^\dim \xcoord
\simeq \int_{-\infty}^{\infty} e^{i(\Jcoord-\Jcoord') \cdot \theta/\hbar} d^\dim \theta/(2\pi)^\dim = \delta^\dim(\Jcoord-\Jcoord') \hbar^\dim.
\end{align} 
For the continuous spectrum, corrections due to the Maslov index might be considered negligible. 
However, one can still account for the Maslov index by considering each $\Jcoord_i$ to  have a lower bound of  $\mu_i \hbar /4$.
 
In configuration space, completeness follows a similar argument using the stationary phase approximation.
The result is 
\begin{align}  
\int {\phi_\Jcoord}(\xcoord_1) {\phi^*_{\Jcoord}(\xcoord_0)} d^\dim \mu[\Jcoord]/\hbar^\dim &=  \int e^{i\left(\WavePhase (\tcoord, \xcoord_1, \Jcoord)-i\WavePhase (\tcoord, \xcoord_0, \Jcoord)\right)/\hbar}\left( \left.\det \partial \pcoord /\partial \Jcoord \right|_{\xcoord_1} \right)^{1/2}\left(\left. \det \partial \pcoord /\partial \Jcoord \right|_{\xcoord_0} \right)^{1/2} d^\dim \mu[\Jcoord]/(2\pi\hbar )^\dim
\\
&\simeq  \int e^{i\pcoord\cdot (\xcoord_1-\xcoord_0)/\hbar} d^\dim\mu[\pcoord]/(2\pi \hbar)^d  =\delta^\dim(\xcoord_1-\xcoord_0) 
\end{align}  
for $\xcoord_1-\xcoord_0$ restricted to the region of interest.

\subsection{Quantum Tunneling}
Quantum tunneling arises because the exact solution to the original higher order PDE must be continuous for continuous initial data. 
In order for the JWKB eigenfunctions to also have this property, one must generically glue a number of smoothly embedded Lagrangian manifolds together in order to satisfy both the boundary conditions and continuity.
In general, the method for accomplishing this is called the \emph{complex JWKB method} \cite{Froman65book, Knoll76ap, Maslov94book} 
 or the \emph{exact WKB method} \cite{Balian74,Voros83apt,Sueishi20jhep}  and requires complexifying the notion of phase space.
In full generality, this requires a complexification of the underlying phase space manifold as well as analytic continuation of the equations of motion.  
Yet, for many situations of interest, the wavefunction decays rapidly in the imaginary dimensions and only a perturbative analysis is required, which Maslov called ``the method of Lagrangian manifolds with complex germ"  \cite{Maslov94book}.
While we will not describe these methods in detail, such constructions are needed for the semiclassical description of quantum tunneling.

Classical mechanics is typically formulated for real coordinates, and, hence, regions where the momenta become complex are discarded as unacceptable classically forbidden regions. 
However, 
the mathematical solution to a completely integrable classical  system is usually
defined through analytic continuation of the equations of motion to the complex domain.
This is because solving for the momenta in terms of the action coordinates requires extending the momentum coordinates to the complex domain -- simply because only the complex numbers are algebraically complete.
These complex solutions of the classical equations of motion are required for  complex JWKB analysis.

\section{Examples \label{sec:examples}}

The simple examples in this section are used to remind the reader of the lessons learned from 1D semiclassical theory.
The boundary conditions for bound eigenstates should be chosen so that the approximate wavefunction is continuous and vanishes at spatial infinity. 
This requires the momenta and group velocities to become complex in classically forbidden regions where the wavefunction becomes evanescent.
For a  potential well, the momentum must become imaginary at classical turning points, and, since the action will remain constant and real, in turn, the conjugate angle must become imaginary. 

As explained previously, the KvH spectrum also includes the classical spectrum, $\nu\cdot \hbar \omega$. However, since these modes are well understood and should be removed from the semiclassical spectrum, they will not be included in this section.

\subsection{Harmonic Oscillator \label{sec:sho}}
\def\HermiteH{H}
\def\nuindex{n}
In this section, we review the construction of the JWKB eigenfunctions for the harmonic oscillator.
First the eigenfunctions are constructed in phase space and, then, the usual JWKB eigenfunctions in configuration space are constructed.

\subsubsection{Quantum Harmonic Oscillator}
Consider the quantum harmonic oscillator with Hamiltonian 
\begin{align}
\Hhat = \phat^2/2\mass + \mass \omega^2 \qhat^2/2.
\end{align}
As is well known, the quantum eigenfunctions 
\begin{align}
\phi_\nuindex(\tcoord,\xcoord) = e^{-\xcoord^2/4\xcoord_0^2-i\nuindex \omega \tcoord/\hbar} \HermiteH_\nuindex(\xcoord/\sqrt{2}\xcoord_0) /\sqrt{2^\nuindex \nuindex!}
\end{align}
are defined in terms of the Hermite polynomials $\HermiteH_\nuindex(x)$ where $\xcoord_0^2=\hbar/\mass\omega$.
A general quantum mechanical solution is in the form of a superposition of eigenstates
\begin{align}
\waveFunc^{QM} (\tcoord,\xcoord) =\sum_\nuindex \waveFunc_\nuindex \basisFunc_\nuindex(\tcoord,\xcoord).
\end{align}

\begin{figure}[tbp]
\centering
\includegraphics[width=3in]{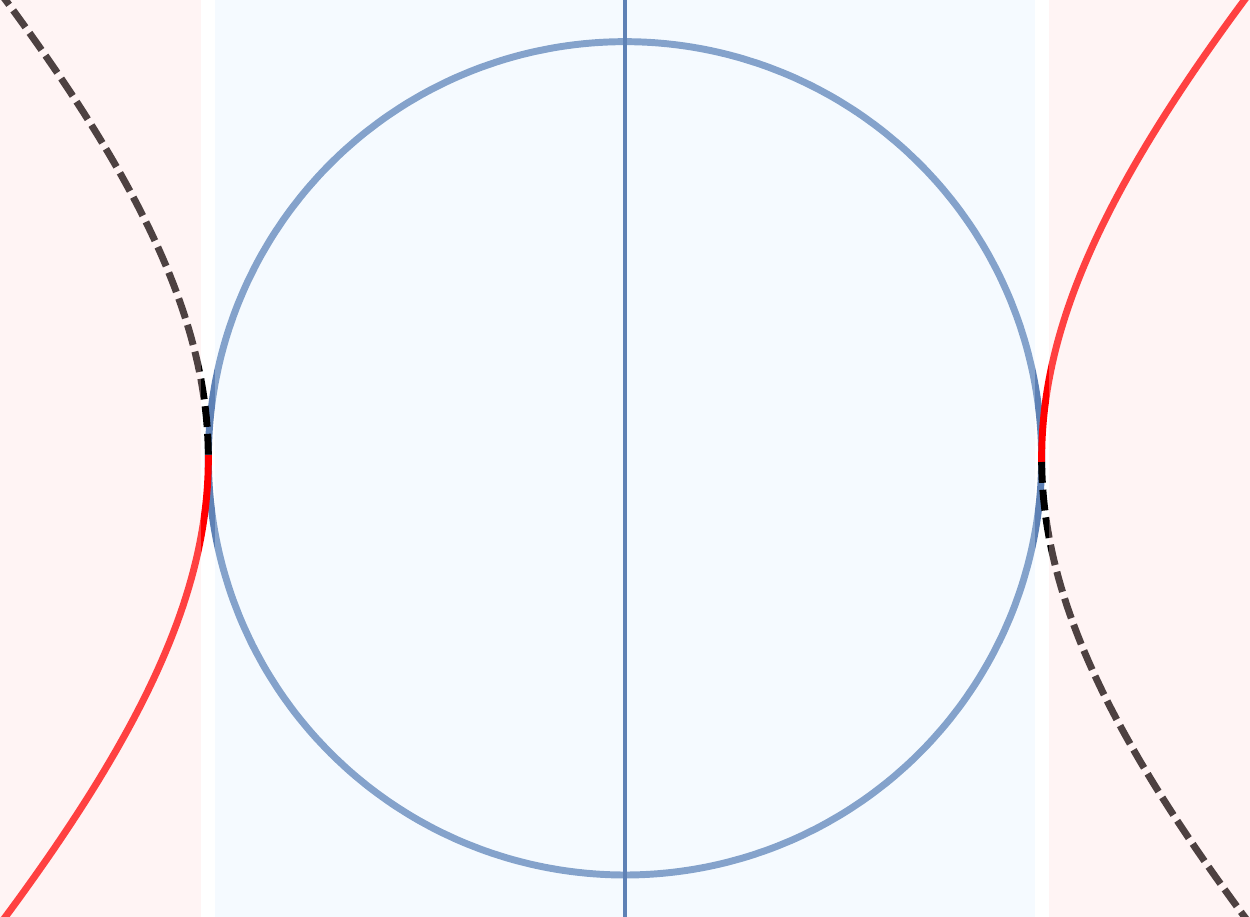} 
\includegraphics[width=3in]{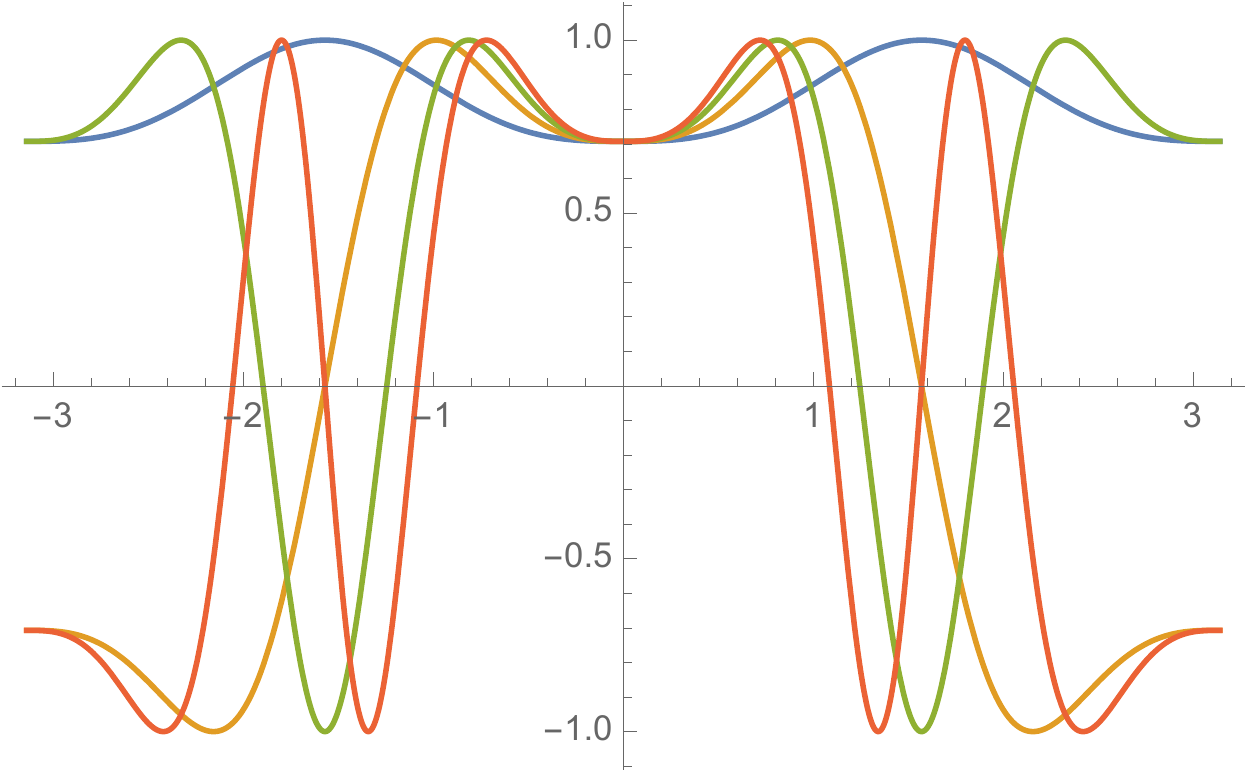}
\caption{(Left) In phase space, each  harmonic oscillator JWKB eigenfunction is supported on a Lagrangian submanifold. For the classical allowed region $\mass\omega  \xcoord^2\leq 2\Jcoord$ (shaded blue), the submanifold is a circle with $\pcoord\in \Reals$. For the classically forbidden region $\mass\omega  \xcoord^2\geq 2\Jcoord$ (shaded red), the submanifold is defined by hyperbolae with $-i\pcoord\in \Reals$.  The wavefunction only has support on the decaying branches (solid red), not the growing branches (dashed black). (Right) Harmonic oscillator JWKB eigenfunctions as a function of angle $\angle=\tan^{-1}({\pcoord/\mass\omega\xcoord})$ for $n=0,1,2,3$.  
Here, no amplitude factor is applied.
}
\label{fig:wkb-ps_support}
\end{figure}

\subsubsection{Hamilton-Jacobi Equation}
To construct the JWKB eigenfunctions, one must solve the HJE 
\begin{align}
\WavePhase(\tcoord,\xcoord;  \xcoord_0,\tcoord_0)
&=\int_{\tcoord_0,\xcoord_0}^{\tcoord,\xcoord}  \pcoord d\xcoord -\Energy d\tcoord  = \left.\RouthianFunc - \Energy \tcoord \right|_{\tcoord_0,\xcoord_0}^{\tcoord,\xcoord}
\end{align}
which  requires the solutions to the classical equations of motion.
The phase space eigenfunctions have support on the Lagrangian submanifold shown on the left of Fig.~\ref{fig:wkb-ps_support}. The classically allowed region is defined by the circle $ (\xcoord/\xcoord_0)^2+(\pcoord/\pcoord_0)^2=2\Jcoord/\hbar$ (blue circle) where  $\xcoord_0^2=\hbar/\mass\omega$ and $\pcoord_0^2=\hbar \mass \omega$.  The classically forbidden region (shaded red) is defined by the hyperbolae $ (\xcoord/\xcoord_0)^2-(\pcoord/\pcoord_0)^2=2\Jcoord/\hbar$.  

In the classically allowed region, the solution is
\begin{align}
\xcoord&=\sqrt{2\Jcoord/\mass\omega} \cos{(\theta)} & \pcoord &=  \sqrt{2\Jcoord \mass \omega  } \sin{(\theta)} \\
\theta&=\theta_0+\omega \tcoord & \Jcoord&=\Jcoord_0
\end{align}
which yields the energy $\Energy= \Jcoord \omega$. 
Hamilton's principle function is 
\begin{align}
\RouthianFunc = \Jcoord\angle+\pcoord \xcoord/2=\Jcoord\left(\angle + \sin{(2\angle)}/2\right). 
\end{align}
As a function of $\angle$, there is only one branch of $\RouthianFunc(\angle)$ which steadily increases in the clockwise direction.
However, as illustrated in Fig.~\ref{fig:wkb-ps_support}, the two solution branches for the momentum
\begin{align}
 \pcoord_\pm = \pm(2 \Jcoord\mass\omega -\mass^2 \omega^2 \xcoord^2  )^{1/2}
\end{align}  
 yield two solution branches for the phase, which defines the Lagrangian submanifold $\pcoord=\partial_\xcoord\WavePhase$.
Thus, as a function of $\xcoord$, there are two branches of the principle function $\pm\RouthianFunc(\xcoord)$, corresponding to $\pm\pcoord$.
In order to match the required periodicity conditions, it is useful to define the change in Hamilton's principle function 
\begin{align}
\RouthianFunc_\pm = 
\left. \left(  \Jcoord \theta + \pcoord   \xcoord /2   \right)\right|^\xcoord _{\xi_\pm} 
\end{align}
from each of the two classical turning points $ \xi_\pm=\pm \sqrt{2\Jcoord/\hbar}$.
With this definition $\RouthianFunc_\pm>0$ ($<0$) in the clockwise (counterclockwise) direction.
In this case, the $\pm\pcoord$ branches are $\pm\RouthianFunc_-(\xcoord)$ from the leftmost turning point to the right and $\pm \RouthianFunc_+(\xcoord)$ from the rightmost turning point to left.
Note that many authors use a convention for the Routhian (e.g. \cite{Grunwald71ajp}) that is even on the two branches. This is perfectly acceptable for configuration space eigenfunctions which are determined by an even cosine function. However, that convention does not give the correct phase space eigenfunctions because they also depend on the odd sine function. 

Determination of the eigenfunction over the entire range of $\xcoord\in\left(-\infty,+\infty\right)$  also requires treating the classically forbidden region where $\xcoord^2 > 2J/\mass \omega$ and $\pcoord^2 <0$.
Clearly, the momentum must become imaginary, which is possible if the conjugate angle $\theta=-i\vartheta$ also becomes imaginary.
This yields the solutions
\begin{align}
\xcoord_\pm&=\pm \sqrt{2  \Jcoord  /\mass\omega} \cosh{(\vartheta)} & p &=  i \sqrt{2  \Jcoord  \mass \omega  }  \sinh{(\vartheta)} \\
\vartheta&=\vartheta_0+i\omega \tcoord & \Jcoord&=\Jcoord_0
\end{align}
where $\vartheta\geq 0$.
Now, we define the absolute value of the change in action
\begin{align}
 \RouthianForbidden_\pm=\left.  \abs{ \Jcoord \vartheta}   + \abs{ \xcoord_\pm  \pcoord_\pm} /2 \right|^\xcoord_{\xi_\pm} =\Jcoord  \abs{ \vartheta  + \sinh{\vartheta}\cosh{\vartheta}} 
 .
\end{align}
 As illustrated in Fig.~\ref{fig:wkb-ps_support}, there are still two branches of the Lagrangian submanifold for $\pm i\pcoord>0$. However, the wavefunction only has support on branches where it decays.

\begin{figure}[tbp]
\centering
\includegraphics[height=1.75in]{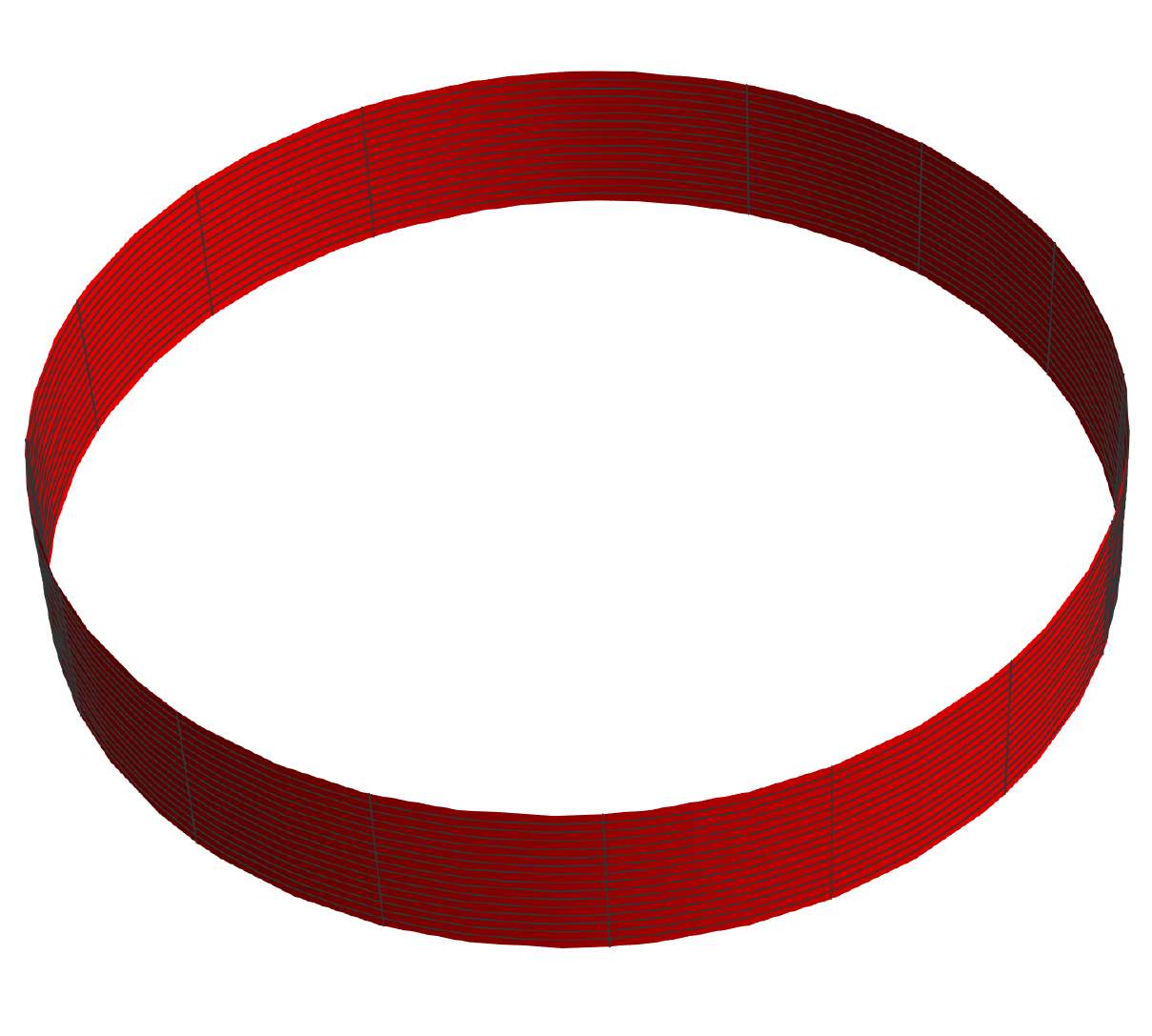} 
\includegraphics[height=1.75in]{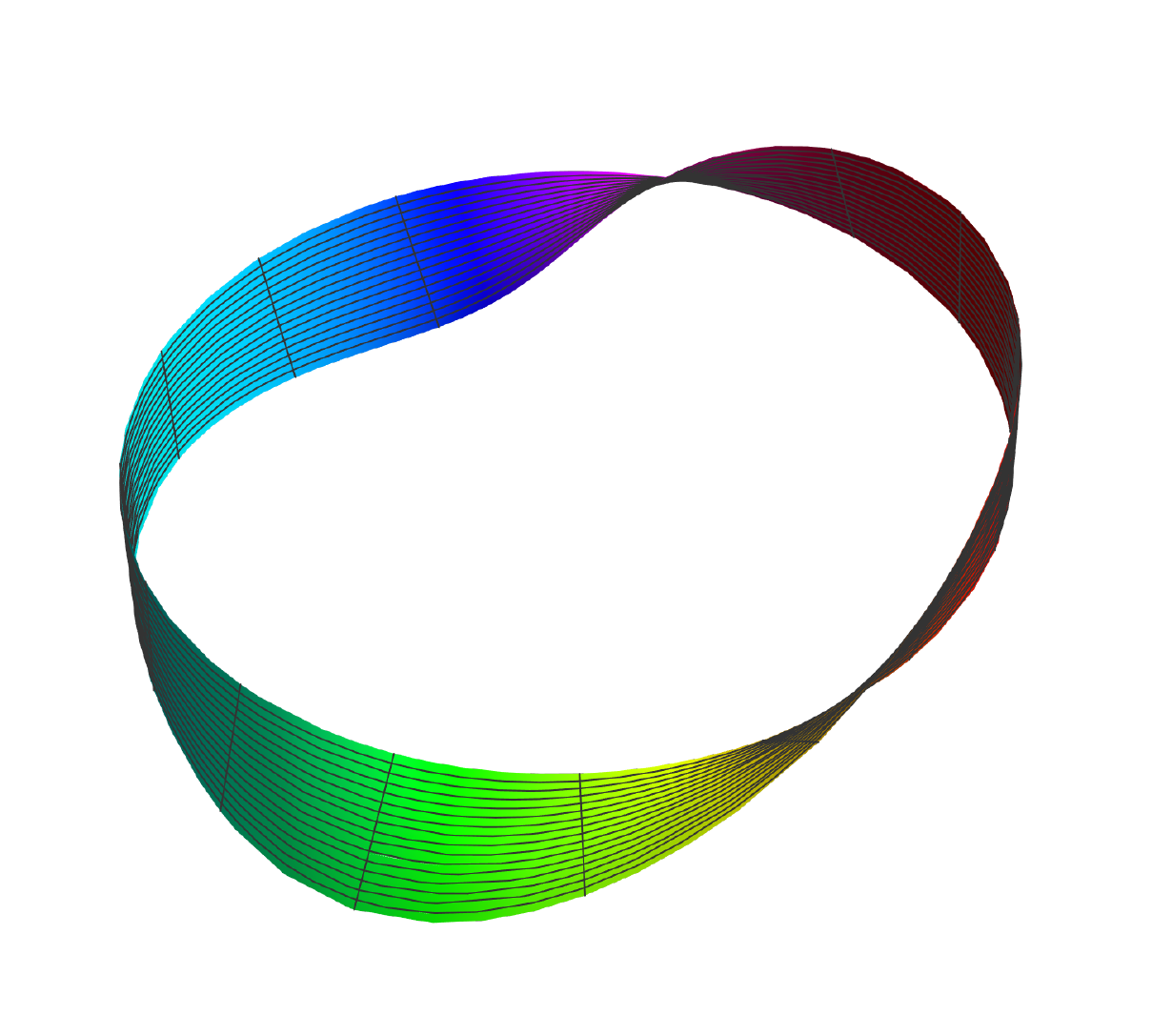}\\
\includegraphics[height=1.75in]{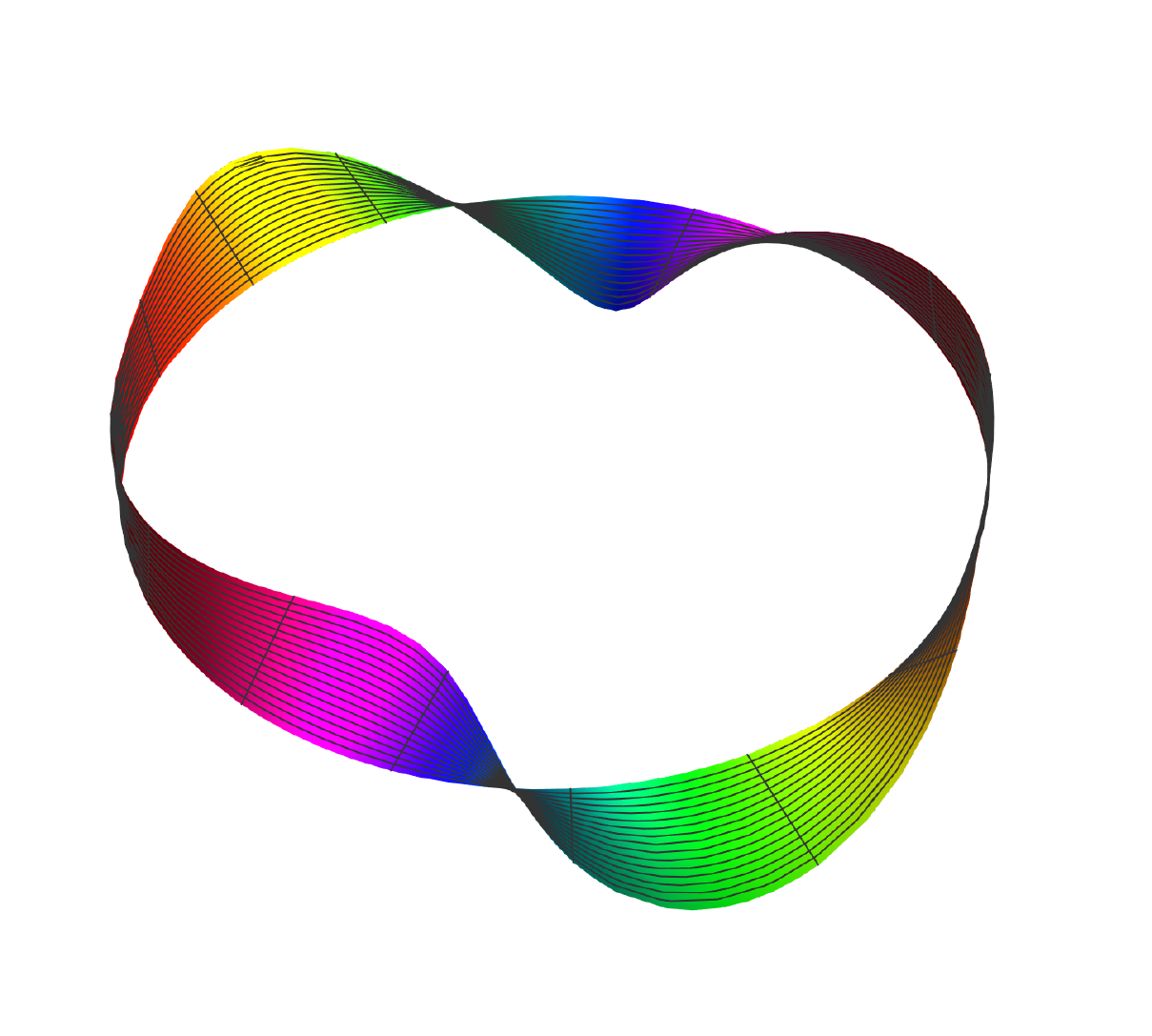} 
\includegraphics[height=1.75in]{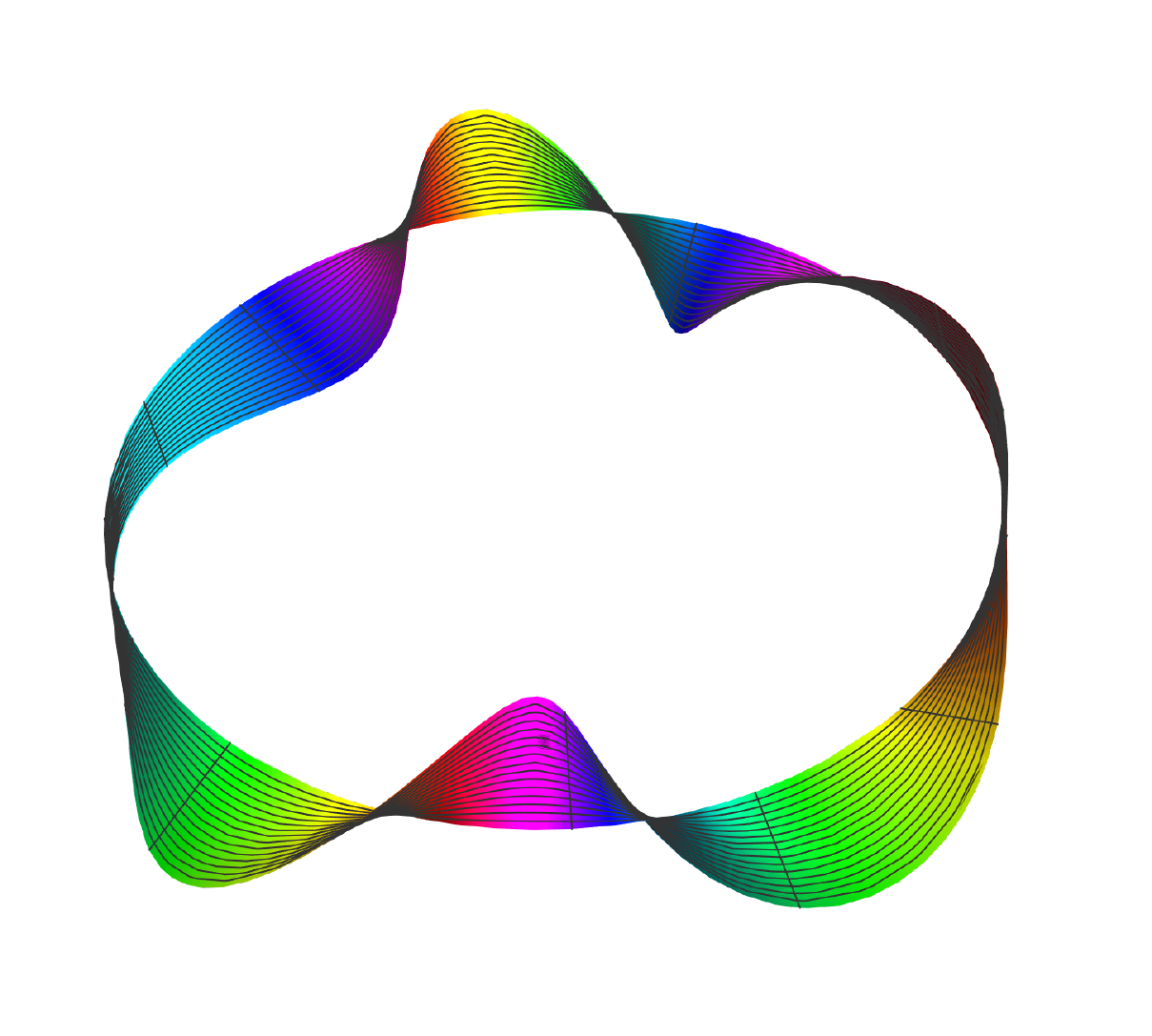}
\caption{ The KvH harmonic oscillator eigenfunctions in phase space  using Bohr-Sommerfeld quantization are continuous: $\nuindex=0$ plotted with finite radius (upper left), $\nuindex=1$ (upper right), $\nuindex=2$ (lower left), $\nuindex=3$ (lower right). For each eigenfunction, the complex phase is plotted as a ribbon the rotates around the ring that defines the Lagrangian submanifold; the phase is also indicated by the hue of the ribbon. 
}
\label{fig:bohr-eigs}
\end{figure}

\subsubsection{Phase Space Eigenfunctions}

In phase space, the semiclassical KvH wavefunctions have the form
\begin{align}
\BasisFunc^\SC_{\Jcoord_0} (\zcoord)  & = \waveFunc_{\Jcoord_0}(\zcoord) \delta(\Jcoord-\Jcoord_0) \abs{\mass\omega/\pcoord}^{1/2}
\\
&= \left[\waveFunc_{\Jcoord_0,+}(\zcoord) \delta(\pcoord-\partial_\xcoord \RouthianFunc(\xcoord,\Jcoord_0)) 
+ \waveFunc_{\Jcoord_0,-}(\zcoord) \delta(\pcoord+\partial_\xcoord \RouthianFunc(\xcoord,\Jcoord_0))
\right]\abs{\pcoord/\mass\omega}^{1/2}.
\end{align}
In this form, phase jumps caused by the square root of $\pcoord$ are absorbed into the complex phase factor of $\waveFunc_{\Jcoord_0}$.
The $\pcoord^{-1/2}$  factor causes the solution to add the phase $-\pi/4$ when proceeding clockwise ($+\pi/4$ when proceeding counterclockwise) for each half turn through a classical turning point.

\begin{figure}[tbp]
\centering
\includegraphics[height=1.5in]{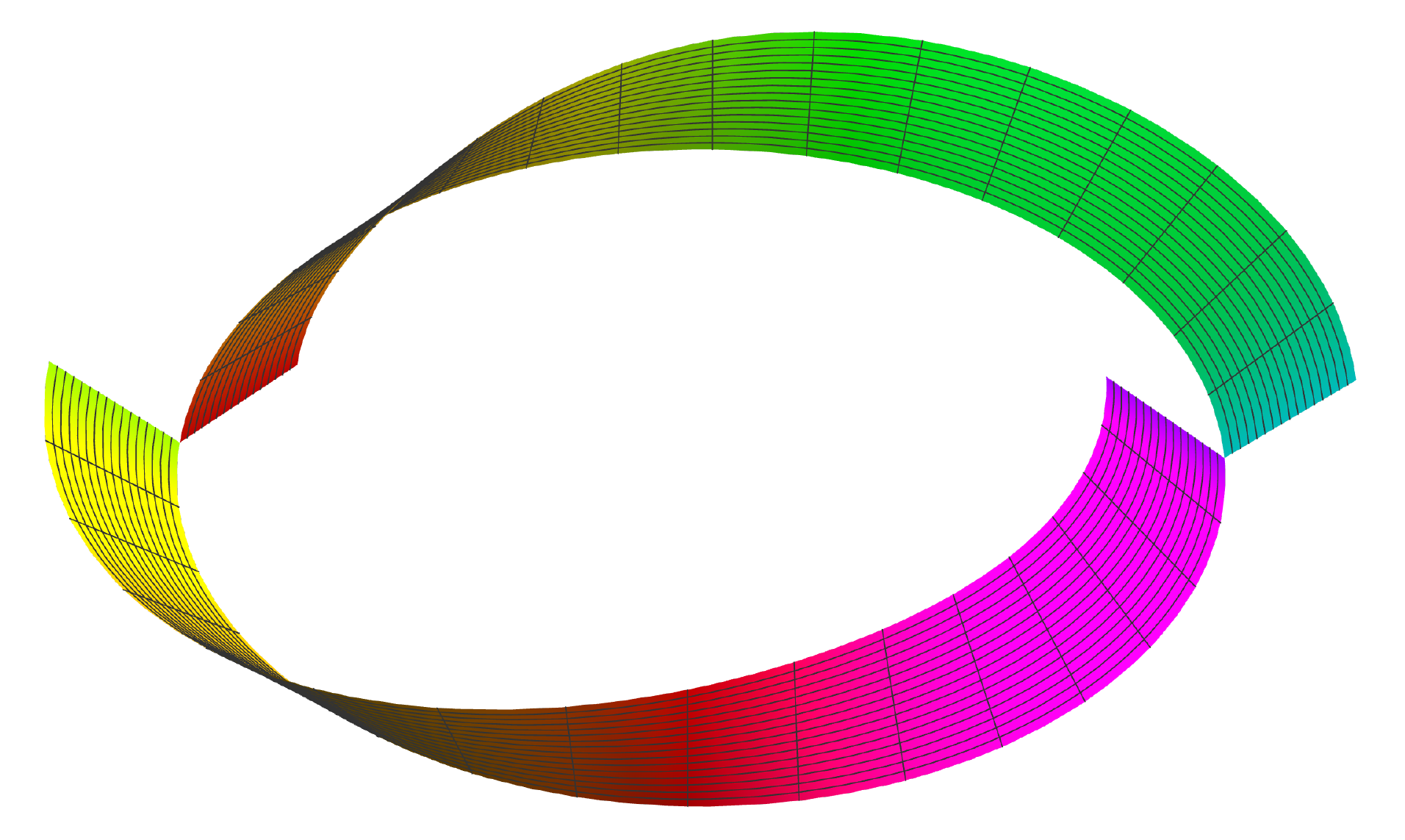} 
\includegraphics[height=1.5in]{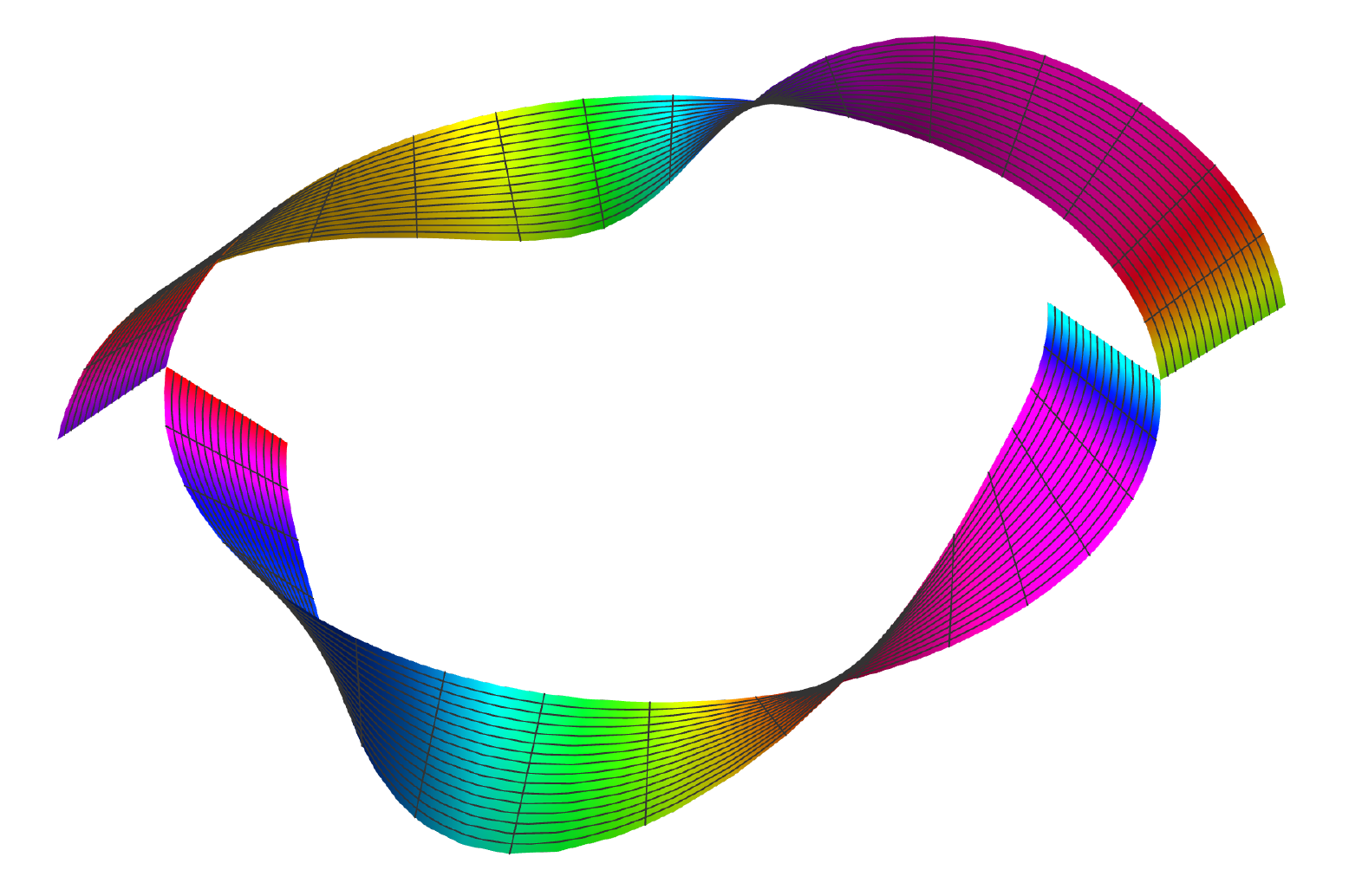}\\
\includegraphics[height=1.5in]{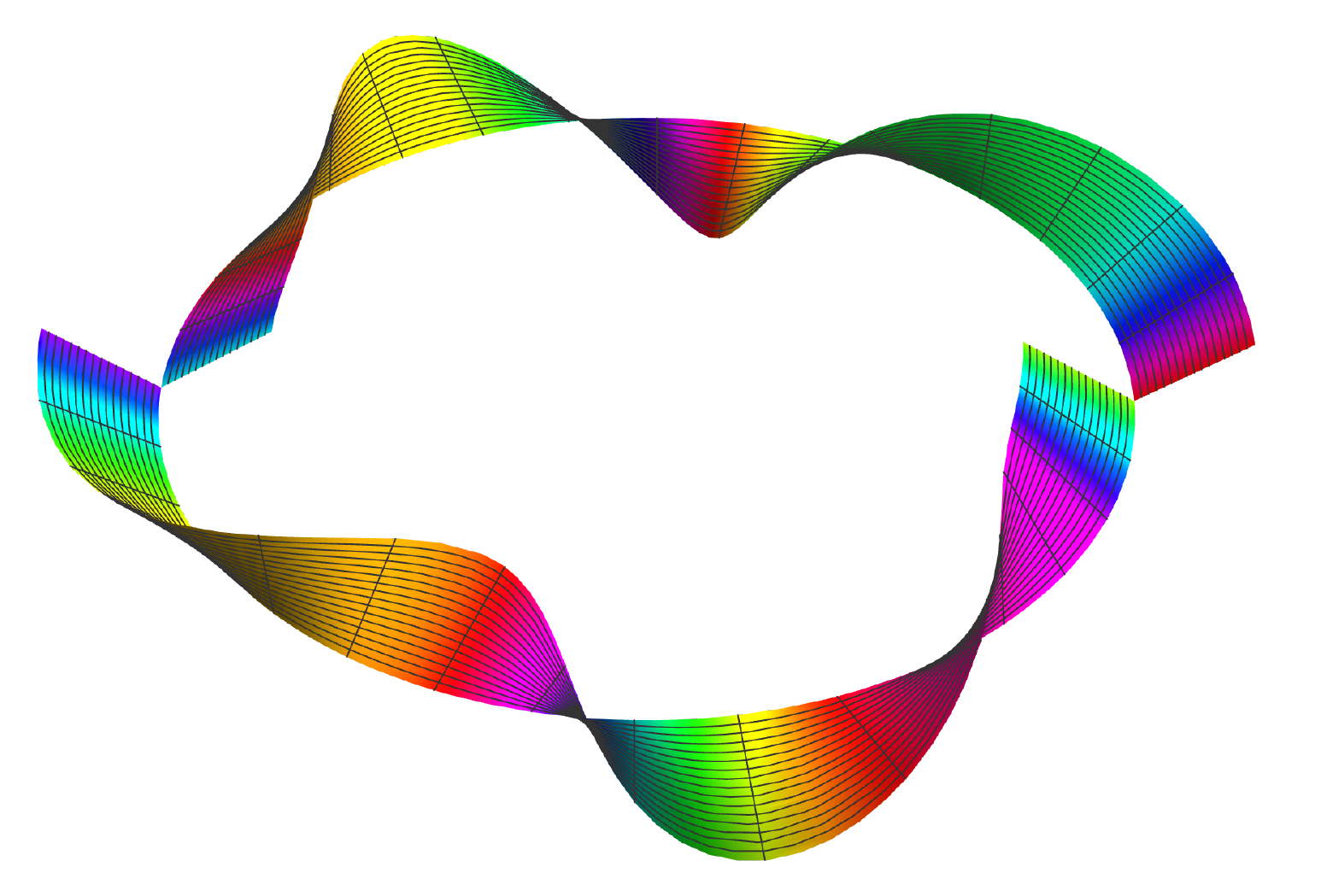} 
\includegraphics[height=1.5in]{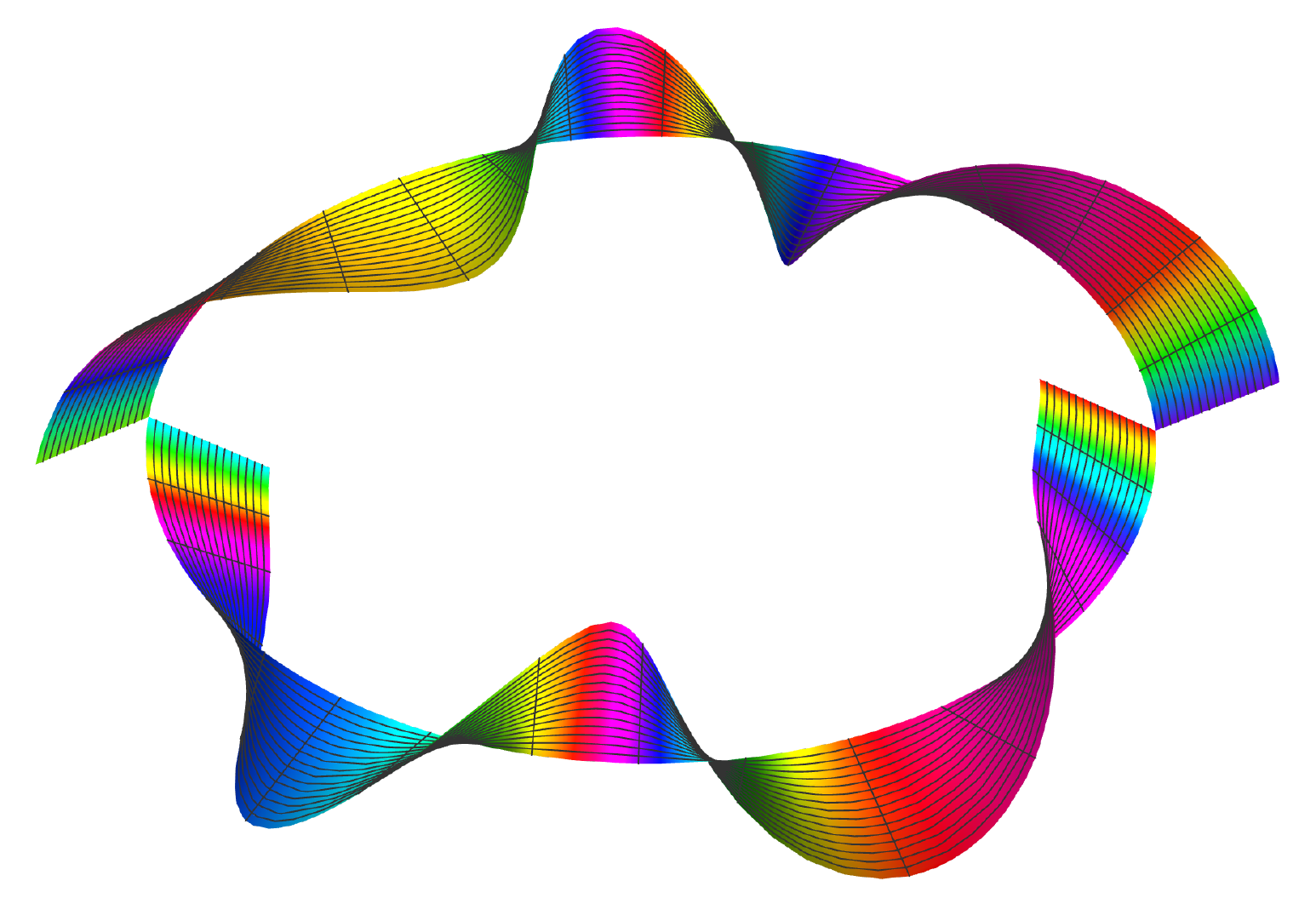}
\caption{ The KvH harmonic oscillator eigenfunctions in phase space  using EBK quantization with  the Keller-Maslov correction: $\nuindex=0$ plotted with finite radius (upper left), $\nuindex=1$ (upper right), $\nuindex=2$ (lower left), $\nuindex=3$ (lower right). For each eigenfunction, the complex phase is plotted as a ribbon the rotates around the ring that defines the Lagrangian submanifold; the phase is also indicated by the hue of the ribbon. Note the discontinuous jump in phase at classical turning points.
}
\label{fig:JWKB-eigs}
\end{figure}

In classically forbidden regions, the wavefunction must decay
\begin{align}
\waveFunc_\Jcoord( \zcoord) =b_\pm e^{- \RouthianForbidden_\pm/ \hbar  }.
\end{align}
In the classically allowed region, the wavefunction is a single exponential that can be written in two different ways
\begin{align}
\waveFunc_\Jcoord( \zcoord) =a_+ e^{ i (\RouthianFunc_+/ \hbar + \pi/4)  }
=a_-  e^{ i (\RouthianFunc_-/ \hbar - \pi/4)  }
\end{align}
where  the additional $\pi/4$ is due to the Maslov index. Due to symmetry $\abs{b_+}=\abs{b_-}=\abs{a_+}=\abs{a_-}$.
In order to ensure continuity, the phase of the wavefunction must satisfy $\oint\dee\log{\psi} /2\pi i=  \nuindex\in \Integers$ around any closed loop.
In this case, $b_-=(-1)^\nuindex b_+$ and $a_-=(- 1)^\nuindex a_+$.  
Here,  the choice $a_+=b_+=1$ is made, so that the wavefunction is always positive for the right turning point $\xi_+$.

For Bohr-Sommerfeld quantization, one neglects the $\pi/4$ phase jumps caused by the square root of $\pcoord$, so that the eigenfunction takes the form $\exp(i\RouthianFunc_\pm/\hbar)$. In this case, the quantization condition is 
$\Jcoord/\hbar = \nuindex\in\Naturals$. The first few eigenfunctions for $\nuindex=0,1,2,3$ are shown for the classically allowed region in Fig.~\ref{fig:bohr-eigs}. 
Technically speaking, $\nuindex=0$ does not map to a discrete eigenfunction without some regularization procedure.  For comparison to the JWKB eigenfunctions, in the figure, the $\nuindex=0$ eigenfunction is considered to have finite radius.

For EBK quantization, the periodicity condition is
$\jindex= \Jcoord / \hbar =\nuindex + 1/2$.
The additional 1/2, corresponding to an additional phase change of $\pi$, results from the Maslov index $\mu=2$.
The phase space eigenfunctions are given by  (compare to \cite{Grunwald71ajp})
\begin{align}
\waveFunc_\Jcoord(\zcoord) =\left\{\begin{array} {lll} 
 a_+ \exp{\left( i\left[ \jindex (\theta+\sin{(2\theta)} /2) 
 +\pi/4\right]\right)}  
  \\
(\sign{x})^{\nuindex} a_+ (\abs{\sinh{\vartheta}}+ \cosh{\vartheta})^{\jindex} 
\exp{ \left(-\jindex \abs{\sinh{(2\vartheta)}/2}\right) }  
\end{array}\right.
\end{align}
in the classically allowed and forbidden regions, respectively.
Note that the amplitude of the wavefunction is continuous across the transition from classically allowed to forbidden. 
In phase space, only the complex phase has a discontinuity.

\begin{figure}[tbp]
\centering
\includegraphics[height=1.925in]{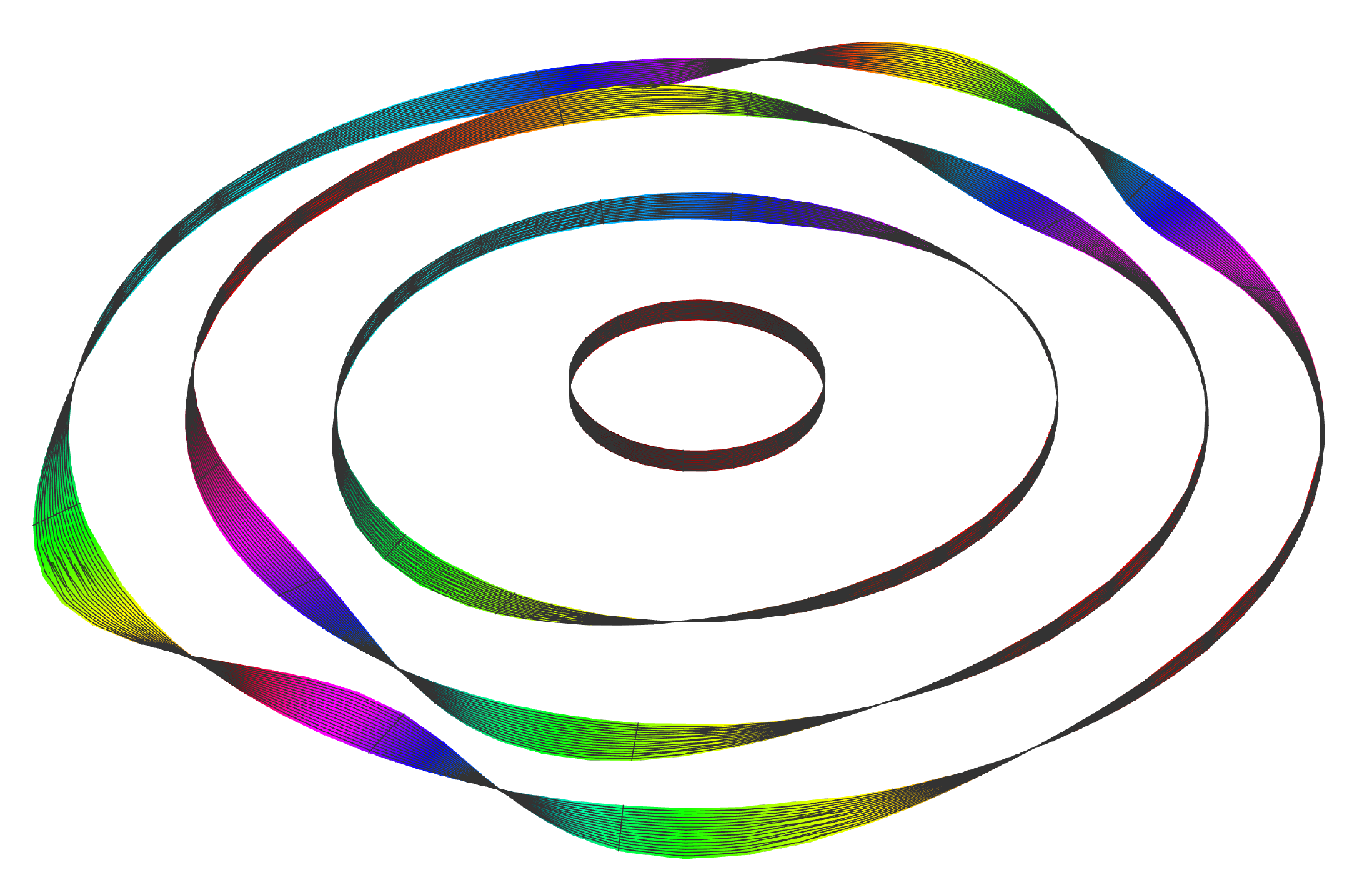} 
\includegraphics[height=1.925in]{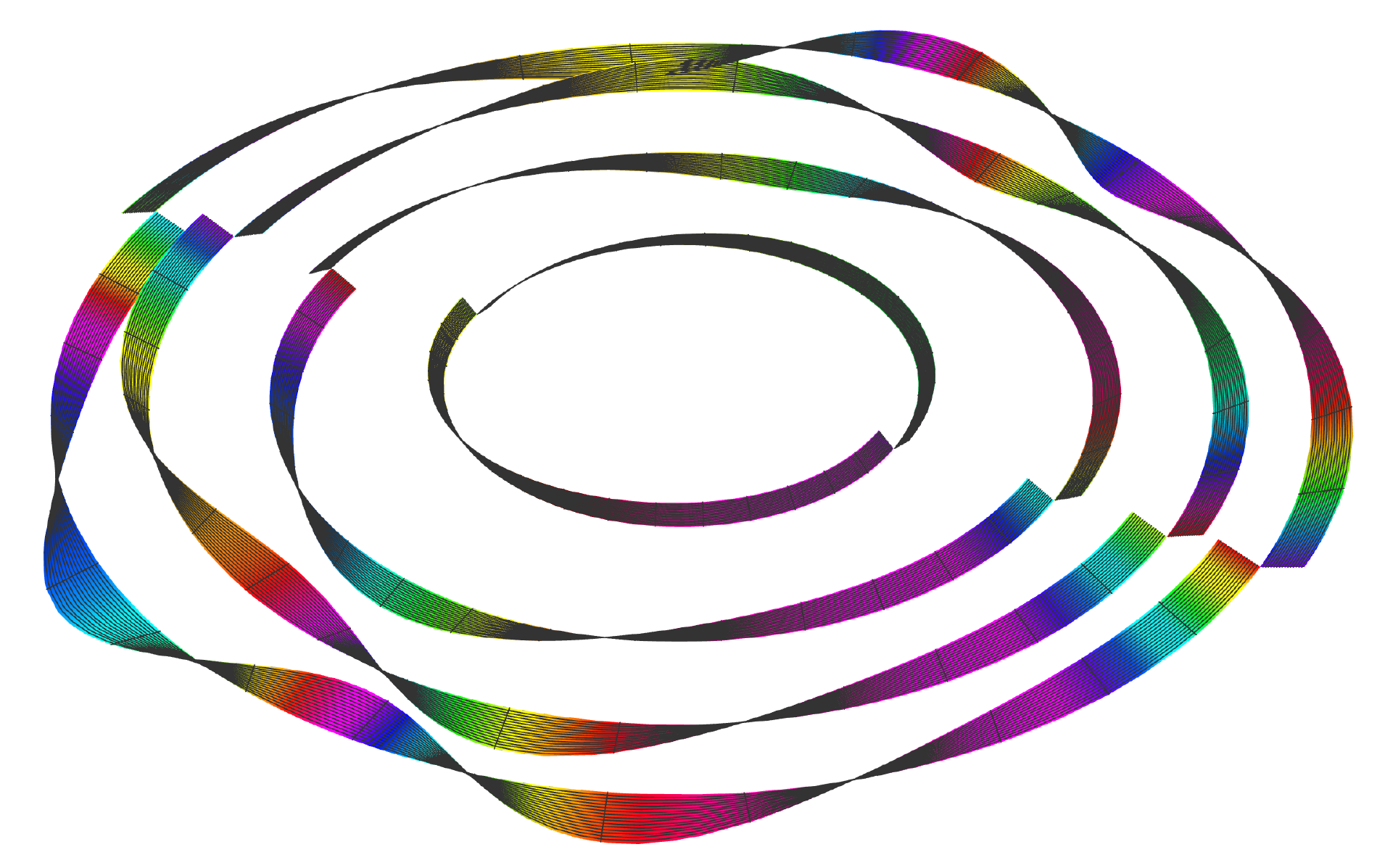} 
\caption{ A uniform superposition of KvH harmonic oscillator eigenfunctions in phase space including $\nuindex=0,1,2,3$: (left) Bohr-Sommerfeld quantization, (right) EBK quantization.
The $\nuindex=0$ Bohr-Sommerfeld eigenfunction is illustrated on a circle of finite radius.}
\label{fig:wkb-ps_super}
\end{figure}

The first few eigenfunctions for $\nuindex=0,1,2,3$ are shown for the classically allowed region in Fig.~\ref{fig:JWKB-eigs}. 
Note the jump in phase at the classical turning points; this jump must be untwisted by the phase along each section.
A uniform superposition of $\nuindex=0,1,2,3$ eigenfunctions is illustrated in Fig.~\ref{fig:wkb-ps_super}.  
In phase space, each eigenfunction has support on a different submanifold, so there can be no interference for different values of $\Jcoord$.
If the KvH equation is treated as a standard PDE, the KvH eigenfunctions take the Bohr-Sommerfeld form. 
If the Keller-Maslov correction is taken into account, the KvH eigenfunctions must have a discrete jump in phase at the classical turning points.

The real part of the phase space eigenfunctions are shown in Fig.~\ref{fig:jwkb-ps_real}, now including both the classically allowed and forbidden regions on the same plot.
The imaginary part of the eigenfunctions is odd across the $\pm\pcoord>0$ branch, so that the overall contribution to the configuration space eigenfunctions is real.
The dependence of the real part of the first few eigenfunctions on the angle $\angle=\tan^{-1}{(\pcoord/\mass \omega \xcoord)}$ is shown on the right of Fig.~\ref{fig:wkb-ps_support}.

\subsubsection{Configuration Space Eigenfunctions}
\begin{figure}[tb]
\centering
\includegraphics[width=3in]{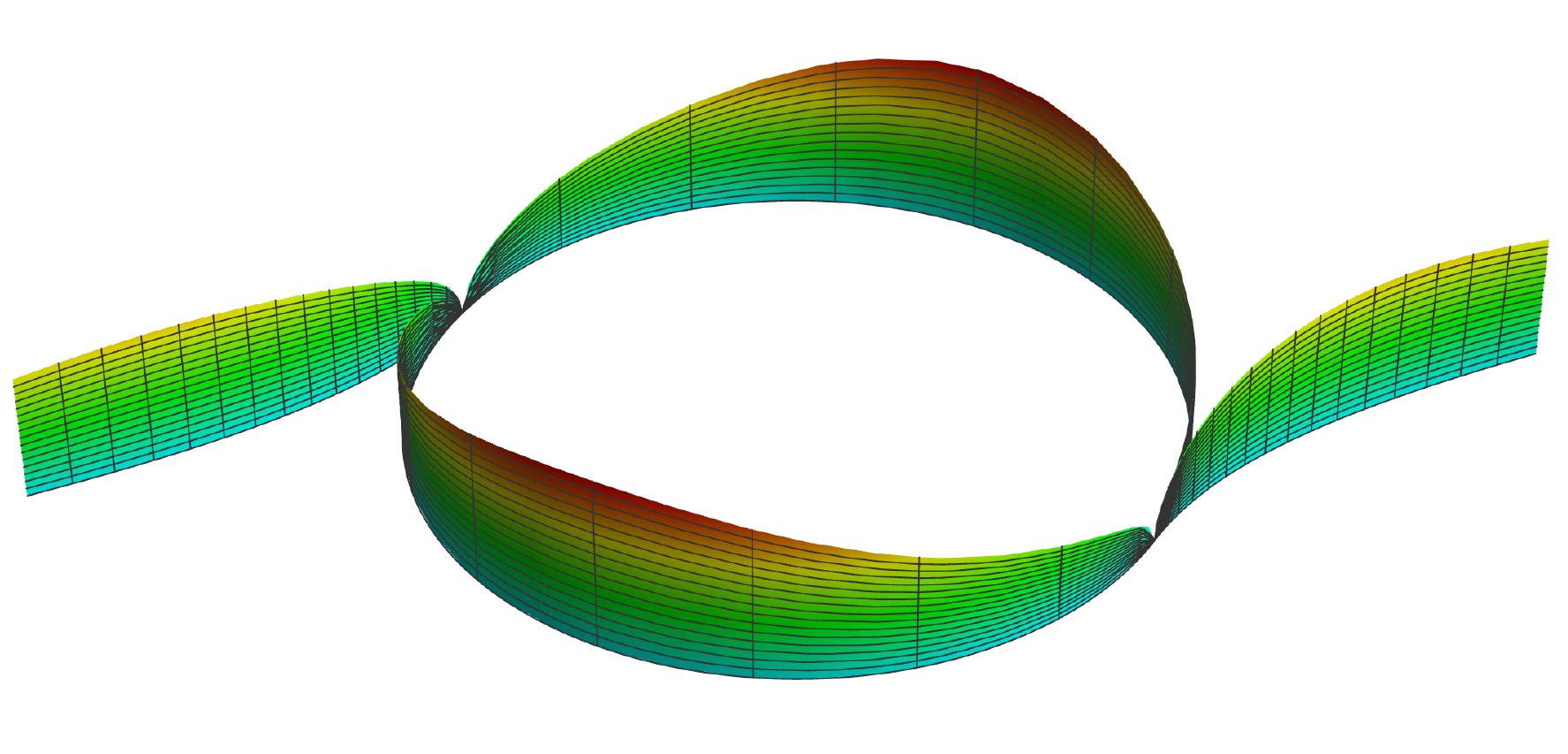} 
\includegraphics[width=3in]{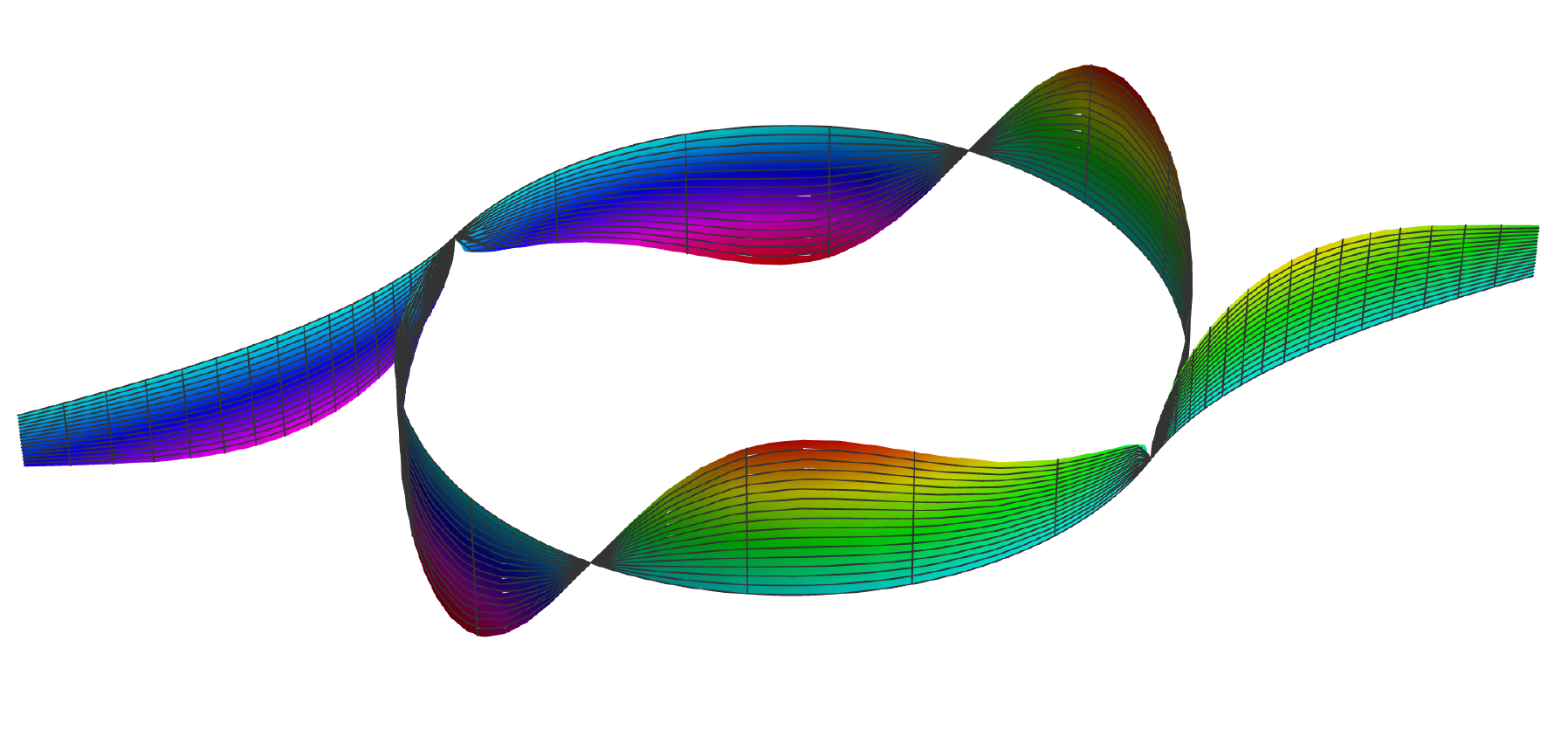}\\
\includegraphics[width=3in]{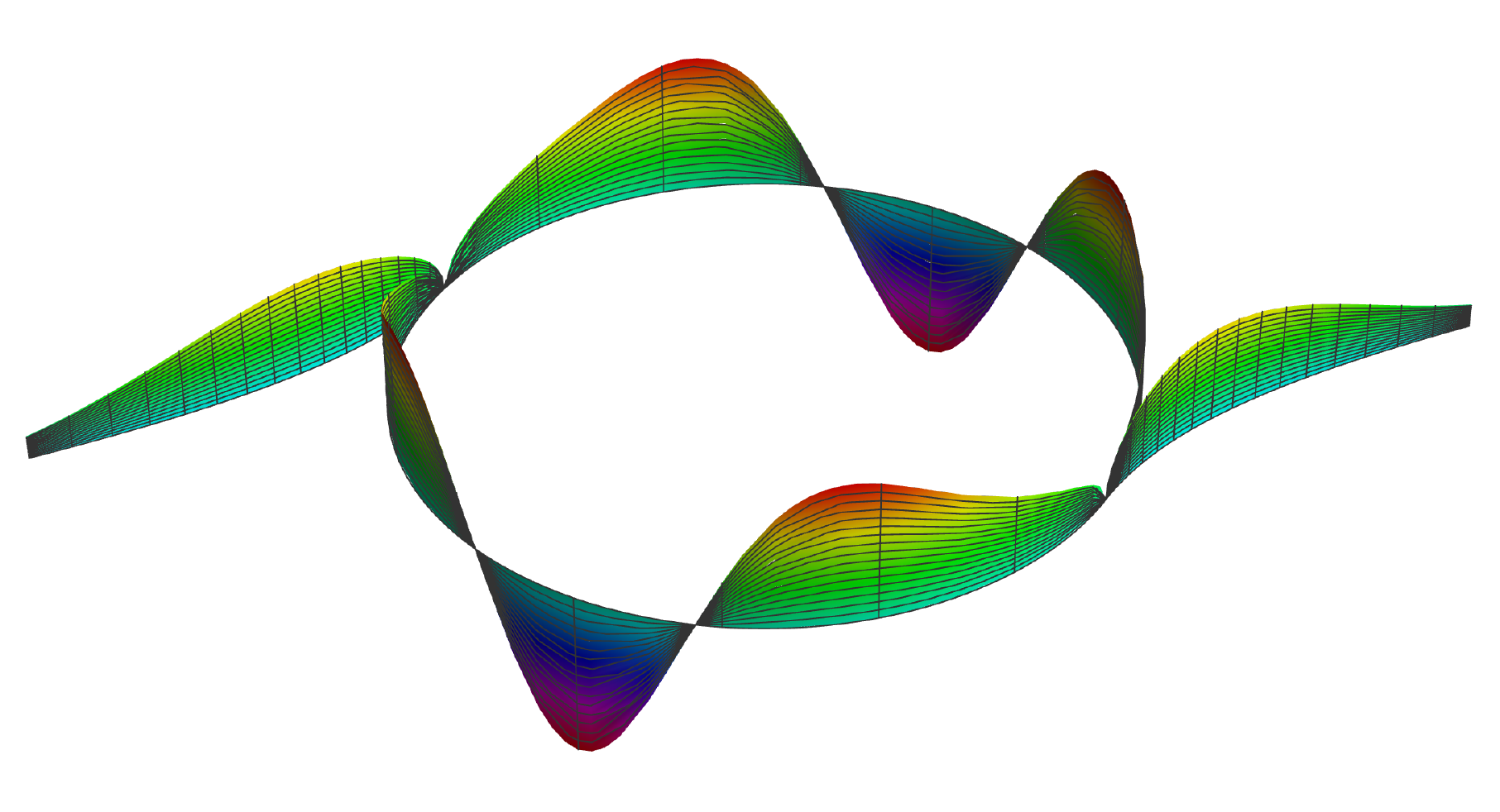} 
\includegraphics[width=3in]{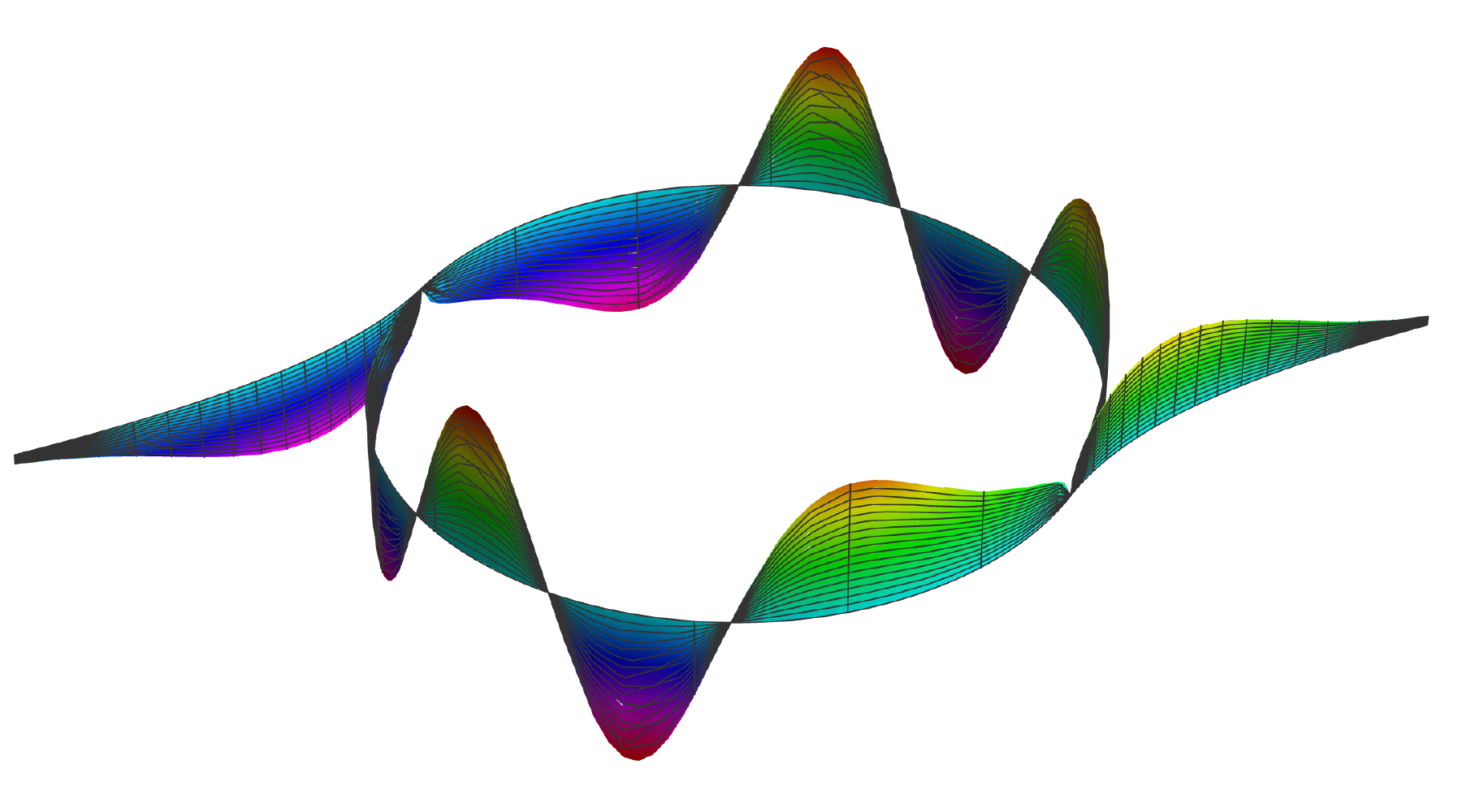}
\caption{  The real part of the phase space harmonic oscillator eigenfunctions with JWKB matching conditions and $\abs{p}^{1/2}$ amplitude factor: $\nuindex=0$ (upper left), $\nuindex=1$ (upper right), $\nuindex=2$ (lower left), $\nuindex=3$ (lower right). For each eigenfunction, the circular submanifold for $(\xcoord/\xcoord_0)^2\leq 2\Jcoord/\hbar$ is plotted vs. $\pcoord$ and the (half) hyperbolic submanifold for $(\xcoord/\xcoord_0)^2> 2\Jcoord/\hbar$ is plotted vs. $-i\pcoord$.
For each plot, the coordinates $\xcoord/\xcoord_0$ and $ \pcoord/\pcoord_0 $ are scaled to $(2\Jcoord/\hbar)^{1/2}$ to focus on the region of interest.
}
\label{fig:jwkb-ps_real}
\end{figure}

Integrating $\BasisFunc^\SC_{\Jcoord_0}$ over momentum space yields the configuration space eigenfunctions. 
The configuration space wavefunction is the sum 
\begin{align}
\basisFunc_{\Jcoord} (\xcoord) 
=\sum_\sigma \waveFunc_{\Jcoord, \sigma}(\zcoord)    \abs{\mass\omega/\pcoord}^{1/2}
=\waveFunc_{\Jcoord }(\zcoord)   \abs{\mass\omega/\pcoord}^{1/2}
\end{align}
where $\sigma$ labels the upper and lower branches of  the  Lagrangian submanifold.
Because there is only a single branch in the forbidden regions, the wavefunction has the same form as before
\begin{align}
\waveFunc_\Jcoord( \xcoord) =(\pm 1)^\nuindex a_+ e^{-\RouthianForbidden_\pm/ \hbar  }.
\end{align}
In the classically allowed region, the wavefunction is the sum over both branches
\begin{align}
\waveFunc_\Jcoord( \xcoord) = 2a_+ \cos{(\RouthianFunc_\pm/ \hbar\pm \pi/4 )}.
\end{align}
Only the real part remains, but it has twice the amplitude that it did in phase space.
In contrast, the Bohr-Sommerfeld eigenfunctions would have the form $\waveFunc_\Jcoord=2a_+ \cos{(\RouthianFunc_\pm/\hbar)}$.
This is also the form for the classical limit, $\hbar\rightarrow 0$, where the semiclassical spectrum becomes continuous.

In order to better understand how this result arises, the real part of the semiclassical KvH phase space eigenfunctions are plotted in Fig.~\ref{fig:jwkb-ps_real}, including both the classically allowed and classically forbidden regions.
For each eigenfunction, the classically allowed region, defined by the circle at center, is plotted vs. $\setlist{\xcoord/\xcoord_0,\pcoord/\pcoord_0} \in \Reals$.
The classically forbidden region, defined by the hyperbolic tails, is plotted vs. $\setlist{\xcoord/\xcoord_0,-i\pcoord/\pcoord_0} \in \Reals$.
Unlike the absolute value of the phase space wavefunction, the real part is discontinuous at the boundary between classically allowed and forbidden regions.

The configuration space JWKB eigenfunctions $\basisFunc_\Jcoord(\xcoord)$ are compared to the exact harmonic oscillator eigenfunctions on the right of Fig.~\ref{fig:wkb-cs}. 
The configuration space eigenfunctions \cite{Grunwald71ajp} are specified by $\jindex=\Jcoord/\hbar=\nuindex+1/2$ and
\begin{align}
\waveFunc_\Jcoord (\xcoord)=\left\{\begin{array} {lll} 
2a_+ \cos{\left(  \jindex[\theta+\sin{(2\theta)} /2 ] 
+\pi/4\right)}  
  \\
( \sign{x})^{\nuindex} a_+ (\abs{\sinh{\vartheta}}+ \cosh{\vartheta})^{\jindex} 
\exp{ \left(-\jindex \abs{\sinh{(2\vartheta)}/2}\right) }  
\end{array}\right.
\end{align}
in the classically allowed and forbidden regions, respectively.
In configuration space, the eigenfunctions are real, so the phase is continuous  across the transition from the classically allowed to forbidden region, but there is a discontinuity in magnitude.

The contribution of the standard phase space KvH eigenfunctions, $\waveFunc_\Jcoord (\xcoord)$, with JWKB matching conditions, is shown in the middle of Fig.~\ref{fig:wkb-cs}. 
In the integrand of Eq.~\ref{eq:KvH_phase-space_projection1}, this contribution is to be multiplied by a $\abs{\pcoord}^{1/2}$ weight factor.
The semiclassical KvH eigenfunctions carry this factor, so the overall dependence of the integrand in Eqs.~\ref{eq:KvH_semiclassical_projection} and \ref{eq:KvH_phase-space_projection1} is shown on the right of Fig.~\ref{fig:wkb-cs}. 
In phase space, the semiclassical eigenfunctions are multiplied by $\abs{\pcoord}^{1/2}$, which generates a zero in the overall contribution to the integrand in Eqs. \ref{eq:KvH_semiclassical_projection} and \ref{eq:KvH_phase-space_projection1}.
The projection to configuration space then applies  $\abs{\pcoord}^{-1}$, for an overall amplitude factor  $\abs{\pcoord}^{-1/2} $, shown on the left of Fig.~\ref{fig:wkb-cs}. 
This is quite noticeable because it generates a weak integrable singularity at the classical turning points and is required for agreement with the exact eigenfunctions far from the turning points.

\subsection{Potential Well \label{sec:potential-well}}

\begin{figure}[tbp]
\centering
\includegraphics[width=2in]{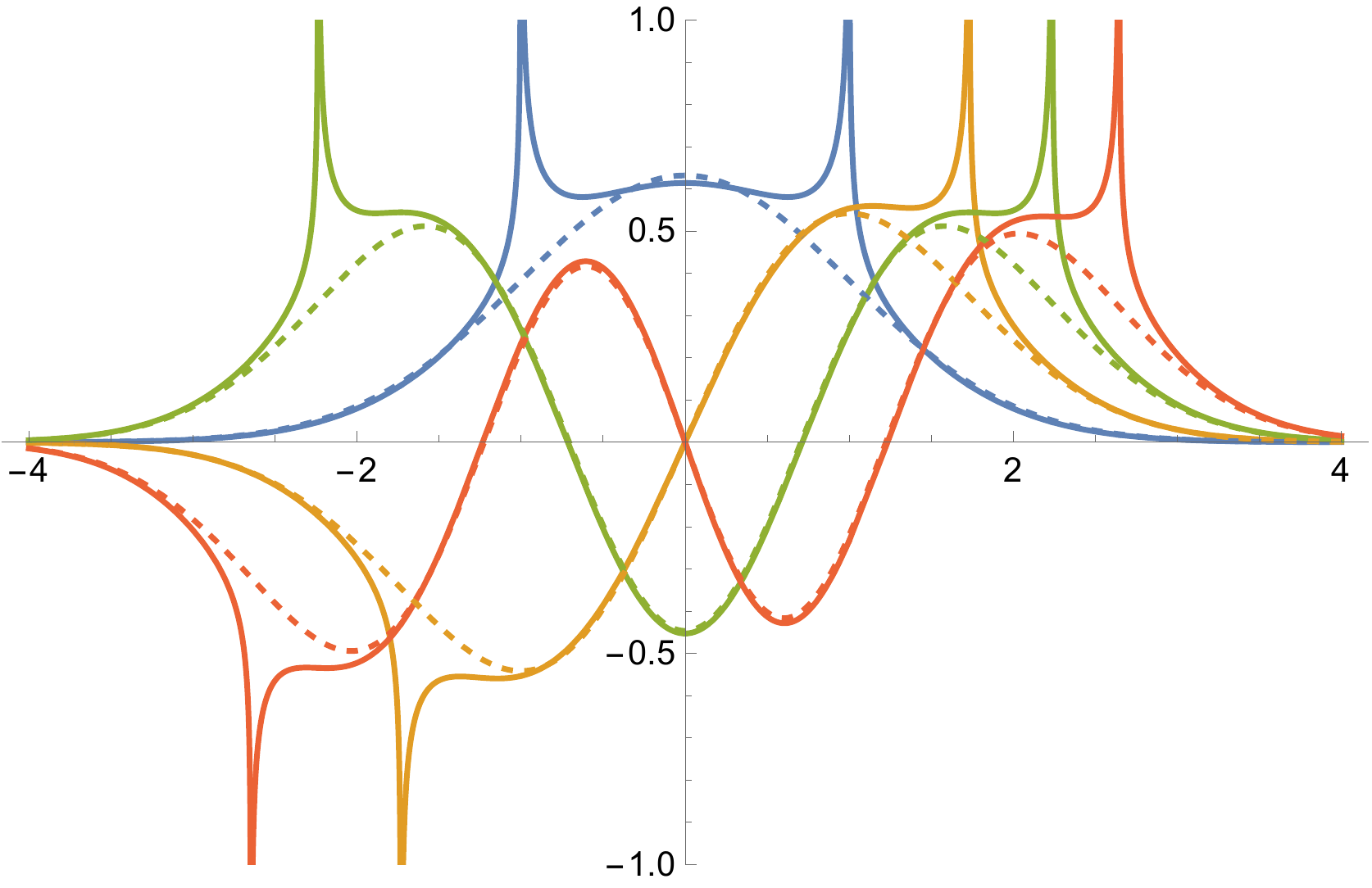}
\includegraphics[width=2in]{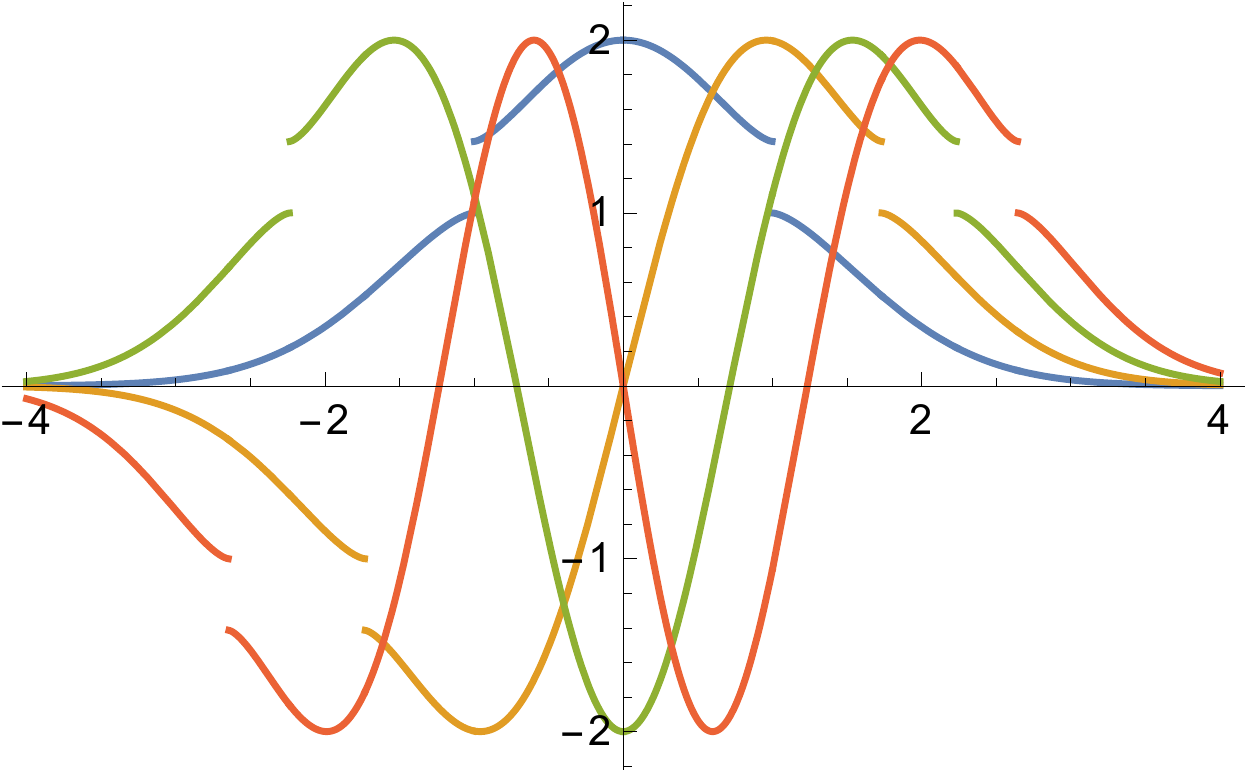} 
\includegraphics[width=2in]{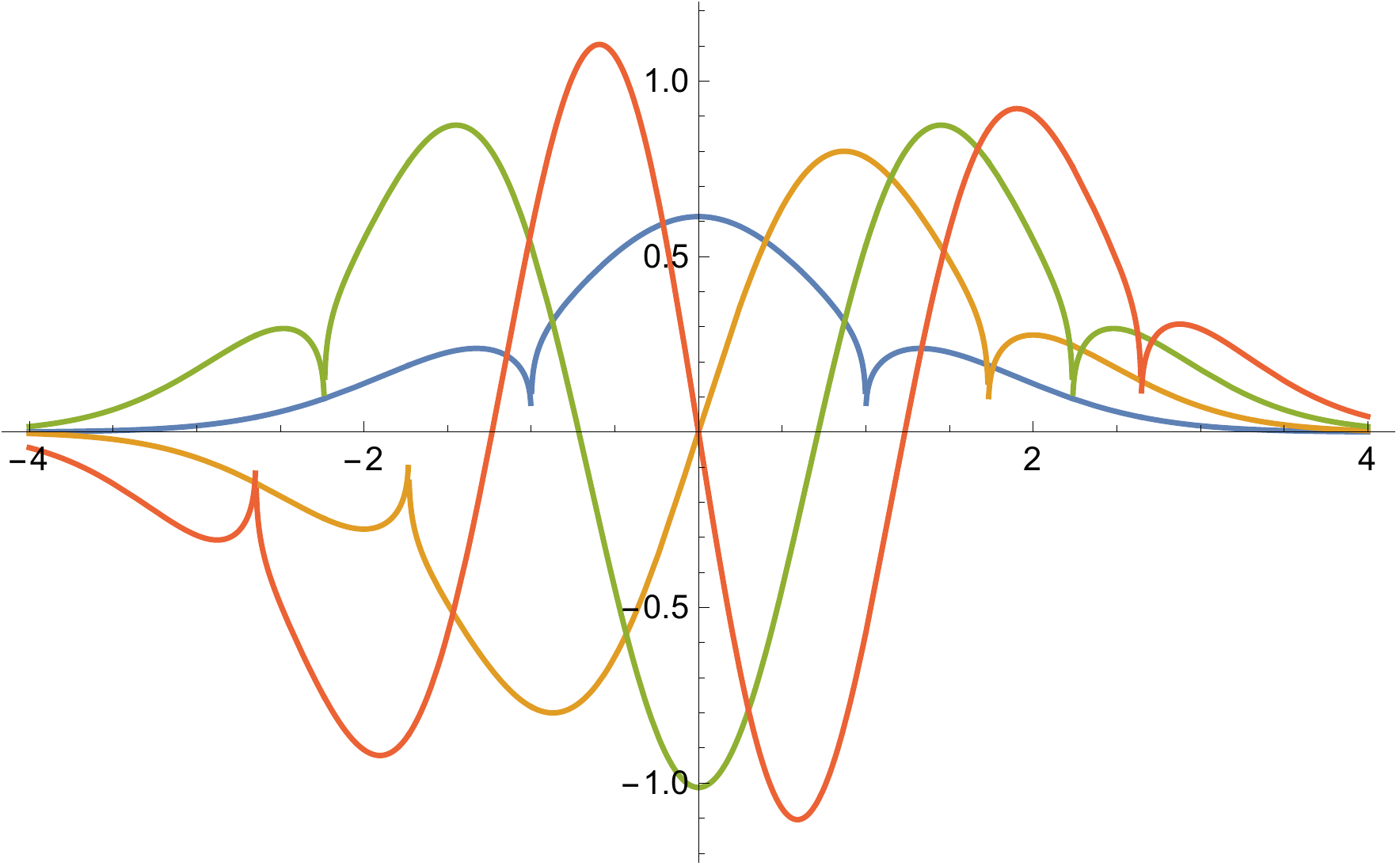} 
\caption{(Left) Comparison of semiclassical harmonic oscillator JWKB
 eigenfunctions (solid) with $\abs{\pcoord}^{-1/2}$ amplitude factor to exact quantum eigenfunctions (dashed) in configuration space for $n=0,1,2,3$ (blue, orange, green, red).
The contribution to the configuration space eigenfunctions: 
(middle) phase space KvH without amplitude factor; 
(right) normalized semiclassical phase space KvH with  $\abs{\pcoord}^{1/2}$ amplitude factor.
 }
\label{fig:wkb-cs}
\end{figure}

In 1D, similar results hold true for any potential energy function $m\omega^2\xcoord^2/2\rightarrow U(\xcoord)$ with similar topology, i.e. having a single potential well.
In this case, there are two classically allowed branches for $\pcoord_\pm(\xcoord,\Jcoord)=\pm(2\mass 
\Energy(\Jcoord) - U(\xcoord))^{1/2}$ and $\RouthianFunc_\pm=\int_{\xi_\mp}^\xcoord \pcoord_\pm\dee\xcoord$ where $\xi_\mp$ are the left/right turning points.
The classically forbidden branches for $\pm\xcoord>0$ are given by $\pcoord_\pm(\xcoord,\Jcoord)=\pm i (2\mass 
\Energy(\Jcoord) - U(\xcoord))^{1/2}$ and $\RouthianForbidden_\pm =-i \int_{\xi_\pm}^\xcoord \pcoord_\pm\dee\xcoord$.

The phase space KvH eigenfunctions have the form 
\begin{align}
\BasisFunc_{\Jcoord_0}(\tcoord, \zcoord) = \waveFunc_\Jcoord e^{i\WavePhase(\tcoord,\xcoord,\Jcoord)/\hbar} \delta^\dim(\Jcoord-\Jcoord_0) 
\end{align}
and the semiclassical KvH eigenfunctions have the form
\begin{align}
\BasisFunc^\mod_{\Jcoord_0}(\tcoord, \zcoord) = \waveFunc_\Jcoord e^{i\WavePhase(\tcoord,\xcoord,\Jcoord)/\hbar} \delta^\dim(\Jcoord-\Jcoord_0) \abs{\partial \pcoord/\partial\Jcoord}^{1/2}
\end{align}
where
\begin{align}
\waveFunc_\Jcoord(\tcoord,\zcoord)=\left\{
\begin{array}{l}
 a_\pm e^{i\left(\RouthianFunc_\pm /\hbar\mp \pi/4\right)}e^{-iE\tcoord/\hbar} \\
 a_\pm e^{- \RouthianForbidden_\pm  /\hbar -i\Energy \tcoord/\hbar  }
\end{array}
\right.
\end{align}
in the classically allowed and forbidden regions, respectively.
The configuration space eigenfunctions in the classically allowed region are given by
\begin{align}
\basisFunc_\Jcoord(\tcoord,\xcoord)=2a_+ \cos{\left(\RouthianFunc_\pm /\hbar\pm \pi/4\right)}e^{-iE\tcoord/\hbar} \abs{\partial \pcoord/\partial\Jcoord}^{1/2}.
\end{align}
In the classically forbidden region, there are two branches for $\pcoord_\pm(\xcoord,\Jcoord)=\pm i (2\mass \Energy(\Jcoord) - U(\xcoord))^{1/2}$ and the configuration space eigenfunctions are given by
\begin{align}
\basisFunc_\Jcoord(\tcoord,\xcoord)=a_\pm e^{- \RouthianForbidden_\pm  /\hbar -i\Energy \tcoord/\hbar  } \abs{\partial \pcoord/\partial\Jcoord}^{1/2}.
\end{align}
The spectrum is discrete due to the EBK quantization condition
\begin{align}
\Jcoord = \oint \pcoord d\xcoord/2\pi = \hbar(\nuindex+1/2).
\end{align}
This leads to energy differences that are approximately $\hbar\omega$, where the classical frequency is $\omega=d\Ham/d\Jcoord =2\pi/\oint d\xcoord/\dot\xcoord$.

In the classical limit $\hbar\rightarrow 0$, the semiclassical spectrum becomes continuous and the Maslov correction can be ignored. Now, the standard phase space KvH eigenfunctions are given by
\begin{align}
\waveFunc_{\Jcoord_0}(\tcoord,\theta,\Jcoord)=a_\pm e^{i  \RouthianFunc_\pm/\hbar - i\Energy \tcoord/\hbar} \delta^\dim  (\Jcoord-\Jcoord_0) ,
\end{align}
the semiclassical eigenfunctions are given by
\begin{align}
\waveFunc^\mod_{\Jcoord_0}(\tcoord,\theta,\Jcoord)=a_\pm e^{i  \RouthianFunc_\pm/\hbar - i\Energy \tcoord/\hbar} \delta^\dim  (\Jcoord-\Jcoord_0) \abs{\partial \pcoord/\partial\Jcoord}^{1/2} ,
\end{align}
and the configuration space eigenfunctions are given by
\begin{align}
\basisFunc_{\Jcoord}(\tcoord,\xcoord)= 2a_+  \cos{ (\RouthianFunc_\pm/\hbar)}e^{ - i\Energy \tcoord/\hbar}\abs{\partial \pcoord/\partial\Jcoord}^{1/2}
\end{align}
in the classically allowed region only. 
This corresponds to the usual eigenfunctions of the standard phase space KvH equation.

\section{Semiclassical Interference and Decoherence \label{sec:interference} }
KvH wavefunctions clearly allow the possibility of forming superposition states and, hence, of observing interference effects.
Thus, in this section we ask the question: \emph{Why would classical probability theory arise from the semiclassical limit of quantum probability theory?}
The answer to this question arises from two separate physical considerations: (1)  semiclassical observables are approximately local in phase space and (2)  semiclassical systems are highly susceptible to decoherence.

\subsection{Local Observables Imply Classical Probability \label{sec:local-observables} }
The Born postulate of quantum mechanics states that the probability of outcome $\ketbra{\zcoord}{\zcoord}$ is given by the trace of this state with the  probability density matrix  $\densityMatrix$. 
Hence, the diagonal part of the density matrix, 
$\prob(\zcoord)=\braOpket{\zcoord} {\densityMatrix} {\zcoord}$, is the classical probability density function (PDF) in a given basis.
For example, if observable $\ObservableOp$ is diagonalized in the $\ket{\zcoord}$ basis, so that $\ObservableOp\ket{\zcoord} = \Observable(\zcoord)\ket{\zcoord}$, then its expectation value
\begin{align} \label{eq:measurement}
\avg{\ObservableOp} := \Tr{(\ObservableOp \densityMatrix )} = \sum_{\ycoord, \zcoord} \braOpket{\zcoord}{\ObservableOp} {\ycoord} \braOpket{\ycoord} {\densityMatrix} {\zcoord}
&= \sum_{ \zcoord  } \braOpket{\zcoord}{\ObservableOp}{\zcoord} \braOpket{\zcoord} {\densityMatrix} {\zcoord}=\sum_\zcoord  \Observable(\zcoord) \prob(\zcoord)
\end{align}  
is only determined by the classical PDF in this basis.
Thus, due to the Born rule, the expectation value of an observable is determined by the rules of classical probability with respect to the eigenstates of the observable.

The key defining characteristic of a classical system is that there is a preferred basis, the phase space basis, $\ket{\zcoord}$, in which all classical observables commute.
Hence, the classical laws of probability in Eq.~\ref{eq:measurement} apply to all observables in the phase space basis.
For the lowest order semiclassical approximation, both position and momentum (Eq.~\ref{eq:HJE-momentum}) are local operators to zeroth order.
The explanation is simply because, in the semiclassical limit, the configuration space basis is composed of extremely large numbers of states, and, so, a macroscopic distance between two wavepackets is easily measurable because there are many localized states ``in between'' the two packets.
From this point of view, the definition of ``macroscopic'' is simply any distance that is much larger that the combined widths of the wavepacket envelopes.
 
Semiclassical wavefunctions  in configuration space certainly display interference effects. 
 The semiclassical eigenfunctions yield a good approximation to the matrix elements of nonlocal operators -- accurate within the appropriate limits of validity. 
One can certainly consider operators that are nonlocal, such as a translation operator, and, hence, are not diagonalized in the position basis.
Simply pushing the semiclassical expansion to higher order forces one to realize that the momentum operator is not diagonalized in the position basis.
Hence, within the framework of semiclassical theory, a wavefunction is necessary because interference effects are measurable in principle.

However, it is important to understand that \emph{if, in practice, the observables are approximated as being local in phase space, then the predictions of the semiclassical wavefunction are completely equivalent to that of a classical PDF in phase space.} 
This is the context in which the phase space KvH equation has typically been applied.

\subsection{Decoherence \label{sec:decoherence}}
Decoherence is yet another reason why a superposition of individual wavepackets might not display interference effects and, instead, follow the laws of classical probability \cite{Zeh70fop, Zurek03rmp}. 
If one repeats the experiment of separating and then recombining two wavepackets in the presence of interactions with the environment, then, in the semiclassical limit, small perturbations to the Hamiltonian can cause order unity changes in phase. 
These perturbations to the phase will become even larger if each wavepacket has many internal degrees of freedom, because each component will have its own interaction with the environment, and the total phase change will be composed of a large sum of terms.  
If the phase is altered to order unity by such interactions, then the phases will appear to be approximately random and interference effects will be washed out.
Hence, uncontrolled interactions with the environment  effectively generate random noise in the evolution of the phase.
In this case, multiple realizations of the same state preparation protocol
will behave as an incoherent mixture.

\section{Conclusion \label{sec:conclusion}}
In summary, we have proven that the asymptotic JWKB theory of partial differential equations (PDEs) results in the semiclassical Koopman-van Hove (KvH) equation on configuration space.
There is a natural injection of configuration space solutions into phase space. 
Hence, all solutions of the semiclassical configuration space KvH equation must satisfy the semiclassical phase space KvH equation.
Because the configuration space solutions also satisfy the Hamilton-Jacobi constraint relations, phase space solutions are generically superpositions of configuration space solutions.
There is a natural projection of phase space solutions onto configuration space.  
Hence, configuration space solutions are defined by a projection of phase space solutions onto configuration space.
However, the semiclassical problem differs from the classical problem in the choice of boundary conditions, the continuity of the eigenfunctions, and the domain of definition.

The analysis shows that the propagator of the semiclassical configuration and phase space KvH equations can be put into correspondence with each other for sufficiently short times.
Moreover, for integrable dynamics where there is a complete solution to the Hamilton-Jacobi equation that is valid for all time, the semiclassical phase space and configuration space KvH eigenstates can be put in one-to-one correspondence.
In general, the KvH eigenspectrum is the Cartesian product of a classical spectrum and a semiclassical spectrum.
However, if the ATE is restricted to be time-independent, or, more accurately, if the form of the asymptotic expansion is determined by the solution of the Hamilton-Jacobi equation, then the classical part of the spectrum is eliminated and the usual JWKB semiclassical spectrum results.

In phase space, the Hamilton-Jacobi constraint implies that semiclassical KvH eigenfunctions are confined to Lagrangian submanifolds that are glued together in a manner consistent with continuity of the wavefunction and with the required boundary conditions.  
This implies that these generalized eigenfunctions are in $L^1$ rather than $L^2$.
This procedure leads to the EBK quantization conditions, which take the Maslov index into account.
In turn this implies that the semiclassical KvH eigenproblem is complete for the configuration space initial value problem, but not for the phase space initial value problem.
In order for the semiclassical version to describe tunneling effects, the domain of the definition of the KvH PDE should actually be considered to be the complexification of the classical phase space manifold.
This complex JWKB methodology allows the wavefunction to remain continuous but become evanescent in classically forbidden regions.

An interesting paradox arises for the phase space formulation in that there are two independent conserved current densities.
We have proven that, while the two forms do not necessarily agree for arbitrary solutions of the phase space KvH equation, they do agree for the semiclassical solutions of the KvH equation, precisely because they satisfy the Hamiton-Jacobi constraint.
Hence, for configuration space solutions, there is perfect agreement between both densities in configuration space.

While the configuration space version only requires $\dim$ dimensions for its description, it is highly nonlinear and the solutions suffer from weak singularities in the vicinity of caustics.
Hence, although the phase space version requires $2\dim$ dimensions, this formulation may be desirable in many circumstances because it is linear and the solutions are well-behaved near caustics. 
For example, since a quantum computer has exponential memory resources and can easily simulate sparse linear equations, it would be more desirable to use the phase space formulation than the configuration space formulation.
Nonetheless, since the phase space KvH eigenstates are singular, stating a precise quantum algorithm for solving the semiclassical KvH eigenproblem requires further consideration.

Although KvH wavefunctions are capable of describing interference effects, if all observables are treated as being local in phase space, then the predictions of the KvH wavefunction are the same as that of the diagonal part of the density matrix in phase space, i.e. the phase space probability distribution function.
In order to observe interference between superpositions of states, then at least some observables must be treated as nonlocal.
This will occur naturally if one computes to higher order in the asymptotic expansion.

If one uses the classical phase space KvH equation in the limit $\hbar\rightarrow 0^+$, then the spectrum will become continuous, the Keller-Maslov correction will become negligible, and tunneling will become infinitely rapid.
In this sense, the KvH equation is the classical limit of the Schro\"dinger equation. 
However, in order to obtain quantitatively accurate semiclassical results, one should use a formulation based on the Hamilton-Jacobi equation that is consistent with semiclassical theory.

\appendix

\section{Derivation of Semiclassical  Action \label{sec:semiclassical_derivation}}

The symmetric Weyl operator ordering prescription \cite{Weyl1927zp} is often used as a key step in making a quantitative connection between quantum and classical mechanics.
This is important for the phase space interpretation of quantum mechanics as well as for deformation quantization \cite{BatesBook, HallBook}.
However, for our purposes, it is sufficient to calculate the action of the Hamiltonian on the asymptotic perturbation expansion of the wavefunction to order $\CO(\epsilon)$.

Assume that the Hamiltonian operator, which is assumed to be Hermitian, is given by a sum of terms of the form
\begin{align}
\HamOp(\tcoord,\xOp,\pOp)=\tfrac{1}{2} \sum_\mindex A_\mindex(\tcoord,\xOp)(\epsilon \pOp)^\mindex B_\mindex(\tcoord,\xOp) +B_\mindex(\tcoord,\xOp)(\epsilon \pOp)^\mindex A_\mindex(\tcoord,\xOp) .
\end{align}
This form is Hermitian, is clearly sufficient for any dependence on $\xOp$, and is clearly sufficient for approximating any desired dependence on $\pOp$, including operator ordering.
Because derivatives are assumed to be weak, they are ordered as $ \epsilon \phat $ in the Hamiltonian.
The leading order term in the asymptotic expansion for $\HamOp \waveFunc$ simply replaces the  operators with their classical commuting symbols: $\xOp\rightarrow \xcoord $ and $\pOp\rightarrow \pcoord=\partial_\xcoord \WavePhase$.
This yields the  leading term
 \begin{align}
\Ham(\tcoord,\xcoord,\pcoord)= \sum_\mindex A_\mindex(\tcoord,\xcoord) B_\mindex(\tcoord,\xcoord)  \pcoord^\mindex  
\end{align}
which is referred to as the ``classical Hamiltonian'' in the main text.

For the next to leading order term, one of the derivatives in $\pOp^\mindex$ will act on one of the other functions of $\xcoord$, including $A_\mindex$, $B_\mindex$, $\WaveAmp$, and $\pcoord=\partial_\xcoord \WavePhase$ itself.
The effect of the derivatives on the product of functions can be determined by using the Liebniz property of the derivative.
The Fa\'a di Bruno formula can be used to systematically determine the effect on the exponential function of the phase. 
For example, one finds that
\begin{align}
\waveFunc^{-1} (\epsilon\phat)^\mindex \waveFunc=\pcoord^\mindex +\epsilon \mindex \pcoord^{\mindex-1} \phat   \log{\WaveAmp} + \tfrac{1}{2}\epsilon  \mindex (\mindex-1)\pcoord^{\mindex-2} \phat\pcoord+\dots
\end{align}
and, hence, that, for each term in the Hamiltonian,
\begin{multline}
\tfrac{1}{2} \waveFunc^{-1} \left[A_\mindex (\epsilon\phat)^\mindex B_\mindex+B_\mindex (\epsilon\phat)^\mindex A_\mindex \right]  \waveFunc =
\\ A_\mindex B_\mindex\pcoord^\mindex 
+ \tfrac{1}{2}\mindex \WaveAmp^{-2}(-i\epsilon\hbar\partial_\xcoord)\cdot  \left(\WaveAmp^2    A_\mindex B_\mindex  \pcoord^{\mindex-1} \right)+\dots.
\end{multline}
To first order, the asymptotic series for the Hamiltonian simply yields
\begin{align}
\waveFunc^* \HamOp(\tcoord,\xOp,\pOp)\waveFunc =\abs{\waveFunc}^2\left[  \Ham-\tfrac{1}{2}i\epsilon\hbar  \WaveAmp^{-2}\partial_\xcoord\cdot  \left(\WaveAmp^2 \partial_\pcoord\Ham \right) +\dots  \right].
\end{align}
Thus, the asymptotic series for the Schr\"odinger equation 
\begin{multline}
0=\waveFunc^* \left[\HamOp(\tcoord,\xOp,\pOp)-i\hbar\partial_\tcoord\right]\waveFunc \\
=\abs{\waveFunc}^2\left\{\left[\partial_\tcoord\WavePhase+\Ham\right]-\tfrac{1}{2}i\epsilon\hbar \WaveAmp^{-2} \left[\partial_\tcoord \WaveAmp^2+ \partial_\xcoord\cdot  \left(\WaveAmp^2 \partial_\pcoord\Ham \right)\right] +\dots  \right\}
\end{multline} 
leads to
the Hamilton-Jacobi equation (HJE) at zeroth order and to the amplitude transport equation (ATE) at first order.
Note that  this derivation holds for any complex $\WavePhase$ and $\WaveAmp$ that satisfy the ordering assumptions. 
In particular, it does not require the assumption that $\WavePhase$ and $\WaveAmp$ are real. 

In order to derive the variational principle, we now assume that  $\WavePhase$ and $\WaveAmp$ are real.
When this form for the Hamiltonian is inserted into the variational principle for the original PDE Eq.~\ref{eq:quantum-action-integral} the result is 
\begin{align}
\WaveActionPrinciple[\WavePhase,\WaveAmp]=\int \left\{-\WaveAmp^2\left[\partial_\tcoord\WavePhase+\Ham\right] +\tfrac{1}{2}i\epsilon\hbar \left[\partial_\tcoord \WaveAmp^2+ \partial_\xcoord\cdot  \left(\WaveAmp^2 \partial_\pcoord\Ham \right)\right] \right\} d^\dim\xcoord  d\tcoord.
\end{align}
Because the first order term is a space-time divergence, for suitable boundary conditions, e.g. periodic or $\WaveAmp\rightarrow 0$ at infinity, it does not contribute to the  variational principle.
Moreover, since the first order term is precisely the amplitude transport equation (ATE), this term vanishes identically for asymptotic solutions.
Thus, the variational principle reduces to Eq.~\ref{eq:config-space_action}.
  An interesting aspect of this variational principle is that it is the variation of the zeroth order term, which is the HJE, that must give rise to the first order term given by the ATE.
 
\section{Square Root of the Dirac Delta Function \label{sec:square_root_delta}}

In order to define a square integrable eigenfunction that only has support on a Lagrangian submanifold, one needs to generalize the definition of distributions so that the ``square root of the delta function'' has meaning. This is simply a technical device for describing a narrow ``bump-like'' eigenfunction that can be normalized. 
The square root of the delta function is useful whenever one would like to map from a generalized eigenfunction in the continuous spectrum to a square-integrable eigenfunction in the discrete spectrum.
This is because it maps the standard orthonormality relation for the continuous spectrum to the orthonormality relation for the discrete spectrum.  

The Dirac delta function \cite{Dirac1927prsa} is so singular that it took quite some time for the Schwartz theory of distributions \cite{SchwartzBook} to put this important concept on solid mathematical footing.
Nonetheless, the great utility of this idea was immediately apparent and has given it a ubiquitous role in modern mathematical physics.
Unfortunately, it is well known that there are paradoxes that arise if one attempts to develop an algebra of distributions that allows multiplication or composition of distributions.

Egorov~\cite{Egorov1990rms} developed a simple and consistent theory of generalized functions that meets the needs of our purposes here.
Rather than taking a limit that is potentially ill-defined, Egorov defines a generalized function as an equivalence class of sequences of functions $\setlist{\func_\kindex}$ for $\kindex \in \Naturals$. 
The strong equivalence relation, $\setlist{\func_\kindex}=\setlist{\func'_\kindex}$, is defined by the requirement that there exists an $\kindex_* \in \Naturals$ such that $\func_\kindex=\func'_\kindex$ for $\kindex \geq \kindex_*$. 
He proves that generalized functions can be differentiated, integrated, composed, etc.
A weak equivalence relation between generalized functions, $\setlist{\func_\kindex} \approx \setlist{\func'_\kindex}$, is defined as $\lim_{\kindex\rightarrow \infty} (\func_\kindex -\func'_\kindex)=0$. 

Using the notion of weak equality one can define a family of generalized functions, $w^{a}(x)$, parameterized by $a\in(0,1)$, that satisfy
\begin{align}
w^{a} (x)&\approx 0 & w^{a} (x) w^{1-a}(x) \approx \delta(x)
.
\end{align}
Because composition of generalized functions is well-defined, these functions are simply defined as taking the $a$th power of any sequence of functions that is weakly equivalent to the delta function.
If the sequence $\setlist{w_\kindex}$ is weakly equivalent to the delta function, $\delta(x)$, then sequence $\setlist{w_\kindex^a}$, i.e. $w_\kindex$ to the $a$th power, is weakly equivalent to the desired solution, which we denote $\delta^a(x)$.
In particular, the square root of the delta function $\delta^{1/2}(x)$ is defined in this manner.

The delta function itself can be defined as the sequence $\setlist{\Delta_\kindex}$ where
\begin{align}
\Delta_{\kindex}(\xcoord) =  
\kindex  \Theta(\abs{\xcoord}- 1/2\kindex) 
\end{align}
where $\Theta(\xcoord)$ is the unit step function.
Taking a power of the delta function then yields the sequence $\setlist{\Delta_\kindex^a}$ 
\begin{align}
 \Delta^a_{\kindex}(\xcoord) 
&= \kindex^a   \Theta^a(\abs{\xcoord}- 1/2\kindex) = \kindex^a   \Theta(\abs{\xcoord}- 1/2\kindex) .
\end{align}
Note that when $a\in(0,1)$, \emph{these functions are less singular than the Dirac delta function} because their height grows as $k^a<k$.
In particular, the square root of the delta function is the sequence $\setlist{\Delta_\kindex^{1/2}}$
\begin{align}
\Delta_\kindex^{1/2}(\xcoord)
&= \kindex^{1/2}  \Theta(\abs{\xcoord}- 1/2\kindex) .
\end{align}
However, this sequence of functions is not differentiable.

A differentiable model for the delta function is given by the sequence of Gaussian functions $\setlist{\Gaussian_\kindex}$
\begin{align}
\Gaussian_{\kcoord} (\xcoord):=\kcoord e^{-(\kcoord \xcoord)^2/2 }/(2\pi )^{1/2} .
\end{align}
Powers of the delta function are
simply defined as the sequence $\setlist{\Gaussian^a_\kindex(x)}$ where 
\begin{align}
\Gaussian^a_\kindex(x) =  \kindex^a e^{-a(\kcoord\xcoord)^2/2}/(2\pi  )^{a/2}.
\end{align}
In particular, the square root of the delta function is defined by the sequence $\setlist{\Gaussian^{1/2}_\kindex(x)}$ where
\begin{align}
\Gaussian^{1/2}_\kindex(x)   =  \kindex^{1/2} e^{-(\kcoord\xcoord)^2/4} /(2\pi )^{1/4}.
\end{align}
This model of the square root of the delta function is simply the ground state of the harmonic oscillator.

A useful consequence of these definitions is that products of powers of delta functions can be treated in an intuitive manner, because, for each term in the sequence  
\begin{align}
 w^a_{\kindex} (\xcoord) w^b_{\kindex} (\xcoord-\ycoord)
 =  w^{a+b} _{\kindex} (\xcoord)   \delta_{\xcoord,\ycoord},
\end{align}
where $\delta_{\xcoord,\ycoord}$ is the Kronecker delta function, as long as $0<a+b<1$.
This implies the weak equivalence relation
\begin{align} \label{eq:delta_product}
 \delta^a(\xcoord) \delta^b(\xcoord-\ycoord) \approx \delta^{a+b} (\xcoord)\delta_{\xcoord,\ycoord} .
\end{align}
Well-defined expressions occur when the sum satisfies $a+b=1$ because, then, the right hand side is simply the Dirac delta function.
The same is true for (countable) products of delta functions whose powers lie between zero and one, iff the powers sum to unity.
Rigorously speaking, such expressions must always yield a full delta function that resides within an integrand; i.e. they are only rigorous after the full delta function is integrated over.

For each term in the sequence modeling $\delta^a(\xcoord)$, integration of a product of a function times a positive power of a delta function over a domain that contains the marked point produces a result
\begin{align}
 \int   \pdf(\xcoord) w^a_{\kindex}  (\xcoord-\ycoord) \del \xcoord \propto \pdf(\ycoord)  \kindex^{a-1 } 
\end{align}
 that is singular as $\kindex\rightarrow \infty$ unless $\alpha=1$.
This illustrates the need to introduce the notion of weak equality and shows that one needs to manipulate such  expressions with care.
Using Eq.~\ref{eq:delta_product} for $a+b=1$ yields the useful result
\begin{align}
\int \pdf(\xcoord) \delta^a(\xcoord-\ycoord) \delta^{1-a}(\xcoord-\zcoord) \del\xcoord &\approx \pdf(\ycoord)\delta_{\ycoord,\zcoord}.
\end{align}
For example, the square root of the delta function satisfies the discrete orthonormality relation
\begin{align}
\int  \delta^{1/2}(\xcoord-\ycoord) \delta^{1/2}(\xcoord-\zcoord) \del\xcoord &\approx \int  \delta(\xcoord-\ycoord)\delta_{\ycoord,\zcoord}  \del\xcoord \approx \delta_{\ycoord,\zcoord}
\end{align}
 discussed at the beginning of this section.

Now let us explore the multidimensional case.
Assume that a set of coordinates evolves via the rule $\zcoord=\zeta(\tcoord; \tcoord_0, \zcoord_0)$ and the initial coordinates can be found using the inverse function $\zcoord_0=\zeta^{-1}(\tcoord_0; \tcoord, \zcoord)$.
A delta function that behaves as a scalar function is a function of $\zcoord_0$
\begin{align}
s_{\zcoord_0} (\tcoord,\zcoord) = \delta^\dim(\zcoord_0-\zeta^{-1}(\tcoord_0; \tcoord, \zcoord)) =  \delta^\dim \left(\zcoord-\zeta(\tcoord; \tcoord_0, \zcoord_0))\right)  \det \partial \zcoord/\partial \zcoord_0
\end{align}
because this is only guaranteed to integrate to unity in the initial coordinate system.
A delta function that behaves as a volume form is a function of $\zcoord$
\begin{align}
\pdf_{\tcoord_0,\zcoord_0}(\tcoord,\zcoord) = \delta^\dim \left(\zcoord-\zeta(\tcoord; \tcoord_0, \zcoord_0))\right)= \delta^\dim (\zcoord_0-\zeta^{-1}(\tcoord_0; \tcoord, \zcoord)) \det \partial \zcoord_0/\partial \zcoord.
\end{align}
because this integrates to unity in the final coordinate system. 
One can generate a delta function that behaves as a fractional power of a volume form by multiplying by a fractional power of the determinant 
\begin{align}
\pdf^{(a)}_{\tcoord_0,\zcoord_0}(\tcoord,\zcoord) = 
 \delta^\dim (\zcoord_0-\zeta^{-1}(\tcoord_0; \tcoord, \zcoord)) \left(\det \partial \zcoord_0/\partial \zcoord\right)^a
= \delta^\dim \left(\zcoord-\zeta(\tcoord; \tcoord_0, \zcoord_0))\right) (\det \partial \zcoord/\partial \zcoord_0)^{1-a}.
\end{align}
However, the product of such functions is typically quite singular.

The fractional powers of a delta function also behave as a fractional power of a volume form
\begin{align}
\pdf^a_{\tcoord_0,\zcoord_0}(\tcoord,\zcoord) 
=\left[ \delta^\dim\left(\zcoord-\zeta(\tcoord; \tcoord_0, \zcoord_0))\right)\right]^a 
= \left[\delta^\dim(\zcoord_0-\zeta^{-1}(\tcoord_0; \tcoord, \zcoord))\right] ^a
\left(\det \partial \zcoord_0/\partial \zcoord \right)^a.
\end{align}
The complex phase of this expression, involving a fractional power, has yet to be specified.
For fractional powers of the determinant, one must make a consistent choice of the branch of the complex phase factor.  

If the function space is real, then one can follow the standard convention of using the absolute value, because then, multiplying the fractional powers  of a delta function for  $a$ and $1-a$ generates a true volume form.
 The same will be true for a product of delta functions whose powers sum to one.  Then, the product yields a delta function that will be integrated over with the understanding that the region of integration will have fixed orientation; i.e. a change in sign of the determinant reverses the orientation and the two sign changes cancel. 

For a complex Hilbert space, with $a=1/2$, one can choose either branch of the square root $\left( \det \partial \zcoord_0/\partial \zcoord\right)^{1/2}$, because $\abs{\pdf^{1/2}_{\tcoord_0,\zcoord_0}}^2 $ yields a full delta function.
This is useful for understanding how the Koopman-van Hove equation should be applied to complex analytic Hilbert spaces (see Appendix B of \cite{Joseph23pop}).
Perhaps the notion of $\delta^{1/p}(x)$ could play a useful role in defining the limit of localized yet integrable solutions within $L^p$ spaces.

\section{Yet Another Mapping from  Phase Space to Configuration Space
\label{sec:alternate_mapping_norm}
}
\def\nuindex{\nu}

Is it possible to directly map  the  KvH eigenfunctions in phase space to  configuration space in a manner that makes the \emph{ phase space and configuration space inner products} agree with each other?
In phase space, the orthogonality and completeness relations for $\Jcoord$ always have the form for a continuous spectrum, rather than the form for a discrete spectrum.
In this section, we will show that one can generate the desired orthonormality and completeness relations by using the square root of the delta function, defined in 
~\ref{sec:square_root_delta}, for the dependence of the basis functions on $\Jcoord$.

Since the solution of the KvH equation is meant to represent the square root of a volume form, it may not be surprising that the square root of the delta function would be useful for defining the generalized eigenfunctions.
Because the square root of the delta function satisfies
\begin{align}
\int   \delta^{1/2}(\Jcoord-\Jcoord_0)\delta^{1/2}(\Jcoord-\Jcoord_1) d\Jcoord 
= \int  \delta(\Jcoord-\Jcoord_0)\delta_{\Jcoord_0,\Jcoord_1}d\Jcoord 
= \delta_{\Jcoord_0,\Jcoord_1}
\end{align} 
this definition will yield the discrete orthonormality and completeness relations for $\Jcoord$.
Moreover, when converting the delta function from $\Jcoord$ to $\pcoord$, then the result carries the desired configuration space weight factor 
\begin{align}
 \delta^{\dim/2}(\Jcoord-\Jcoord_0)=\sum_\sigma \delta^{\dim/2}(\pcoord-\partial_\xcoord\WavePhase) \left(  \det \left. \partial \pcoord/\partial \Jcoord \right|_\xcoord  \right)^{1/2}.
\end{align}
If one is willing to accept this formulation, then the combination of the standard phase space KvH equation with the correct JWKB matching conditions will yield the semiclassical spectrum.

The phase space orthonormality relation in Eq.~\ref{eq:phase-space_orthonormality} yields
\begin{align}
\BraKet{\BasisFunc_{\Jcoord_0,\nuindex_0}}{\BasisFunc_{\Jcoord_1, \nuindex_1}} &=  
\int e^{i(\nu_1-\nu_0) \cdot \angle + i(\WavePhaseEig( \tcoord,\xcoord,\Jcoord_1)-\WavePhaseEig( \tcoord,\xcoord,\Jcoord_0))/\hbar} 
\delta^\dim(\Jcoord-\Jcoord_0)\delta^\dim(\Jcoord-\Jcoord_1) d^\dim \Jcoord d^\dim \angle/(2\pi )^{\dim} 
\\
&=  \delta^\dim(\Jcoord_1-\Jcoord_0) 
\int e^{i(\nu_1-\nu_0) \cdot \angle}d^\dim \angle/(2\pi )^{\dim} = \delta^\dim(\Jcoord_1-\Jcoord_0) \delta^\dim_{ \nu_1-\nu_0}
\end{align}
where for each $\nu_i$, $\delta_{ \nu_{1}-\nu_{0}}$ is either $\delta_{ \nu_{1},\nu_{0}}$ or $\delta( \nu_{1}-\nu_{0})$ depending on whether the range of $\theta_i$ is finite or infinite.
This result is too singular to match the configuration space inner product for both the discrete and the continuous spectrum.
Instead, if one defines the phase space eigenfunctions as
\begin{align}
\BasisFunc_{\Jcoord_0,\nuindex} (\tcoord,\zcoord)= e^{i\nuindex \cdot\theta_0+i\WavePhaseEig(\tcoord,\xcoord,\Jcoord )/\hbar} \left[ \delta^\dim\left(\Jcoord- \Jcoord_0\right) \right]^{1/2} /(2\pi )^{\dim/2} .
\end{align} 
then the phase space orthonormality relation becomes
\begin{align}
\BraKet{\BasisFunc_{\Jcoord_0,\nuindex_0}}{\BasisFunc_{\Jcoord_1, \nuindex_1}} 
&=  
\int e^{i(\nu_1-\nu_0) \cdot \angle + i\left(\WavePhaseEig (\tcoord,\xcoord,\Jcoord_1)-\WavePhaseEig (\tcoord,\xcoord,\Jcoord_0)\right)/\hbar} \left[\delta^\dim(\Jcoord-\Jcoord_0)\delta^\dim(\Jcoord-\Jcoord_1) \right]^{1/2} d^\dim \Jcoord d^\dim \angle /(2\pi)^\dim
\\
&=  
\int e^{i(\nu_1-\nu_0) \cdot \angle }  \delta^\dim(\Jcoord-\Jcoord_0)   \delta^\dim_{\Jcoord_1,\Jcoord_0}  d^\dim \Jcoord d^\dim \angle /(2\pi)^\dim
 \\
&=  \delta^\dim_{\Jcoord_1,\Jcoord_0}
\int e^{i(\nu_1-\nu_0) \cdot \angle}d^\dim \angle  /(2\pi)^\dim
= \delta^\dim_{\Jcoord_1,\Jcoord_0} \delta^\dim_{ \nu_1-\nu_0} =\delta^\dim_{\Jcoord_1-\Jcoord_0} \delta^\dim_{ \nu_1,\nu_0}
\end{align}
where for each coordinate $\Jcoord_i$, $\delta_{\Jcoord_1-\Jcoord_0} $ is either $ \delta_{\Jcoord_1,\Jcoord_0} $ or $ \delta {\left((\Jcoord_1-\Jcoord_0)/\hbar\right)} $ depending on whether the range of $\angle_i$ is finite or infinite.
This matches the result of the configuration space inner product.
These basis functions are complete in $\angle$ because of the $\nuindex$ dependence, but completeness in $\Jcoord$ results in the wrong normalization ($\delta^\dim_{\Jcoord_1,\Jcoord_0}$ instead of $\delta^\dim(\Jcoord_1-\Jcoord_0)$).

With these definitions, one can directly define a projection to configuration  
space eigenfunctions by integrating over momentum space with another square root of the delta function
\begin{align} \label{eq:standard_KvH_projection}
 \basisFunc_{\Jcoord_0,\nuindex} (\tcoord,\xcoord)=
 \sum_\sigma \int   \BasisFunc_{\Jcoord_0,\nuindex} (\tcoord,\zcoord)  \left[\delta^\dim\left(\pcoord- \partial_\xcoord \WavePhase_\sigma \right)\right]^{1/2} d^\dim\pcoord.
\end{align}
This is the equivalent of the mapping in Eq.~\ref{eq:KvH_semiclassical_projection} and \ref{eq:KvH_phase-space_projection1}.
To obtain semiclassical eigenstates, one must restrict the eigenfunctions to the $\nu=0$ sector.
If one now defines the phase space eigenfunctions
\begin{align}
\BasisFunc_{\Jcoord_0} (\tcoord,\zcoord)= e^{i\WavePhase (\tcoord,\xcoord,\Jcoord )/\hbar} \left[ \delta^\dim\left(\Jcoord- \Jcoord_0\right) \right]^{1/2} /(2\pi )^{\dim/2} 
\end{align} 
then the orthonormality relation becomes
\begin{align}
\BraKet{\BasisFunc_{\Jcoord_0}}{\BasisFunc_{\Jcoord_1}} 
&=  \delta^\dim_{\Jcoord_1,\Jcoord_0}
\int  d^\dim \angle /(2\pi)^\dim
= \delta^\dim_{\Jcoord_1,\Jcoord_0} \delta^\dim_{ 0} =  \delta^\dim_{ \Jcoord_1-\Jcoord_0}.
\end{align}
These eigenfunctions are complete in $\angle$ because
\begin{align}
\int \BasisFunc^*_{\Jcoord}(\zcoord_0) \BasisFunc_{\Jcoord}(\zcoord_1) d^\dim \pcoord 
&=   
\int e^{i(\WavePhase(\tcoord,\xcoord_1,\Jcoord)-{\WavePhase(\tcoord,\xcoord_0,\Jcoord)}/\hbar}  \delta^\dim(\Jcoord-\Jcoord_1) \delta^\dim_{\Jcoord_1,\Jcoord_0}  d^\dim \pcoord /(2\pi)^\dim
\\
&=   \delta^\dim_{\Jcoord_1,\Jcoord_0}
 e^{i\left(\WavePhase (\tcoord,\xcoord_1,\Jcoord_1)-\WavePhase (\tcoord,\xcoord_0,\Jcoord_0)\right)/\hbar}   \abs{ \det{  \partial \angle/\partial \xcoord }}/(2\pi)^\dim
\end{align}
matches the configuration space result within the stationary phase approximation.

\section*{Acknowledgements}
The author would like to thank J. L. DuBois, V. I. Geyko, F. R. Graziani, S. B. Libby, R. Minich, M. D. Porter,  C. Tronci, and Y. Shi for interesting discussions on this topic. This work, LLNL-JRNL-840538, was performed under the auspices of the U.S. DOE by LLNL under Contract DE-AC52-07NA27344 and was supported by the DOE Office of Fusion Energy Sciences ``Quantum Leap for Fusion Energy Sciences" project FWP-SCW1680.

\section*{References}
\bibliographystyle{iopart-num}
\bibliography{Quantum-Bibliography.bib}{}


\end{document}